\documentclass[aps,prl,reprint,superscriptaddress,amsmath,amssymb,floatfix,nofootinbib]{revtex4-2}
\usepackage{graphicx,booktabs,array,xcolor,bm}
\usepackage{microtype}
\usepackage[colorlinks=true,linkcolor=blue!50!black,citecolor=green!40!black,urlcolor=blue!60!black]{hyperref}
\newcommand{\kB}{k_{\mathrm B}}
\newcommand{\avg}[1]{\left\langle #1\right\rangle}
\makeatletter
\newcommand{\SIr}[1]{\@ifundefined{SIr@#1}{??}{\csname SIr@#1\endcsname}}
\makeatother
\expandafter\gdef\csname SIr@sec:setting\endcsname{1}
\expandafter\gdef\csname SIr@tab:notation\endcsname{1}
\expandafter\gdef\csname SIr@eq:Z\endcsname{1.1}
\expandafter\gdef\csname SIr@sec:thermo\endcsname{2}
\expandafter\gdef\csname SIr@prop:thermo\endcsname{2.1}
\expandafter\gdef\csname SIr@eq:E\endcsname{2.1}
\expandafter\gdef\csname SIr@eq:P\endcsname{2.2}
\expandafter\gdef\csname SIr@prop:virial\endcsname{2.2}
\expandafter\gdef\csname SIr@eq:virial\endcsname{2.3}
\expandafter\gdef\csname SIr@sec:tt\endcsname{3}
\expandafter\gdef\csname SIr@eq:list\endcsname{3.1}
\expandafter\gdef\csname SIr@sec:ttmap\endcsname{3.1}
\expandafter\gdef\csname SIr@eq:coordchain\endcsname{3.2}
\expandafter\gdef\csname SIr@tab:statements\endcsname{2}
\expandafter\gdef\csname SIr@eq:T0\endcsname{3.3}
\expandafter\gdef\csname SIr@eq:contractW\endcsname{3.4}
\expandafter\gdef\csname SIr@eq:T1\endcsname{3.5}
\expandafter\gdef\csname SIr@eq:T2\endcsname{3.6}
\expandafter\gdef\csname SIr@eq:T2tt\endcsname{3.7}
\expandafter\gdef\csname SIr@eq:T3\endcsname{3.8}
\expandafter\gdef\csname SIr@eq:ttU\endcsname{3.9}
\expandafter\gdef\csname SIr@eq:ttP\endcsname{3.10}
\expandafter\gdef\csname SIr@eq:ttPhi\endcsname{3.11}
\expandafter\gdef\csname SIr@eq:ttPsi\endcsname{3.12}
\expandafter\gdef\csname SIr@eq:ttF\endcsname{3.13}
\expandafter\gdef\csname SIr@eq:ttZ\endcsname{3.14}
\expandafter\gdef\csname SIr@tab:tensors\endcsname{3}
\expandafter\gdef\csname SIr@lem:smooth\endcsname{3.1}
\expandafter\gdef\csname SIr@eq:Ubar\endcsname{3.15}
\expandafter\gdef\csname SIr@eq:smoothderiv\endcsname{3.16}
\expandafter\gdef\csname SIr@lem:whiten\endcsname{3.2}
\expandafter\gdef\csname SIr@eq:whitescha\endcsname{3.17}
\expandafter\gdef\csname SIr@eq:update\endcsname{3.18}
\expandafter\gdef\csname SIr@eq:Xhermite\endcsname{3.19}
\expandafter\gdef\csname SIr@lem:wick\endcsname{3.3}
\expandafter\gdef\csname SIr@eq:wick\endcsname{3.20}
\expandafter\gdef\csname SIr@lem:marginal\endcsname{3.4}
\expandafter\gdef\csname SIr@eq:marginal\endcsname{3.21}
\expandafter\gdef\csname SIr@eq:eam\endcsname{3.22}
\expandafter\gdef\csname SIr@lem:split\endcsname{3.5}
\expandafter\gdef\csname SIr@eq:split3\endcsname{3.23}
\expandafter\gdef\csname SIr@eq:remsmall\endcsname{3.24}
\expandafter\gdef\csname SIr@prop:pairrule\endcsname{3.6}
\expandafter\gdef\csname SIr@eq:pair2\endcsname{3.25}
\expandafter\gdef\csname SIr@eq:pair3\endcsname{3.26}
\expandafter\gdef\csname SIr@eq:pairV\endcsname{3.27}
\expandafter\gdef\csname SIr@lem:orbit\endcsname{3.8}
\expandafter\gdef\csname SIr@eq:assemble\endcsname{3.28}
\expandafter\gdef\csname SIr@lem:sepsum\endcsname{3.9}
\expandafter\gdef\csname SIr@eq:binomcore\endcsname{3.29}
\expandafter\gdef\csname SIr@prop:adapted\endcsname{3.10}
\expandafter\gdef\csname SIr@eq:adapted\endcsname{3.30}
\expandafter\gdef\csname SIr@prop:weighted\endcsname{3.11}
\expandafter\gdef\csname SIr@eq:weighted\endcsname{3.31}
\expandafter\gdef\csname SIr@eq:momerr\endcsname{3.32}
\expandafter\gdef\csname SIr@lem:contract\endcsname{3.12}
\expandafter\gdef\csname SIr@alg:main\endcsname{3.5}
\expandafter\gdef\csname SIr@thm:budget\endcsname{3.13}
\expandafter\gdef\csname SIr@eq:fperr\endcsname{3.33}
\expandafter\gdef\csname SIr@sec:results\endcsname{4}
\expandafter\gdef\csname SIr@tab:systems\endcsname{4}
\expandafter\gdef\csname SIr@sec:validation\endcsname{4.2}
\expandafter\gdef\csname SIr@sec:stationarity\endcsname{4.3}
\expandafter\gdef\csname SIr@fig:conv\endcsname{1}
\expandafter\gdef\csname SIr@tab:val\endcsname{5}
\expandafter\gdef\csname SIr@fig:importance\endcsname{2}
\expandafter\gdef\csname SIr@tab:ng\endcsname{6}
\expandafter\gdef\csname SIr@tab:rank\endcsname{7}
\expandafter\gdef\csname SIr@tab:NiAl\endcsname{8}
\expandafter\gdef\csname SIr@tab:TiAl\endcsname{9}
\expandafter\gdef\csname SIr@tab:diag\endcsname{10}
\expandafter\gdef\csname SIr@sec:T3\endcsname{4.5}
\expandafter\gdef\csname SIr@fig:dev\endcsname{3}
\expandafter\gdef\csname SIr@fig:bubble\endcsname{4}
\expandafter\gdef\csname SIr@tab:t3\endcsname{11}
\expandafter\gdef\csname SIr@fig:t3\endcsname{5}
\expandafter\gdef\csname SIr@tab:t3p\endcsname{12}
\expandafter\gdef\csname SIr@tab:ratio\endcsname{13}
\expandafter\gdef\csname SIr@sec:spectrum\endcsname{4.6}
\expandafter\gdef\csname SIr@fig:eig\endcsname{6}
\expandafter\gdef\csname SIr@tab:t3free\endcsname{14}
\expandafter\gdef\csname SIr@tab:t3small\endcsname{15}
\expandafter\gdef\csname SIr@tab:eig\endcsname{16}
\expandafter\gdef\csname SIr@sec:ti\endcsname{5}
\expandafter\gdef\csname SIr@tab:tinodes\endcsname{17}
\expandafter\gdef\csname SIr@tab:ticost\endcsname{18}
\expandafter\gdef\csname SIr@tab:harm\endcsname{19}
\expandafter\gdef\csname SIr@tab:tirerun\endcsname{20}
\expandafter\gdef\csname SIr@sec:sizedense\endcsname{5.2}
\expandafter\gdef\csname SIr@tab:mdsize\endcsname{21}
\expandafter\gdef\csname SIr@fig:ttsize\endcsname{7}
\expandafter\gdef\csname SIr@sec:cellsize\endcsname{5.3}
\expandafter\gdef\csname SIr@tab:local\endcsname{22}
\expandafter\gdef\csname SIr@fig:ttlocal\endcsname{8}
\expandafter\gdef\csname SIr@sec:loc3\endcsname{5.4}
\expandafter\gdef\csname SIr@tab:loc3\endcsname{23}
\expandafter\gdef\csname SIr@sec:fast\endcsname{5.5}
\expandafter\gdef\csname SIr@tab:fast\endcsname{24}
\expandafter\gdef\csname SIr@sec:md65k\endcsname{6}
\expandafter\gdef\csname SIr@tab:md65k\endcsname{25}
\expandafter\gdef\csname SIr@fig:m65\endcsname{9}
\expandafter\gdef\csname SIr@tab:pivot\endcsname{26}
\expandafter\gdef\csname SIr@tab:sweep\endcsname{27}
\expandafter\gdef\csname SIr@tab:rankcap\endcsname{28}
\expandafter\gdef\csname SIr@tab:seedmean\endcsname{29}
\expandafter\gdef\csname SIr@tab:pnext\endcsname{30}
\expandafter\gdef\csname SIr@tab:thermNiAl\endcsname{31}
\expandafter\gdef\csname SIr@tab:thermTiAl\endcsname{32}
\expandafter\gdef\csname SIr@tab:elstat\endcsname{33}
\expandafter\gdef\csname SIr@tab:elNiAl\endcsname{34}
\expandafter\gdef\csname SIr@tab:elTiAl\endcsname{35}
\expandafter\gdef\csname SIr@tab:elB\endcsname{36}
\expandafter\gdef\csname SIr@tab:elpot\endcsname{37}
\expandafter\gdef\csname SIr@sec:bulk\endcsname{7}
\expandafter\gdef\csname SIr@tab:bulk\endcsname{38}
\expandafter\gdef\csname SIr@tab:tial108\endcsname{39}
\expandafter\gdef\csname SIr@app:A\endcsname{8}
\expandafter\gdef\csname SIr@eq:A11dir\endcsname{8.1}
\expandafter\gdef\csname SIr@eq:A11emb\endcsname{8.2}
\expandafter\gdef\csname SIr@thmA:NiAl\endcsname{8.1}
\expandafter\gdef\csname SIr@eq:A11det\endcsname{8.3}
\expandafter\gdef\csname SIr@eq:A11Z\endcsname{8.4}
\expandafter\gdef\csname SIr@eq:A11bound\endcsname{8.5}
\expandafter\gdef\csname SIr@eq:A11P\endcsname{8.6}
\expandafter\gdef\csname SIr@remA:NiAl\endcsname{8.2}
\expandafter\gdef\csname SIr@app:B\endcsname{9}
\expandafter\gdef\csname SIr@secB:expansion\endcsname{9.1}
\expandafter\gdef\csname SIr@lemB:stein\endcsname{9.1}
\expandafter\gdef\csname SIr@lemB:chaos\endcsname{9.2}
\expandafter\gdef\csname SIr@thmB:GB\endcsname{9.3}
\expandafter\gdef\csname SIr@eq:B-FGB\endcsname{9.1}
\expandafter\gdef\csname SIr@thmB:scha\endcsname{9.4}
\expandafter\gdef\csname SIr@propB:HF\endcsname{9.5}
\expandafter\gdef\csname SIr@propB:W12\endcsname{9.6}
\expandafter\gdef\csname SIr@thmB:bubble\endcsname{9.7}
\expandafter\gdef\csname SIr@eq:B-Fb\endcsname{9.2}
\expandafter\gdef\csname SIr@propB:T3\endcsname{9.8}
\expandafter\gdef\csname SIr@eq:T3ab\endcsname{9.3}
\expandafter\gdef\csname SIr@eq:T3c\endcsname{9.4}
\expandafter\gdef\csname SIr@secB:zero\endcsname{9.4}
\expandafter\gdef\csname SIr@tab:zero\endcsname{40}
\expandafter\gdef\csname SIr@tab:zeroV\endcsname{41}
\expandafter\gdef\csname SIr@app:C\endcsname{10}
\expandafter\gdef\csname SIr@propC:int\endcsname{10.1}
\expandafter\gdef\csname SIr@eq:C-int\endcsname{10.1}
\expandafter\gdef\csname SIr@alg:C1\endcsname{10.2}
\expandafter\gdef\csname SIr@alg:C2\endcsname{10.3}
\expandafter\gdef\csname SIr@propC:bilinear\endcsname{10.2}
\expandafter\gdef\csname SIr@tab:Ctensors\endcsname{42}
\expandafter\gdef\csname SIr@propC:binom\endcsname{10.3}
\expandafter\gdef\csname SIr@fig:ttprofile\endcsname{10}
\expandafter\gdef\csname SIr@tab:cost\endcsname{43}
\expandafter\gdef\csname SIr@tab:code\endcsname{44}

\begin{document}

\title{TT-SCHA for Thermodynamics of Binary Alloys}

\author{Boian S. Alexandrov}
\email[Corresponding author: ]{balexan1@unm.edu}
\affiliation{Chemical and Biological Engineering, University of New Mexico, Albuquerque, NM, USA}

\author{Dimiter Petsev}
\email{dimiter@unm.edu}
\affiliation{Chemical and Biological Engineering, University of New Mexico, Albuquerque, NM, USA}


\begin{abstract}
The free energy of a crystal is an integral over the positions of all its atoms, in $3N$ dimensions. We evaluate it without random sampling for the ordered alloys B2 NiAl and $\gamma$-TiAl with embedded-atom potentials, using a self-consistent harmonic approximation (SCHA) corrected to third order in temperature, calculated via thermal averages of tensor-train (TT) integrals built from one atom and its neighbors, which we call TT-SCHA. The result is for an infinite crystal without sampling error; its numerical error is set by the starting pivots of the interpolation (up to 9\,MPa in the pressure, reduced by averaging three runs; SM Table~S\SIr{tab:pivot}). Against the molecular dynamics of 65\,536 atoms (Frenkel--Ladd free energy at 600\,K, integrated energy above), TT-SCHA free energy agrees to 0.13\,meV per atom at 600--1000\,K, using 420--500 times fewer core-seconds for one run. At zero pressure, results agree with published calculations that use the same embedded-atom potential: the published elastic constants of $\gamma$-TiAl (to 1\,GPa, by a static lattice calculation) and its thermal expansion from Monte Carlo sampling (to 0.012 percentage points).
\end{abstract}

\maketitle

\emph{Introduction: } The non-ideal thermodynamic properties of strongly interacting condensed-matter systems follow from the configurational integral $Z=\int e^{-U/\kB T}\,d^{3N}\!x$, where $U(\mathbf{x})$ is the potential energy of the $N$ atoms. The excess free energy is $F^{\mathrm{ex}}=-\kB T\ln(Z/V^N)$, and the other properties follow from its derivatives. The high dimensionality of $Z$, however, presents a challenge that is insurmountable for direct quadrature. With only ten grid points in each of the $3N$ dimensions, it would need $10^{3N}$ evaluations. MD and Monte Carlo are noisy, and a free energy requires thermodynamic integration over a chain of simulations, as in the Frenkel--Ladd method~\cite{FrenkelLadd1984}. For crystals, the alternative is the self-consistent harmonic approximation (SCHA)~\cite{Hooton1955,Werthamer1970}: coupled springs whose stiffness is the curvature of the true potential averaged over the harmonic distribution itself. Modern versions use stochastic sampling~\cite{Errea2014,Monacelli2021} or MD~\cite{Hellman2011}.

For calculation of the configurational integral without random sampling, the recent tensor train (TT) approach offers another route~\cite{Truong2025}, which we call here direct TT-cross. A TT writes a function of many variables as a product of small matrices, one per variable; their size, the rank, measures how strongly the variables are coupled, and the cost is linear in the number of variables when the rank is small. TT-cross builds the train from a few function values~\cite{OseledetsTyrtyshnikov2010,Savostyanov2011,Oseledets2011}. For crystals of identical atoms (Cu, Ar, Sn) the Boltzmann factor has rank one or two, and the integral could be computed in this way~\cite{Truong2025}. Ordered alloys do not fit that picture, because the two elements occupy interlocking sublattices with different vibrations, which the reference must follow. A direct TT-cross gives the internal energy to within 1\,meV per atom in NiAl and 3\,meV in $\gamma$-TiAl (rank one, 100--800\,K), since the energy is set by equipartition and hardly depends on the coupling between modes, but not reliably the pressure: at rank one it is off by $-6$ to $-7$\,\% (Cartesian axes) or $+14$ to $+16$\,\% (Hessian eigenbasis) for NiAl and by $+1$ to $+2$\,\% for TiAl. The pressure is the volume derivative of the curvature at thermal amplitude, so it depends on the anharmonic coupling. In tensor terms, the rank across a cut of the TT counts the independent couplings between the coordinates on its two sides. In the normal-mode basis, the anharmonic part couples every pair, about $d^2$ weak quartic terms; the Boltzmann factor is the exponential of their sum, and the rank needed grows with their number. For 54-atom NiAl at 300\,K with adaptively chosen pivots, the direct TT-cross in the Hessian eigenbasis converges with rank (pressure error $+11.5$\,\% at rank 1, $+2.0$\,\% at rank 6), but the pivot matrices become ill-conditioned (condition number $5\times10^{5}$ at rank 6), a $+2$\,\% bias remains when the pivots of the reference volume are reused, while with pivots rebuilt at each volume the pressure scatters by 4 to 15\,\% between seeds, and in Cartesian axes the train diverges at every rank (SM Tables~S\SIr{tab:zero}, S\SIr{tab:zeroV}); with fixed thermal pivots, ranks 2 to 8 gave $-3.9$ to $7.6$\,GPa against $1.96$\,GPa from MD. The cumulant expansion avoids this: each term is a Gaussian average of a derivative of order two to four, a polynomial whose rank is set by few coupling blocks, which for an embedded-atom potential involves one atom and its neighbors and is an exact contraction of a low-rank train.

Here we show that the tensor-train calculation works for ordered binary alloys, combined with a self-consistent Gaussian reference and a local formulation, which we call TT-SCHA. We report three results: (i) The absolute free energy of B2 NiAl and $\gamma$-TiAl, and the entropy, energy and pressure calculated by TT-SCHA, agree with 65\,536-atom MD with Frenkel--Ladd free energies to 0.13\,meV per atom in $F$ at 600--1000\,K, in $S$ and $U$ to 0.002\,$\kB$ and 0.23\,meV, and in the pressure to 6\,MPa ($\gamma$-TiAl) and 20\,MPa (NiAl), with no sampling. (ii) The result is the infinite crystal, free of sampling and finite-size error, because every quantity is built from one atom and its neighbors. (iii) A scan of 600--1000\,K, each temperature at three volumes, costs $0.8\times10^3$--$1.1\times10^3$ core-seconds, 420--500 times fewer than the 65\,536-atom reference (140--165 times for three starting-pivot runs). The derivations, tests and data tables are in the Supplemental Material (SM)~\cite{SM}.

\emph{Method: }Let $u\in\mathbb R^{d}$ be the displacements of the atoms from their lattice sites. A Gaussian trial potential $U_\Phi=U_0+\tfrac12u^{\mathsf T}\Phi u$ gives, through the Gibbs--Bogoliubov inequality~\cite{Bogoliubov1958}, the bound $F\le F_{\rm GB}=F_\Phi+\avg{U-U_\Phi}_\Phi$. The bound is stationary when $\Phi=\Phi^\star=\avg{\nabla^2U}_{\Phi^\star}$, which is the SCHA, and we solve it by iteration. Its covariance is the thermal ellipsoid of the single well; in coordinates aligned with it the Gaussian is a rank-one tensor train, and all trains below are built in them. The remainder is expanded in the cumulants of $W=U-U_{\Phi^\star}$,
\begin{equation}
F=F_{\rm GB}+F_b+\left(F_{4a}+F_{4b}+F_{4c}\right)+O(T^4),\label{eq:F}
\end{equation}
where $F_b=-\tfrac{\kB T}{12}\sum X^2$ is the cubic (bubble) term, $X$ being the Gaussian average of the third derivatives of $U$ in coordinates scaled by the thermal width, and $F_{4a},F_{4b},F_{4c}$ are the three terms of order $T^3$, which contain fourth derivatives and third derivatives twice or four times (SM Secs.~S\SIr{secB:expansion}, S\SIr{app:B}). Every ingredient is a Gaussian average of a derivative of $U$ of order two, three, or four. The pressure follows from $F$ at $V(1\pm1\%)$, the entropy and energy from its temperature dependence.

\emph{Why can the averages be computed deterministically: }For an embedded-atom potential~\cite{Daw1984}, $U=\tfrac12\sum_{i\neq j}\varphi(r_{ij})+\sum_iF_i(\rho_i)$ with $\rho_i=\sum_jf(r_{ij})$, the Gaussian averages split exactly (SM Lemma~S\SIr{lem:split}). The pair part involves only the three-dimensional relative displacement of two atoms, whose covariance is a $3\times3$ block of that of the Gaussian, and is a tensor-product Gauss--Hermite quadrature, 24 nodes per axis. The embedding part is a function of the displacements of the neighbors of each atom; once the pair-like part of $F_i(\rho_i)$ is removed, it has a low rank in coordinates adapted to the local electron density (rank 6, and 10 for its volume derivative), and it is built by TT-cross. The Hermite moments of a train up to fourth order, which are the Gaussian averages of its derivatives, are computed exactly from the cores in one pass, at a cost linear in the number of cores (SM Lemma~S\SIr{lem:contract}, Secs.~S\SIr{sec:tt}, S\SIr{app:C}).

\emph{Local formulation: }In a perfect crystal all atoms of one sublattice have the same environment, so we store $\Phi^\star$ as force constants between one atom of each sublattice and its neighbors within five lattice spacings, and obtain the covariance of the Gaussian by a Fourier transform over wave vectors. The embedding train is built for one atom of each sublattice. The number of atoms enters only through the wave-vector grid: the wall-clock time of the self-consistent Gaussian is 69--72\,s on one core for cells from 250 to 65\,536 atoms, against a time growing as $N^2$ for the dense formulation (SM Sec.~S\SIr{sec:ti}). The third and fourth derivatives are treated in the same way, as vertices $V(\mathbf k_1j_1,\mathbf k_2j_2,\mathbf k_3j_3)$ with $\mathbf k_1+\mathbf k_2+\mathbf k_3=0$ integrated over the Brillouin zone with Sobol nets, with TT-cross pivots reused between iterations (SM Secs.~S\SIr{sec:ti}, S\SIr{sec:bulk}). Atoms related by lattice translations have equal averages, so the number of integrals is independent of the cell size (SM Lemma~S\SIr{lem:orbit}): the result is the infinite crystal at a cost independent of $N$.

\emph{What is proved: }The SM proves the steps that are not numerical. (i) For B2 NiAl the harmonic integral factorizes over the two sublattices, $Z_{\rm h}=e^{-\beta U_0}Z_AZ_B\det(\mathbb 1-M)^{-1/2}$, where $M$ is the coupling between them (SM Theorem~S\SIr{thmA:NiAl}); the harmonic energy is then additive and the pressure is not, which isolates the coupling pressure. (ii) The embedded-atom splitting, the exact Hermite contraction, and the reduction to translation orbits used above (SM Lemmas~S\SIr{lem:split}, S\SIr{lem:contract}, S\SIr{lem:orbit}). (iii) The calculation contains no sampling error, and if the self-consistent map is a contraction with constant $L<1$ the error of its fixed point is at most that of one evaluation divided by $1-L$ (SM Theorem~S\SIr{thm:budget}). Not proved: the low rank of the remainder (checked numerically), the value of $L$ (measured, about 0.09) and the convergence of the series in $T$.

\emph{Systems and references: }We study B2 NiAl (128-atom cell, embedded-atom potential of Purja Pun and Mishin~\cite{PurjaPunMishin2009}) and $\gamma$-TiAl (L1$_0$, 108 atoms, potential of Zope and Mishin~\cite{ZopeMishin2003}) at fixed volume below melting. The references are LAMMPS~\cite{LAMMPS} NVT MD, Hamiltonian Monte Carlo of the exact Boltzmann distribution, thermodynamic integration, and, for the main comparison, MD of 65\,536 atoms with a Frenkel--Ladd free energy at 600\,K and, above it, the integrated MD energy (SM Sec.~S\SIr{sec:md65k}).

\emph{Comparison with 65\,536-atom MD: }In 65\,536 atoms the finite-size error of the sampling is negligible, so this tests the absolute free energy. On the TT side $F$ is the absolute free energy, Eq.~(\ref{eq:F}); $S=-\partial F/\partial T$ and $U=F+TS$ follow from its temperature dependence alone (Table~\ref{tab:cmp}; SM Fig.~S\SIr{fig:m65}). Values are means of three runs (SM Table~S\SIr{tab:seedmean}). The free energy agrees to 0.11--0.13\,meV per atom ($3\times10^{-5}$ of its value) at all five temperatures for both alloys. The entropy and energy agree to 0.002\,$\kB$ and 0.23\,meV for both alloys, within the limits fixed beforehand (0.005\,$\kB$, 0.5\,meV). The pressure agrees to within 6\,MPa for $\gamma$-TiAl. For NiAl it differs by $-5$ to $-20$\,MPa, growing with temperature; a rough next-order estimate reduces this to $-2$ to $-8$\,MPa, and the cause is not established.

\begin{table}[tbp]\centering\small
\caption{\textbf{Agreement with, and cost relative to, 65\,536-atom MD, 600--1000\,K (Frenkel--Ladd free energy at 600\,K).} Largest absolute difference TT$-$MD over the five temperatures, mean of three runs from different starting pivots. Cost: core-seconds for the whole range (MD: read from the LAMMPS run logs, 64 cores, including the volume run, the Frenkel--Ladd integration and the five NVT runs; TT: from the run logs, three volumes per temperature).}\label{tab:cmp}
\footnotesize\setlength{\tabcolsep}{4pt}
\begin{tabular}{@{}lcc@{}}\toprule
 & NiAl & $\gamma$-TiAl\\\midrule
$|\Delta F|$ (meV/atom) & 0.13 & 0.13\\
$|\Delta S|$ ($\kB$/atom) & 0.0012 & 0.0021\\
$|\Delta U|$ (meV/atom) & 0.23 & 0.14\\
$|\Delta P|$ (MPa) & 20 & 6\\\midrule
MD cost (core-s, from logs) & $4.63\times10^{5}$ & $4.11\times10^{5}$\\
TT cost, one run (core-s) & $1.11\times10^{3}$ & $0.83\times10^{3}$\\
ratio, one run & 420 & 500\\
ratio, three runs & 140 & 165\\
\bottomrule
\end{tabular}
\end{table}

\begin{figure*}[tbp]\centering
\includegraphics[width=0.68\textwidth]{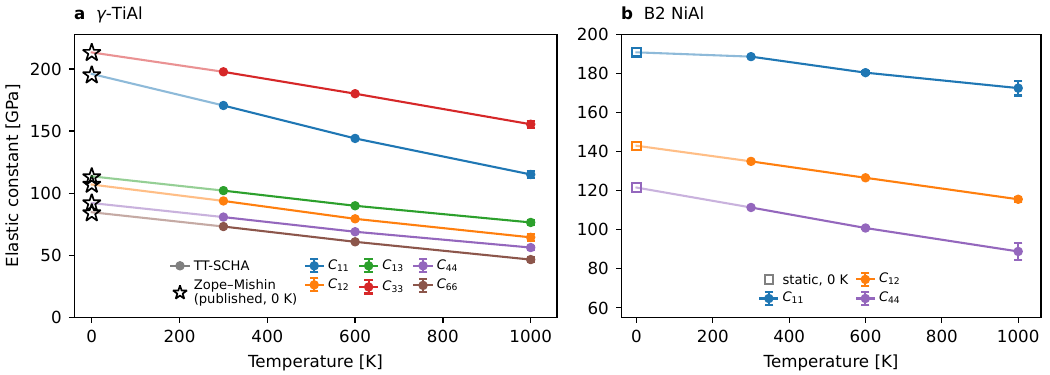}
\caption{\textbf{Elastic constants from the tensor-train free energy.} (a) The six constants of $\gamma$-TiAl at 300, 600 and 1000\,K (circles; error bars are the standard deviation over pivot sets, three at 300 and 600\,K, nine at 1000\,K) and at 0\,K from our static lattice calculation (line ends at 0\,K), with the published zero-temperature values of the same potential, Zope and Mishin~\cite{ZopeMishin2003} (stars). (b) The three constants of B2 NiAl, with our static values at 0\,K (squares); no published values are used.}\label{fig:elastic}
\end{figure*}

\emph{Zero pressure and published calculations: }Pressures at several volumes give the classical zero-pressure expansion, modulus and heat capacities (SM Tables~S\SIr{tab:thermNiAl}, S\SIr{tab:thermTiAl}). For $\gamma$-TiAl the expansion relative to 400\,K is 0.314, 0.662 and 1.038\,\% at 600, 800 and 1000\,K, against 0.32, 0.67 and 1.05\,\% from Monte Carlo sampling with the same potential~\cite{ZopeMishin2003}. Strained cells give the elastic constants: a static (zero-temperature) lattice calculation of ours reproduces the six published constants of this potential to 1.0\,GPa (Fig.~\ref{fig:elastic}a, SM Table~S\SIr{tab:elstat}), and with the tensor train $\gamma$-TiAl $C_{11}$ falls from 171 to 115\,GPa between 300 and 1000\,K (SM Tables~S\SIr{tab:elNiAl}, S\SIr{tab:elTiAl}); at 1000\,K one run scatters by up to 5\,GPa and the mean of nine is good to 1.7\,GPa.

\begin{figure*}[tbp]\centering
\includegraphics[width=0.60\textwidth]{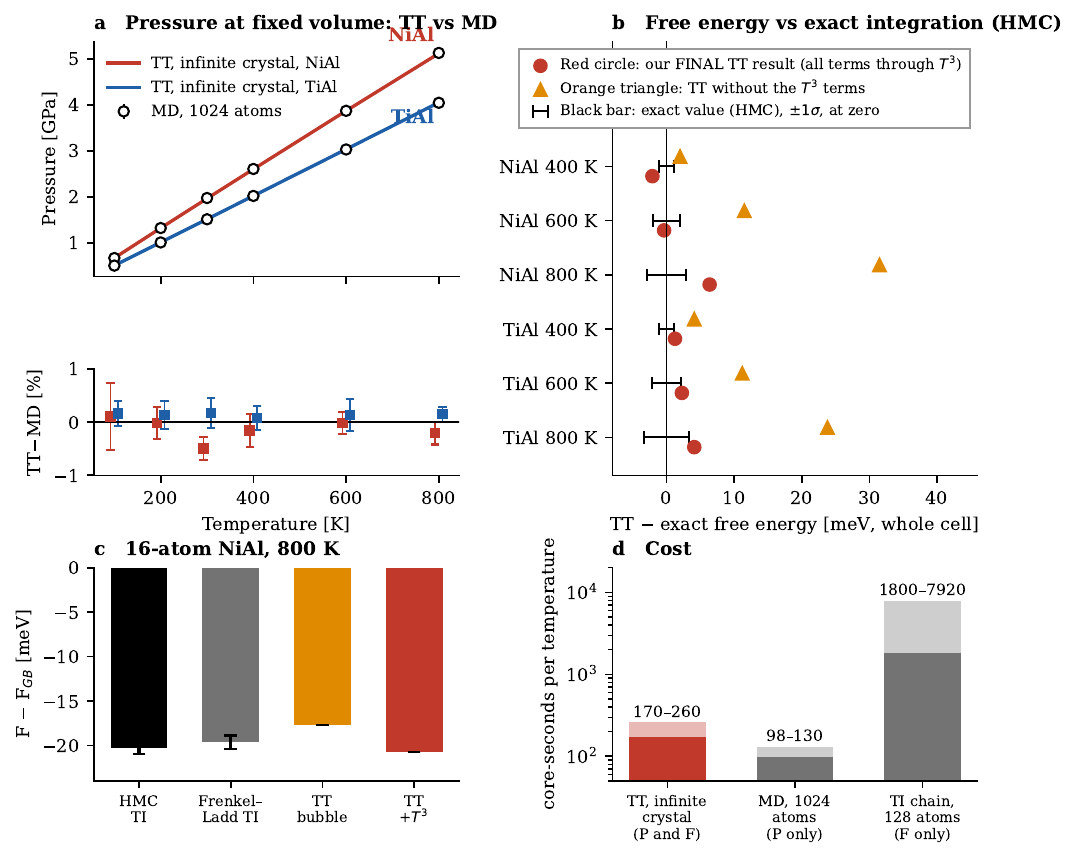}
\caption{\textbf{Tensor-train results against MD, Monte Carlo and thermodynamic integration.} (a) Pressure of NiAl and TiAl at fixed volume, 100 to 800\,K: lines, TT for the infinite crystal; circles, MD of 1\,024 atoms ($\pm2\sigma$); below, relative difference with $2\sigma$ error bars. (b) TT free energy minus the exact free energy of the 128-atom NiAl and 108-atom TiAl cells from thermodynamic integration with Hamiltonian Monte Carlo (error bar $\pm1\sigma$): orange triangles, Gaussian plus cubic term; red circles, our final TT result, with the three terms of order $T^3$. (c) Free energy of the 16-atom NiAl cell at 800\,K relative to the Gaussian: Hamiltonian Monte Carlo, Frenkel--Ladd, and TT. (d) Core-seconds per temperature for TT, MD (1\,024 atoms) and a thermodynamic-integration chain.}\label{fig:headline}
\end{figure*}

\emph{Cost: }The MD campaign (volume run, Einstein-spring constants, ten Frenkel--Ladd nodes and NVT runs at 600--1000\,K) took 2.0\,h (NiAl) and 1.8\,h (TiAl) on 64 cores: $4.63\times10^5$ and $4.11\times10^5$ core-seconds, read from the LAMMPS run logs (SM Sec.~S\SIr{sec:md65k}). The TT calculation took $1.11\times10^3$ and $0.83\times10^3$ core-seconds for one run (420 and 500 times fewer) and three times that for three starting pivots (140 and 165 times fewer). It also reproduces the pressure of 1\,024-atom MD within 0.5\,\% (NiAl) and 0.17\,\% (TiAl) (Fig.~\ref{fig:headline}a, SM Table~S\SIr{tab:bulk}).

\emph{Tests of the terms: }The tests below use cells small enough for exact sampling. The Gaussian with the cubic term alone misses the exact free energy of the 128- and 108-atom cells by 2 to 31\,meV per cell; with the three terms of order $T^3$ the difference falls to within about two standard errors at all temperatures for both alloys (Fig.~\ref{fig:headline}b, SM Table~S\SIr{tab:t3}). In a 16-atom cell an independent Frenkel--Ladd chain agrees (Fig.~\ref{fig:headline}c).

\emph{Limits: }The method expands around one crystal structure and describes solids well below melting; it excludes phase transitions, melting and strongly relaxed defects; it is shown for two embedded-atom potentials, and the low-rank argument does not cover other forms. The expansion stops at order $T^3$, so its error grows with temperature (NiAl pressure at 1000\,K: 20\,MPa, 0.8\,\%). The series is asymptotic and we have no radius of convergence; as a practical guard we use the ratio of the $T^3$ terms to the bubble, 0.08--0.13 (NiAl) and 0.06--0.09 ($\gamma$-TiAl) at 600--1000\,K, growing with temperature (SM Table~S\SIr{tab:ratio}); we claim nothing where it is not small. The order $T^4$ is not computed (SM Table~S\SIr{tab:pnext}). The TT-cross depends on its starting pivots (SM Table~S\SIr{tab:pivot}): the pressure scatters by 2.4--3.5\,MPa (at most 8.9\,MPa) and the Gaussian free energy by at most 0.013\,meV per atom, and neither more sweeps nor a higher rank reduces it (SM Tables~S\SIr{tab:sweep}, S\SIr{tab:rankcap}); we average three runs. Thermodynamic integration in the 128-atom NiAl cell at 800\,K first gave two chains 15\,meV per cell apart; a harmonic crystal with a known answer traced this to two sampling artifacts (Outlook). Corrected, the chains agree ($+1.8\pm3.8$\,meV per cell) and lie $3.3\pm1.5$\,meV per cell below the TT value, the sign and size of the next-order estimate ($-2$ to $-3$\,meV per cell) and of the pivot scatter (about 1.7\,meV per cell); the two are not separated (SM Sec.~S\SIr{sec:ti}).

\emph{Outlook: }Having no sampling, the tensor-train result also checks sampled thermodynamic integration: its disagreement with our Langevin-based chain exposed two sampling artifacts (the momentum-zeroing thermostat option and the fixed trajectory length).

\begin{acknowledgments}
Prepared with the assistance of Claude Sonnet 5.5 (Anthropic).
This work received no funding.
\end{acknowledgments}

\section*{Data availability}
The data supporting the findings of this article and the Python implementation are available from the corresponding author upon request. The files are listed in Supplemental Material Table~S\SIr{tab:code}.

\end{document}


\title{Supplemental Material for: TT-SCHA for Thermodynamics of Binary Alloys}


\author[1]{Boian S. Alexandrov\thanks{Corresponding author: balexan1@unm.edu}}

\author[1]{Dimiter Petsev\thanks{dimiter@unm.edu}}


\affil[1]{\orgdiv{Chemical and Biological Engineering}, \orgname{University of New Mexico}, \city{Albuquerque},  \state{NM}, \country{USA}}



\date{November 2026}

\maketitle
\thispagestyle{empty}
\begin{abstract}\noindent\small
This Supplemental Material accompanies the Letter. It gives the definitions (Sections~\ref{sec:setting} and~\ref{sec:thermo}), the tensor-train construction and its error budget (Section~\ref{sec:tt}), the numerical protocol, validation tests and results for B2 NiAl and $\gamma$-TiAl (Section~\ref{sec:results}), the free-energy chains, the cell-size study and the local formulation (Section~\ref{sec:ti}), the comparison with 65\,536-atom molecular dynamics (Section~\ref{sec:md65k}), the thermodynamic-limit integrals (Section~\ref{sec:bulk}), and the derivations and algorithms specific to this work (Appendices~\ref{app:A}--\ref{app:C}), with the data tables.
\end{abstract}
{\setlength{\parskip}{0pt}\setcounter{tocdepth}{2}\tableofcontents}
\newpage

\section{Setting and notation}\label{sec:setting}

The main symbols are collected in Table~\ref{tab:notation}.

\begin{table}[!htbp]\centering
\renewcommand{\arraystretch}{1.15}
\caption{Notation. Symbols used throughout this Supplemental Material and their meaning. Further symbols are defined where they first appear.}\label{tab:notation}
\begin{tabular}{@{}ll@{}}
\toprule
Symbol & Meaning \\
\midrule
$N$ & number of atoms in the periodic supercell (54, 108, 128 or 256 here)\\
$V$ & volume of the supercell; $V_0$ = zero-pressure volume at $T=0$ \\
$x_0(V)\in\R^{3N}$ & lattice positions (the energy minimum), scaled isotropically with $V$\\
$u\in\R^{3N}$ & displacements of all atoms from their lattice sites\\
$U(x)$ & potential energy (embedded-atom model, evaluated by LAMMPS)\\
$U_0(V)=U(x_0(V))$ & static lattice energy;\quad $\Delta U(u;V)=U(x_0(V)+u)-U_0(V)$\\
$\beta=1/(\kB T)$ & inverse temperature ($\kB = 8.617\times10^{-5}$\,eV/K)\\
$H=\nabla^2U(x_0)$ & Hessian (matrix of second derivatives), $3N\times3N$\\
$H=Q\Lambda Q^{\top}$ & its eigen-decomposition; $\lambda_k$ = eigenvalues, columns of $Q$ = normal modes\\
$d=3N-3$ & number of modes after removing the 3 rigid translations\\
$y=Q^{\top}u$ & displacement written in the mode (eigen) coordinates\\
$\avg{\,\cdot\,}_\mu$ & average with respect to the probability density $\mu$\\
$\mathcal N(0,C)$ & Gaussian (normal) density with mean $0$ and covariance matrix $C$\\
\bottomrule
\end{tabular}
\end{table}

The object of study is the \emph{configurational integral}, written over displacements from the lattice:
\begin{equation}\label{eq:Z}
Z(V,T)\;=\;e^{-\beta U_0(V)}\int_{\R^{3N}} e^{-\beta\,\Delta U(u;V)}\,\mathrm du .
\end{equation}
The three rigid translations of the whole crystal leave $\Delta U$ unchanged. They contribute a factor that does not depend on $V$ at fixed $N$ (up to the centre-of-mass term $\kB T/V$, which is of relative order $1/N$), and they are removed throughout.

\section{Thermodynamics from the configurational integral}\label{sec:thermo}

\begin{proposition}[Energy and pressure]\label{prop:thermo}
With $F=-\kB T\ln Z$,
\begin{align}
\avg{U} &= -\frac{\partial \ln Z}{\partial\beta}, \qquad E=\avg{U}+\tfrac{3N-3}{2}\,\kB T, \label{eq:E}\\[2pt]
P &= \kB T\,\frac{\partial\ln Z}{\partial V} \;=\; -\Big\langle \frac{\partial U\big(x_0(V)+u\big)}{\partial V}\Big|_{u}\Big\rangle, \label{eq:P}
\end{align}
where the average is over the Boltzmann density $e^{-\beta U}/Z$. The kinetic energy is that of the $3N-3$ internal degrees of freedom; the three rigid translations of the crystal are removed, as in the molecular-dynamics reference. The volume derivative in \eqref{eq:P} is taken with the displacements $u$ held fixed while the lattice is scaled.
\end{proposition}

\begin{proof}
\eqref{eq:E} is the standard identity for $\partial_\beta \ln\int e^{-\beta U}$; the kinetic energy $\tfrac32N\kB T$ comes from the momentum integral, which does not depend on $V$. For \eqref{eq:P}, differentiate \eqref{eq:Z} under the integral sign. The integration domain of $u$ does not depend on $V$, because the Boltzmann factor is negligible far from the lattice. \end{proof}

\begin{proposition}[Equivalence with the virial pressure computed by LAMMPS]\label{prop:virial}
Let $P_{\rm vir}=-\frac{1}{3V}\avg{\mathcal W}$ with $\mathcal W$ the \emph{virial}, the derivative of $U$ with respect to a uniform dilation of all positions together with the periodic cell (for pair potentials, the sum $\sum r_{ij}\cdot\nabla_{ij}U$ over pairs). Then, exactly,
\begin{equation}\label{eq:virial}
P \;=\; \frac{(N-1)\,\kB T}{V} + P_{\rm vir}.
\end{equation}
This is the pressure reported by LAMMPS, whose kinetic term is $\mathrm{dof}\cdot\kB T/(3V)$ with $\mathrm{dof}=3N-3$ degrees of freedom (the $3N$ coordinates less the three rigid translations). The virial route to the pressure is classical \cite{Wallace1972}.
\end{proposition}

\begin{proof}
Write $x=\lambda x_0(V_0)+u$ with $\lambda=(V/V_0)^{1/3}$. Scaling the whole configuration (lattice and displacements) gives $\mathrm dU/\mathrm d\lambda|_{x/\lambda}=\mathcal W/\lambda$. Scaling only the lattice differs from this by the displacement term, so $\partial_\lambda U|_u=\lambda^{-1}(\mathcal W-u\cdot\nabla U)$, and hence $\partial_V U|_u=(3V)^{-1}(\mathcal W-u\cdot\nabla U)$. Integration by parts in the $d$ non-translational directions gives $\avg{u\cdot\nabla U}=d\,\kB T$, since the boundary terms vanish. Substitution into \eqref{eq:P} gives \eqref{eq:virial}. \end{proof}

\begin{remark}
Proposition~\ref{prop:virial} makes the tensor-train pressure directly comparable with the LAMMPS pressure. Both are exact expressions for the same thermodynamic derivative, so that any disagreement is an approximation error of the method and not a difference of definitions.
\end{remark}


\section{Tensor-train integration of the Gaussian averages}\label{sec:tt}

Appendix~\ref{app:B} reduces the whole calculation to a short list of averages over one Gaussian, the trial density $\mu=\mathcal N(0,C)$ with $C=\kB T\,\Phi^{+}=SS^{\top}$:
\begin{equation}\label{eq:list}
\avg{U}_\mu,\qquad
\Big\langle \frac{\partial U}{\partial V}\Big|_u\Big\rangle_\mu,\qquad
\Phi^{*}=\avg{\nabla^2U}_\mu,\qquad
\Psi^{*}=\avg{\nabla^3U}_\mu .
\end{equation}
The first two give the energy and the pressure, the third drives the SCHA iteration (Theorem~\ref{thmB:scha}), and the fourth gives the cubic correction (Theorem~\ref{thmB:bubble}). In stochastic implementations of SCHA these averages are estimated by Monte Carlo sampling of the trial Gaussian \cite{Errea2014,Monacelli2021}. In this section they are instead evaluated by deterministic tensor-train (TT) integration; we call the resulting method TT-SCHA. The section consists of twelve statements, listed in Table~\ref{tab:statements} and preceded by a summary of the tensors involved (Section~\ref{sec:ttmap}).

\begin{table}[!htbp]\centering
\renewcommand{\arraystretch}{1.1}
\caption{The statements of Section~\ref{sec:tt} and their role. Left: the lemma, proposition or theorem and its short name; right: what it contributes to the deterministic evaluation of the four Gaussian averages \eqref{eq:list}. Read from top to bottom, the table follows the logic of the construction: all averages are derivatives of one smoothed potential; the embedded-atom energy splits exactly into pair terms, integrated in three dimensions, and a small many-body remainder, represented as a tensor train; and the total error is bounded.}\label{tab:statements}
\begin{tabular}{@{}p{0.39\textwidth}p{0.55\textwidth}@{}}
\toprule
statement & role\\
\midrule
Lemma~\ref{lem:smooth} (smoothed potential) & all four quantities are derivatives of one smooth function $\bar U$\\
Lemma~\ref{lem:whiten} (whitened SCHA) & SCHA condition and bubble as Hermite coefficients\\
Lemma~\ref{lem:wick} (Wick form of the bubble) & the bubble can be assembled in Cartesian coordinates\\
Lemma~\ref{lem:marginal} (marginal reduction) & a term that depends on few coordinates needs a low-dimensional integral\\
Lemma~\ref{lem:split} (exact EAM splitting) & embedded-atom energy $=$ pair part $+$ small many-body remainder\\
Proposition~\ref{prop:pairrule} (pair rule) & pair part $=$ rank-one TT (tensor Gauss--Hermite) in 3-D\\
Lemma~\ref{lem:orbit} (translation orbits) & only pairs and sites of one unit cell are integrated\\
Lemma~\ref{lem:sepsum} (TT rank of a function of a separable sum) & why the many-body remainder has low TT rank\\
Proposition~\ref{prop:adapted} (density-adapted coordinates) & the coordinates in which Lemma~\ref{lem:sepsum} applies\\
Proposition~\ref{prop:weighted} (Gaussian-weighted cross) & how the TT is built, and its error in the right norm\\
Lemma~\ref{lem:contract} (exact Hermite contraction) & moments of order $0$--$3$ from the TT cores\\
Theorem~\ref{thm:budget} (deterministic error budget) & what the total error consists of\\
\bottomrule
\end{tabular}
\end{table}

Throughout, $u=S\xi$ with $\xi\sim\mathcal N(0,I_d)$ (the whitened SCHA coordinates of Section~\ref{app:B}), $\mathrm{He}_n$ are the probabilists' Hermite polynomials, the orthogonal polynomials of the standard normal weight ($\mathrm{He}_1(x)=x$, $\mathrm{He}_2(x)=x^2-1$, $\mathrm{He}_3(x)=x^3-3x$), and $\mathrm{He}_\alpha(\xi)=\prod_k\mathrm{He}_{\alpha_k}(\xi_k)$ for a multi-index $\alpha$.

\subsection{The tensors of the computation and their contractions}\label{sec:ttmap}

Before the individual statements are given, we list all tensors formed in the computation, the function each of them samples, and the contraction that maps it to a physical quantity. Appendix~\ref{app:C} gives the general TT definitions, the cross algorithm and the proofs; Appendix~\ref{app:B} gives the Gaussian identities used below.

\paragraph{Coordinates.} There are three changes of variables, all linear:
\begin{equation}\label{eq:coordchain}
\underbrace{u\in\R^{3N}}_{\text{Cartesian}}
\;\xrightarrow{\;\xi=S^{+}u\;}\;
\underbrace{\xi\in\R^{d}}_{\text{whitened modes}}
\;\xrightarrow{\;\eta=V_m^{\top}\xi\;}\;
\underbrace{\eta\in\R^{d}}_{\text{density-adapted, site }m},
\qquad
u\;\xrightarrow{\;\omega=L_p^{-1}(u_j-u_i)\;}\;\underbrace{\omega\in\R^{3}}_{\text{pair }p=(i,j,R)} .
\end{equation}
Here $d=3N-3$, $S=Q\,\mathrm{diag}(\sigma)$ with $\sigma_k=(\beta\phi_k)^{-1/2}$ from $\Phi=Q\,\mathrm{diag}(\phi)Q^{\top}$, $V_m$ is the orthogonal eigenbasis of $\avg{\nabla^2_\xi\rho_m}$ (Proposition~\ref{prop:adapted}), and $L_p$ is the Cholesky factor of the $3\times3$ covariance of $u_j-u_i$ (Proposition~\ref{prop:pairrule}). ($L_p$ is lower triangular with $L_pL_p^{\top}$ equal to that covariance.) $V_m$ is orthogonal and $L_p$ whitens the pair covariance, so in each of $\xi$, $\eta$ and $\omega$ the trial density is the \emph{same} product of one-dimensional standard Gaussians.

\paragraph{Tensor 0: the Gaussian weight (rank one, exact).} On a product grid $\{x^{(k)}_{i},w^{(k)}_{i}\}$ of Gauss--Hermite nodes and probability weights (weights that sum to one on each axis),
\begin{equation}\label{eq:T0}
\mathsf W(i_1,\dots,i_d)=w^{(1)}_{i_1}w^{(2)}_{i_2}\cdots w^{(d)}_{i_d},
\qquad
\operatorname{rank}_{\rm TT}\mathsf W=1 .
\end{equation}
This is the product form of a Gaussian in the eigenbasis of its stiffness matrix, in discrete form. The harmonic partition function is the closed-form contraction of the continuous version, $Z_\Phi=e^{-\beta U_0}\prod_{k=1}^{d}(2\pi/\beta\phi_k)^{1/2}$, and every average in \eqref{eq:list} is a contraction of some tensor $\mathsf A$ with $\mathsf W$, i.e.\ the sum of the entries of $\mathsf A$ weighted by those of $\mathsf W$:
\begin{equation}\label{eq:contractW}
\avg{f}_\mu\;\approx\;\langle\mathsf A,\mathsf W\rangle=\sum_{i_1,\dots,i_d}\mathsf A(i_1,\dots,i_d)\,w^{(1)}_{i_1}\cdots w^{(d)}_{i_d},\qquad \mathsf A(\mathbf i)=f\big(x^{(1)}_{i_1},\dots,x^{(d)}_{i_d}\big).
\end{equation}
If $\mathsf A$ is a tensor train with cores $\mathcal G_k$, then $\langle\mathsf A,\mathsf W\rangle=E_1E_2\cdots E_d$ with $E_k=\sum_iw^{(k)}_i\mathcal G_k(i)$ (Proposition~\ref{propC:int}): a product of $d$ small matrices instead of a sum over $\prod_kn_k$ terms.

\paragraph{Tensor 1: the pair tensors (three-dimensional, full).} For each of the $n_{\rm orb}n_{\rm nb}$ representative pairs $p=(i,j,R)$ ($2\times168=336$ for NiAl, $2\times134=268$ for TiAl; $n_{\rm orb}=2$ site orbits, i.e.\ two sets of translation-equivalent sites, and $n_{\rm nb}$ neighbours per site),
\begin{equation}\label{eq:T1}
\mathsf P_p(g_1,g_2,g_3)=\psi_p\big(L_p\,\omega^{(g)}\big),\qquad
\nabla\mathsf P_p(g_1,g_2,g_3)=\nabla\psi_p\big(L_p\,\omega^{(g)}\big)\in\R^3,\qquad
\omega^{(g)}=\big(x_{g_1},x_{g_2},x_{g_3}\big),
\end{equation}
an $n_g\times n_g\times n_g$ array ($24^3=13\,824$ entries). It is small enough to be stored in full. Its contractions with the rank-one weight $w_{g_1}w_{g_2}w_{g_3}$ and with the Hermite factors $\omega_a$, $\omega_a\omega_b-\delta_{ab}$ give $\avg{\psi_p}$, $\avg{\nabla^2\psi_p}$, $\avg{\nabla^3\psi_p}$ and $\avg{\partial_V\psi_p}$ by \eqref{eq:pair2}--\eqref{eq:pairV}. This is the ``rank-one TT in three dimensions'' of Proposition~\ref{prop:pairrule}: the weight is rank one, the integrand is not compressed.

\paragraph{Tensor 2: the site remainder (high-dimensional, compressed).} For each orbit representative $m$ (two per structure),
\begin{equation}\label{eq:T2}
\mathsf F_m(i_1,\dots,i_d)=\tilde F_m\Big(\rho_m\big(x_0+S\,V_m\,\eta_{\mathbf i}\big)-\bar\rho_m\Big),
\qquad \eta_{\mathbf i}=\big(x^{(1)}_{i_1},\dots,x^{(d)}_{i_d}\big),
\end{equation}
with $n_k=9$ nodes on the first ten coordinates and $n_k=5$ on the others. For 128-atom NiAl ($d=381$) this tensor has $9^{10}\cdot5^{371}\approx10^{269}$ entries. It is never formed. The weighted cross (Proposition~\ref{prop:weighted}, \hyperref[alg:C1]{Algorithm~C.1}) returns a tensor train
\begin{equation}\label{eq:T2tt}
\mathsf F_m(\mathbf i)\;\approx\;\tilde{\mathsf F}_m(\mathbf i)=\mathcal G^{(m)}_1(i_1)\,\mathcal G^{(m)}_2(i_2)\cdots\mathcal G^{(m)}_d(i_d),\qquad\mathcal G^{(m)}_k(i)\in\R^{r_{k-1}\times r_k},\quad r_k\le r_{\max},
\end{equation}
from about $4\,d\,r^2n^2$ evaluations of \eqref{eq:T2} at points chosen by the maximum-volume principle. With $r_{\max}=6$ the train stores $\sum_kr_{k-1}n_kr_k\approx7\times10^{4}$ numbers. Lemma~\ref{lem:sepsum} and Proposition~\ref{prop:adapted} explain why such small ranks suffice, and Fig.~\ref{fig:ttprofile} in Appendix~\ref{app:C} shows the singular values that are discarded, i.e.\ the weights of the successive rank-one terms in the singular value decomposition (SVD) of the matrices obtained by splitting the variables into two groups.

\paragraph{Tensor 3: the site pressure integrand (compressed).} The volume derivative of the remainder at fixed $u$ and fixed $\bar\rho$ is
\begin{equation}\label{eq:T3}
\mathsf G_m(\mathbf i)=\Big(F_m'\big(\rho_m(\mathbf i)\big)-F_m'(\bar\rho_m)\Big)\,\frac{\partial\rho_m}{\partial V}\Big|_u(\mathbf i),
\end{equation}
sampled at the same points and compressed in the same way ($r_{\max}=10$). Only its mean $\langle\tilde{\mathsf G}_m,\mathsf W\rangle$ is needed.

\paragraph{Contractions.} Let $\mathcal H^{(m)}_n$ be the $n$-th Hermite moment tensor of the train \eqref{eq:T2tt}, i.e.\ $\mathcal H^{(m)}_0=\langle\tilde{\mathsf F}_m,\mathsf W\rangle$, $(\mathcal H^{(m)}_1)_k=\avg{\mathrm{He}_1(\eta_k)\tilde{\mathsf F}_m}$, $(\mathcal H^{(m)}_2)_{kl}=\avg{\mathrm{He}_{kl}(\eta)\tilde{\mathsf F}_m}$, $(\mathcal H^{(m)}_3)_{klq}=\avg{\mathrm{He}_{klq}(\eta)\tilde{\mathsf F}_m}$. All of them are products of the transfer matrices of Lemma~\ref{lem:contract}. With $T_m=S^{+\top}V_m$ (so that $\nabla_u=T_m\nabla_\eta$), the four quantities of \eqref{eq:list} are
\begin{align}
\avg{U}\;&=\;U_{\rm const}+\sum_{p}\,\nu_p\,\langle\mathsf P_p,\mathsf W_3\rangle+\sum_m|\mathcal O_m|\,\mathcal H^{(m)}_0,\label{eq:ttU}\\
\Big\langle\frac{\partial U}{\partial V}\Big|_u\Big\rangle\;&=\;\sum_p\nu_p\,\frac{1}{3V}\big\langle\nabla\mathsf P_p\cdot R_p,\mathsf W_3\big\rangle+\sum_m|\mathcal O_m|\,\langle\tilde{\mathsf G}_m,\mathsf W\rangle,\label{eq:ttP}\\
\Phi^{*}\;&=\;\sum_{t\in\mathcal T}P_t\Big[\sum_p\Pi_p\big[\avg{\nabla^2\psi_p}\big]+\sum_m T_m\,\mathcal H^{(m)}_2\,T_m^{\top}\Big]P_t^{\top},\label{eq:ttPhi}\\
\Psi^{*}\;&=\;\sum_{t\in\mathcal T}P_t^{\otimes3}\Big[\sum_p\Pi^{(3)}_p\big[\avg{\nabla^3\psi_p}\big]+\sum_m\mathcal H^{(m)}_3\times_1T_m\times_2T_m\times_3T_m\Big],\label{eq:ttPsi}
\end{align}
where $\mathsf W_3$ is the three-dimensional weight, $\nu_p$ counts the translates of pair $p$ in the supercell, $\Pi_p$, $\Pi^{(3)}_p$ place the $3\times3$ (or $3\times3\times3$) blocks with the signs of Proposition~\ref{prop:pairrule}, $\times_n$ is the mode-$n$ product (multiplication of the $n$-th index of a tensor by a matrix), and $U_{\rm const}=\sum_i[F_i(\bar\rho_i)-F_i'(\bar\rho_i)\bar\rho_i]$. (The constant does not enter \eqref{eq:ttP}: its $V$-derivative through $\bar\rho$ cancels exactly against the $\bar\rho$-dependence of the other two parts, because the splitting \eqref{eq:split3} holds for \emph{every} $\bar\rho$.) The physical outputs are then
\begin{align}
\text{SCHA update:}\quad&\Phi\;\leftarrow\;\mathcal S\big[\Phi^{*}\big]\quad\text{(Theorem~\ref{thmB:scha})},\nonumber\\
\text{free energy:}\quad&F_{GB}=\frac{\kB T}{2}\sum_{k=1}^{d}\ln\frac{\beta\phi_k}{2\pi}+\avg{U}-\frac{d}{2}\kB T\quad\text{(Theorem~\ref{thmB:GB})},\label{eq:ttF}\\
\text{pressure:}\quad&P_{\rm HF}=-\Big\langle\frac{\partial U}{\partial V}\Big|_u\Big\rangle\quad\text{(Proposition~\ref{propB:HF})},\nonumber\\
\text{bubble:}\quad&F_b=-\frac{\beta}{12}\sum_{abc,a'b'c'}\Psi^{*}_{abc}\,C_{aa'}C_{bb'}C_{cc'}\,\Psi^{*}_{a'b'c'}\quad\text{(Lemma~\ref{lem:wick})}.\nonumber
\end{align}
The ideal-gas part $(N-1)\kB T/V$ of the pressure is contained in $P_{\rm HF}$. The last contraction is again with a rank-one Gaussian: in the whitened basis $C=SS^{\top}$ becomes the identity, and, with $\|\cdot\|_F$ the Frobenius norm (the square root of the sum of the squared entries), $F_b=-\tfrac{\kB T}{12}\|\beta\,\Psi^{*}\times_1S^{\top}\times_2S^{\top}\times_3S^{\top}\|_F^2$. The configurational integral itself is finally
\begin{equation}\label{eq:ttZ}
-\kB T\ln Z\;=\;F_{GB}+F_b+\mathcal O(T^3),\qquad
Z=Z_\Phi\,\big\langle e^{-\beta(U-U_\Phi)}\big\rangle_\mu ,
\end{equation}
the rank-one factor $Z_\Phi$ times a residual whose logarithm is expanded to second order (Theorem~\ref{thmB:bubble}). Table~\ref{tab:tensors} summarises the four tensors and the contractions.

\begin{table}[!htbp]\centering
\renewcommand{\arraystretch}{1.12}
\small
\caption{The tensors formed in a TT-SCHA calculation, with sizes for 128-atom B2 NiAl. Columns: the tensor and its defining equation; the variables on which it depends ($\xi$: whitened mode coordinates, $\eta$: density-adapted coordinates, $\omega$: whitened relative displacement of a pair; $d$ is the number of variables); its number of entries (the full site tensors, $\approx10^{269}$ entries, are never formed); how it is stored (``TT rank 1: a product of one-dimensional factors; ``full 3-D array: all entries stored; ``TT, $r\le6$: tensor train with ranks at most 6); and the quantities obtained by contracting it, i.e.\ by summing its entries against the Gaussian weights and Hermite polynomials.}\label{tab:tensors}
\begin{tabular}{@{}lllll@{}}
\toprule
tensor & variables & size (NiAl, 128 atoms) & format & contracted into\\
\midrule
$\mathsf W$ \eqref{eq:T0} & $\xi$ or $\eta$, $d=381$ & $\prod_kn_k$ & TT rank 1, exact & every average; $Z_\Phi$\\
$\mathsf P_p$ \eqref{eq:T1} & $\omega\in\R^3$ & $24^3$, $\times336$ pairs & full 3-D array & $\avg{\psi},\avg{\nabla^{2,3}\psi},\avg{\partial_V\psi}$\\
$\mathsf F_m$ \eqref{eq:T2} & $\eta$, $d=381$ & $\approx10^{269}$, $\times2$ sites & TT, $r\le6$ & $\mathcal H_0,\dots,\mathcal H_3$\\
$\mathsf G_m$ \eqref{eq:T3} & $\eta$, $d=381$ & $\approx10^{269}$, $\times2$ sites & TT, $r\le10$ & remainder pressure\\
$\Psi^{*}$ & Cartesian, $3N=384$ & $384^3$ & dense, local & $F_b$ (Wick, with $C^{\otimes3}$)\\
\bottomrule
\end{tabular}
\end{table}

\noindent The rest of this section proves that each of these steps is exact or has a controlled error. The general TT facts (unfoldings, the construction of a TT by successive singular value decompositions, the error bounds of skeleton and maximum-volume approximations, the cross interpolation formula, and the recursions behind Lemma~\ref{lem:contract}) are collected with proofs in Appendix~\ref{app:C}. Section~\ref{app:A} contains the sublattice factorisation for NiAl (Theorem~\ref{thmA:NiAl}), and the standard Gaussian identities (Wick--Isserlis, Stein, Price, Hermite chaos), SCHA and the bubble are derived in Appendix~\ref{app:B}.


\subsection{A single smoothed function determines all averages}

\begin{lemma}[Smoothed potential]\label{lem:smooth}
Let $U$ be continuous with at most polynomial growth, and let
\begin{equation}\label{eq:Ubar}
\bar U(x) \;=\; \big\langle U(x+u)\big\rangle_{u\sim\mathcal N(0,C)} .
\end{equation}
Then $\bar U$ is infinitely differentiable on the non-translational subspace, even if $U$ itself is only once differentiable (as the tabulated LAMMPS potentials are). For every multi-index $\alpha$, with derivatives taken along the columns of $S$,
\begin{equation}\label{eq:smoothderiv}
\partial^\alpha_\xi\,\bar U(x_0+S\xi)\Big|_{\xi=0}
\;=\; \big\langle \mathrm{He}_\alpha(\xi)\;U(x_0+S\xi)\big\rangle_{\xi\sim\mathcal N(0,I)} .
\end{equation}
In particular
\[
\Phi^{*}=\nabla^2\bar U(x_0),\qquad
\Psi^{*}=\nabla^3\bar U(x_0),\qquad
\Big\langle\frac{\partial U}{\partial V}\Big|_u\Big\rangle_\mu=\frac{\partial \bar U}{\partial V}\Big|_{C},
\]
where the last derivative scales the lattice with $C$ held fixed. The averages $\avg{\nabla^2U}$ and $\avg{\nabla^3U}$ are well defined even where $U$ has no classical second or third derivative: they are the derivatives of $\bar U$.
\end{lemma}

\begin{proof}
Write $\bar U(x_0+S\eta)=\int U(x_0+S\xi)\,\varphi(\xi-\eta)\,\mathrm d\xi$ with $\varphi$ the standard Gaussian density. All derivatives in $\eta$ fall on $\varphi$, which is smooth, and $-\partial_{\eta_k}\varphi(\xi-\eta)=\partial_{\xi_k}\varphi(\xi-\eta)$. Rodrigues' formula, which generates the Hermite polynomials by differentiating the Gaussian, $(-\partial)^{\alpha}\varphi=\mathrm{He}_\alpha\,\varphi$, gives \eqref{eq:smoothderiv} at $\eta=0$. If $U$ is smooth enough, integration by parts turns the right-hand side back into $\avg{\partial^\alpha U}$, which is the generalised Stein identity (Lemma~\ref{lemB:stein}). The volume statement follows by differentiating \eqref{eq:Ubar} under the integral sign, which is allowed because the Gaussian tails dominate. \end{proof}

\begin{lemma}[Whitened form of SCHA and of the bubble]\label{lem:whiten}
Let $h(\xi)=\beta\,U(x_0+S\xi)$. Then
\begin{enumerate}[label=(\alph*)]
\item the SCHA fixed point $\Phi=\Phi^{*}(\Phi)$ is equivalent to
\begin{equation}\label{eq:whitescha}
\big\langle(\xi_k\xi_l-\delta_{kl})\,h(\xi)\big\rangle=\delta_{kl}\qquad(k,l=1,\dots,d);
\end{equation}
\item the update of the fixed-point iteration is
\begin{equation}\label{eq:update}
\Phi_{\rm new}=S^{+\top}\big(\kB T\,A\big)S^{+},\qquad A_{kl}=\big\langle(\xi_k\xi_l-\delta_{kl})\,h\big\rangle,\qquad S^{+}=\mathrm{diag}(\sigma)^{-1}Q^{\top};
\end{equation}
\item the dimensionless cubic tensor of Theorem~\ref{thmB:bubble} is the third Hermite coefficient,
\begin{equation}\label{eq:Xhermite}
X_{klm}=\big\langle \mathrm{He}_{klm}(\xi)\;h(\xi)\big\rangle,
\end{equation}
where $\mathrm{He}_{klm}=\xi_k\xi_l\xi_m$ for distinct indices, $(\xi_k^2-1)\xi_m$ if $k=l\neq m$, and $\xi_k^3-3\xi_k$ if $k=l=m$.
\end{enumerate}
\end{lemma}

\begin{proof}
By Lemma~\ref{lem:smooth}, $A=\beta S^{\top}\Phi^{*}S$. Since $S^{\top}\Phi S=\mathrm{diag}(\sigma_k^2\phi_k)=\kB T\,I$, the condition $\Phi^{*}=\Phi$ is $A=I$, which is (a). For (b), $S^{+\top}S^{\top}\Phi^{*}SS^{+}$ is the projection of $\Phi^{*}$ onto the non-translational subspace, and the acoustic sum rule makes it equal to $\Phi^{*}$. For (c), apply \eqref{eq:smoothderiv} with $|\alpha|=3$ to $h$. The multi-index with $\alpha_k=2$, $\alpha_m=1$ gives $\mathrm{He}_2(\xi_k)\mathrm{He}_1(\xi_m)$, and so on. \end{proof}

\begin{lemma}[Wick form of the bubble]\label{lem:wick}
The sum in Theorem~\ref{thmB:bubble} does not depend on the basis:
\begin{equation}\label{eq:wick}
\sum_{k,l,m=1}^{d}X_{klm}^{2}
\;=\;\beta^{2}\sum_{a,b,c}\ \sum_{a',b',c'}\Psi^{*}_{abc}\,\Psi^{*}_{a'b'c'}\;C_{aa'}\,C_{bb'}\,C_{cc'},
\qquad
F_b=-\frac{\beta}{12}\,\big\langle\Psi^{*},(C\otimes C\otimes C)\,\Psi^{*}\big\rangle .
\end{equation}
Here $a,b,c$ run over the $3N$ Cartesian coordinates.
\end{lemma}

\begin{proof}
By Lemma~\ref{lem:whiten}(c) and Lemma~\ref{lem:smooth}, $X_{klm}=\beta\sum_{abc}\Psi^{*}_{abc}S_{ak}S_{bl}S_{cm}$. Squaring and summing over $k,l,m$ produces $\sum_kS_{ak}S_{a'k}=C_{aa'}$ in each of the three index pairs. \end{proof}

\noindent Lemma~\ref{lem:wick} has an algorithmic consequence. $\Psi^{*}$ is \emph{local} in Cartesian coordinates, since a cubic force constant couples only atoms within the cutoff of a common atom, and it is invariant under lattice translations. $X$ in the mode basis is dense. The bubble can therefore be assembled term by term in Cartesian coordinates and contracted with $C$ at the end.

\subsection{From $3N$ dimensions to three: marginals and the EAM split}

\begin{lemma}[Marginal reduction]\label{lem:marginal}
Let $g(u)=f(Pu)$ with $P:\R^{3N}\to\R^{m}$ linear. Then for $u\sim\mathcal N(0,C)$ the average of $g$ requires only the marginal distribution of $Pu$, i.e.\ the Gaussian distribution of the $m$ variables $Pu$ alone:
\begin{equation}\label{eq:marginal}
\avg{g}_{\mathcal N(0,C)}=\avg{f}_{\mathcal N(0,\Gamma)},\qquad
\avg{\nabla^n g}=\big(P^{\top}\big)^{\otimes n}\avg{\nabla^n f}_{\mathcal N(0,\Gamma)},\qquad
\Gamma=PCP^{\top}.
\end{equation}
A term of the energy that depends on the displacements of two atoms only through their difference therefore needs a three-dimensional integral, whatever the size of the crystal.
\end{lemma}

\begin{proof}
If $u\sim\mathcal N(0,C)$ then $Pu\sim\mathcal N(0,PCP^{\top})$. The derivative formula is the chain rule, $\nabla^n g=(P^{\top})^{\otimes n}(\nabla^nf)\circ P$, followed by the first statement. \end{proof}

For the embedded-atom model \cite{Daw1984},
\begin{equation}\label{eq:eam}
U=\sum_i F_{i}(\rho_i)+\frac12\sum_{i\neq j}\phi_{ij}(r_{ij}),\qquad
\rho_i=\sum_{j\neq i}f_{j}(r_{ij}),\qquad
r_{ij}=|R_{ij}+w_{ij}|,\quad w_{ij}=u_j-u_i ,
\end{equation}
where $R_{ij}$ is the lattice vector between the sites, including periodic images, $\phi_{ij}$ is the pair potential, $f_j(r)$ is the electron density that an atom of the species of $j$ contributes at distance $r$, so that $\rho_i$ is the density at atom $i$, and the embedding function $F_i$ of the species of atom $i$ gives the energy of embedding that atom in the density $\rho_i$. The pair terms are functions of one relative displacement, so Lemma~\ref{lem:marginal} applies to them directly. The embedding terms are not: $F_i(\rho_i)$ depends on all neighbours of $i$ at once.

\begin{lemma}[Exact EAM splitting]\label{lem:split}
For any constants $\bar\rho_i$,
\begin{align}
U \;&=\; \underbrace{\sum_i\big[F_i(\bar\rho_i)-F_i'(\bar\rho_i)\,\bar\rho_i\big]}_{\text{constant}}
\;+\;\underbrace{\sum_{(i,j)}\psi_{ij}(w_{ij})}_{\text{pair part}}
\;+\;\underbrace{\sum_i \tilde F_i(\rho_i-\bar\rho_i)}_{\text{embedding remainder}},\label{eq:split3}\\[2pt]
\psi_{ij}(w)\;&=\;\tfrac12\,\phi_{ij}(|R_{ij}+w|)+F_i'(\bar\rho_i)\,f_j(|R_{ij}+w|),\qquad
\tilde F(\delta)=F(\bar\rho+\delta)-F(\bar\rho)-F'(\bar\rho)\,\delta ,\nonumber
\end{align}
where $(i,j)$ runs over ordered pairs. With the choice $\bar\rho_i=\avg{\rho_i}_\mu$, which the pair rule below computes to quadrature accuracy, the remainder is of second order in the density fluctuation $\delta\rho_i=\rho_i-\avg{\rho_i}$:
\begin{gather}
0\le \big\langle\tilde F_i\big\rangle \le \tfrac12\,\sup|F_i''|\;\operatorname{Var}(\rho_i)
\qquad\text{(if $F_i$ is convex on the range of $\rho_i$)},\label{eq:remsmall}\\
\big\langle\tilde F_i\big\rangle=\tfrac12F_i''(\bar\rho_i)\operatorname{Var}\rho_i+\mathcal O\big(\avg{|\delta\rho_i|^3}\big).\nonumber
\end{gather}
\end{lemma}

\begin{proof}
Sum the definitions: $\sum_i[F_i(\bar\rho_i)-F_i'\bar\rho_i]+\sum_{(i,j)}F_i'(\bar\rho_i)f_j+\sum_i[F_i(\rho_i)-F_i(\bar\rho_i)-F_i'(\bar\rho_i)(\rho_i-\bar\rho_i)]=\sum_iF_i(\rho_i)$, because $\sum_jf_j=\rho_i$. The pair potentials are unchanged. \eqref{eq:remsmall} is Taylor's theorem with remainder, using $\avg{\delta\rho_i}=0$; the $\mathcal O$ form assumes $F''$ Lipschitz, i.e.\ $|F''(a)-F''(b)|\le c\,|a-b|$ for some constant $c$, which holds for the underlying smooth embedding function but only up to the small jumps of $F''$ at the table nodes for the LAMMPS table. \end{proof}

\noindent The pair part contains the full pair potential and the part of the embedding energy that is linear in the density, i.e.\ the familiar ``effective pair potential'' of the EAM, but evaluated at the thermally averaged density. Only the curvature of $F$ acting on density fluctuations is left as a true many-body term. For NiAl at 300\,K this remainder carries $0.5\%$ of $\Phi^{*}$, $0.7\%$ of the pressure and $10^{-4}$ of $\sum X^2$ (Section~\ref{sec:validation}). It is small, but not negligible at the target accuracy.

\begin{proposition}[Pair rule: a rank-one tensor in three dimensions]\label{prop:pairrule}
For the ordered pair $(i,j)$ let $\Gamma_{ij}=C_{jj}+C_{ii}-C_{ij}-C_{ji}$ (the $3\times3$ covariance of $w_{ij}$), $\Gamma_{ij}=LL^{\top}$ its Cholesky factor, and $w=L\omega$ with $\omega\sim\mathcal N(0,I_3)$. In $\omega$ the Gaussian weight is the product $\varphi(\omega_1)\varphi(\omega_2)\varphi(\omega_3)$, a rank-one tensor (a Gaussian is a product of one-dimensional Gaussians in the eigenbasis of its covariance, here the $3\times3$ marginal). With $n_g$ Gauss--Hermite nodes per axis, $\{\omega^{(g)},W_g\}$, and writing $\nabla\psi=\psi'(r)\,\hat r$,
\begin{align}
\avg{\psi}&\approx\sum_g W_g\,\psi\big(L\omega^{(g)}\big),\qquad
\avg{\nabla^2\psi}\approx L^{-\top}\sum_gW_g\,\omega^{(g)}\,\nabla\psi\big(L\omega^{(g)}\big)^{\!\top},\label{eq:pair2}\\
\avg{\nabla^3\psi}_{abc}&\approx\sum_{a',b'}\big(L^{-\top}\big)_{aa'}\big(L^{-\top}\big)_{bb'}\sum_g W_g\,\big(\omega^{(g)}_{a'}\omega^{(g)}_{b'}-\delta_{a'b'}\big)\,\partial_c\psi\big(L\omega^{(g)}\big),\label{eq:pair3}\\
\Big\langle\frac{\partial\psi}{\partial V}\Big|_w\Big\rangle&\approx\frac{1}{3V}\sum_gW_g\,\nabla\psi\big(L\omega^{(g)}\big)\cdot R_{ij},\label{eq:pairV}
\end{align}
and \eqref{eq:pair3} is symmetrised over its three indices. These averages enter $\Phi^{*}$ and $\Psi^{*}$ through Lemma~\ref{lem:marginal} with $P$ = ``$u_j-u_i$'': a $+$ sign for each index on atom $j$ and a $-$ sign for each index on atom $i$. For example, $\Phi^{*}_{jj}\mathrel{+}=\avg{\nabla^2\psi}$ and $\Phi^{*}_{ij}\mathrel{-}=\avg{\nabla^2\psi}$.
\end{proposition}

\begin{proof}
First-order Stein in $\omega$, applied to the vector function $\nabla_w\psi(L\omega)$, gives $\avg{\omega\,(\nabla_w\psi)^{\top}}=L^{\top}\avg{\nabla^2_w\psi}$, which is \eqref{eq:pair2}. Second-order Stein, applied to $\partial_c\psi$, gives \eqref{eq:pair3}. The volume derivative at fixed $w$ scales $R_{ij}$ by $(V/V_0)^{1/3}$, so $\partial_VR_{ij}=R_{ij}/(3V)$. \end{proof}

\begin{remark}[Gradients instead of second derivatives]
\eqref{eq:pair2}--\eqref{eq:pair3} use only $\psi'$, never $\psi''$ or $\psi'''$. The LAMMPS tables are $C^1$ piecewise cubics whose second derivative jumps at every table node, i.e.\ at every point at which LAMMPS tabulates the functions (spacing $6\times10^{-4}$\,\AA). A rule that samples $\psi''$ pointwise would inherit that noise, while the Stein forms integrate it out. What remains is $\psi'$ itself, which is continuous but has small kinks at the table nodes and larger third-derivative jumps of $\psi$ at the knots of the original spline from which the tables were generated. These limit the Gauss--Hermite convergence to an algebraic rate, i.e.\ to an error that decreases as a power of $1/n_g$ rather than exponentially. Table~\ref{tab:ng} confirms this: $\Phi^{*}$ is converged to $10^{-4}$ at $n_g=8$, and the pressure to about $1$\,bar at $n_g=24$.
\end{remark}

\begin{lemma}[Translation orbits]\label{lem:orbit}
Let $\mathcal T$ be the group of lattice translations of the supercell, acting on atoms by permutations $t$, and suppose $C$ is $\mathcal T$-invariant (which the symmetrisation $\mathcal S$ enforces). Then
\begin{enumerate}[label=(\alph*)]
\item the pair averages of Proposition~\ref{prop:pairrule} are equal for $(i,j,R)$ and $(t(i),t(j),R)$;
\item the embedding remainders of $i$ and $t(i)$ have the same averages, and their derivative tensors are related by the permutation $P_t$.
\end{enumerate}
Hence only the pairs whose first atom is an \emph{orbit representative}, i.e.\ one chosen atom from each set of atoms related by lattice translations (a translation orbit), are integrated, and one embedding site per orbit. For B2 and L1$_0$ there are two orbits. With the neighbour list cut at $r_c+1.2$\,\AA{} this gives $2\times168=336$ pairs for NiAl, of which $88$ per atom lie inside the cutoff, and $2\times134$ for $\gamma$-TiAl. These numbers are \emph{independent of the supercell size}. The contributions are assembled as
\begin{equation}\label{eq:assemble}
\Phi^{*}=\sum_{(i,j)}\Pi_{ij}\big[\avg{\nabla^2\psi_{ij}}\big]+\sum_{m\in\text{reps}}\frac{|\mathcal O_m|}{|\mathcal T|}\sum_{t\in\mathcal T}P_t\,\Phi^{(m)}_{\rm rem}\,P_t^{\top},
\end{equation}
and likewise for $\Psi^{*}$ with $P_t^{\otimes3}$. Here $\Pi_{ij}$ places a $3\times3$ block with the signs of Proposition~\ref{prop:pairrule}, and $|\mathcal O_m|$ is the size of the orbit of $m$.
\end{lemma}

\begin{proof}
A translation maps the pair $(i,j)$ to $(t(i),t(j))$ with the same lattice vector $R$. Invariance of $C$ gives $\Gamma_{t(i)t(j)}=\Gamma_{ij}$, and the integrand depends only on $R$, $\Gamma$ and the species. The same argument applies to the remainder of a site and its whole neighbourhood. \end{proof}

\subsection{The many-body remainder as a low-rank tensor train}

The embedding remainder of site $m$ is a function of all $d$ whitened coordinates, $\tilde F_m\big(\delta\rho_m(\xi)\big)$. The density $\rho_m$ is a sum of 88 (NiAl) or 78 (TiAl) neighbour terms, each depending on a different three-dimensional projection of $\xi$. A tensor train in the mode coordinates $\xi$ has large ranks for such a function, because the anharmonic couplings are all-to-all in these coordinates. The next lemma identifies the coordinates in which it does not.

\begin{lemma}[TT rank of a function of a separable sum]\label{lem:sepsum}
Let $s(\eta)=\sum_{k=1}^{d}h_k(\eta_k)$ and $G(\eta)=F\big(s(\eta)\big)$.
\begin{enumerate}[label=(\alph*)]
\item If $F$ is a polynomial of degree $p$, then $G$ has an \emph{exact} tensor train with all ranks $\le p+1$. The cores are the upper-triangular $(p+1)\times(p+1)$ matrices
\begin{equation}\label{eq:binomcore}
\big[\mathcal G_k(\eta_k)\big]_{ab}=\binom{b}{a}\,h_k(\eta_k)^{\,b-a}\quad(0\le a\le b\le p),
\qquad
G(\eta)=e_0^{\top}\,\mathcal G_1(\eta_1)\cdots\mathcal G_d(\eta_d)\;\mathbf c ,
\end{equation}
where $F(s)=\sum_bc_bs^b$ and $e_0=(1,0,\dots,0)^{\top}$.
\item If $F$ is analytic in the Bernstein ellipse $E_\varrho$ of the interval $[s_{\min},s_{\max}]$ spanned by $s$ on the grid (the ellipse in the complex plane with foci at the ends of the interval and sum of semi-axes equal to $\varrho>1$ times its half-length), with $|F|\le M$ there, then for every $p$ there is a tensor train of ranks $\le p+1$ with
\[
\sup_\eta|G(\eta)-G_p(\eta)|\le\frac{2M\varrho^{-p}}{\varrho-1}.
\]
Ranks of order $\log(1/\varepsilon)$ therefore give accuracy $\varepsilon$. The dimension enters only through the range of $s$ and the bound $M$, i.e.\ through the spread of the density, not through the number of coordinates as such. (The tabulated $F$ is a $C^1$ piecewise cubic, not analytic; (b) applies to its smooth underlying function, and in practice the ranks are chosen adaptively.)
\item If the true argument is $s+R$ with a non-separable remainder $R$, then $\|F(s+R)-F(s)\|_{L^2(\mu)}\le\sup|F'|\,\|R\|_{L^2(\mu)}$.
\end{enumerate}
\end{lemma}

\begin{proof}
(a) By induction on $k$, the first row of $\mathcal G_1\cdots\mathcal G_k$ is $\big(\binom{b}{0}s_k^{\,b}\big)_b=(s_k^b)_b$ with $s_k=\sum_{j\le k}h_j$. This follows from the binomial theorem $(s_{k-1}+h_k)^b=\sum_a\binom ba s_{k-1}^ah_k^{b-a}$. Contracting with $\mathbf c$ gives $F(s_d)$. (b) Truncation after degree $p$ of the expansion in Chebyshev polynomials (the orthogonal polynomials that are near-optimal for uniform approximation on an interval) of a function analytic in $E_\varrho$ has uniform error $2M\varrho^{-p}/(\varrho-1)$ \cite{Trefethen2013}, and a polynomial of degree $p$ has ranks $\le p+1$ by (a). (c) is the mean-value theorem. \end{proof}

\begin{proposition}[Density-adapted coordinates]\label{prop:adapted}
Let $c=\avg{\nabla_\xi\rho_m}$ and $B=\avg{\nabla_\xi^2\rho_m}$ (both given by the pair rule to quadrature accuracy, because $\rho_m$ is a sum of pair functions), let $B=V\,\mathrm{diag}(b)\,V^{\top}$, and set $\eta=V^{\top}\xi$, so that $\eta\sim\mathcal N(0,I)$ again. Then
\begin{equation}\label{eq:adapted}
\delta\rho_m(\eta)=\underbrace{\sum_{k=1}^{d}\Big(\tilde c_k\,\eta_k+\tfrac12\,b_k\,(\eta_k^2-1)\Big)}_{s(\eta):\ \text{separable}}\;+\;R(\eta),
\qquad \tilde c=V^{\top}c,
\end{equation}
where $R$ contains only Hermite chaos of order $\ge3$. Hence $\|R\|^2_{L^2}=\operatorname{Var}\rho_m-|\tilde c|^2-\tfrac12|b|^2$, and by Lemma~\ref{lem:sepsum} the remainder $\tilde F_m(\delta\rho_m)$ has low TT rank in the $\eta$ coordinates. We order the coordinates by their contribution $\tilde c_k^2+\tfrac12b_k^2$ to the variance of $\delta\rho_m$.
\end{proposition}

\begin{proof}
By Lemma~\ref{lemB:chaos}, the chaos components of order 1 and 2 of $\rho_m$ are
\[
\sum_k\avg{\partial_k\rho}\,\eta_k\qquad\text{and}\qquad \tfrac12\sum_{k,l}\avg{\partial_k\partial_l\rho}\,\mathrm{He}_{kl}(\eta).
\]
In the eigenbasis of $B$ the second one is diagonal, $\tfrac12\sum_kb_k(\eta_k^2-1)$. The variance identity is the orthogonality of the chaos decomposition. \end{proof}

\noindent For NiAl the variance of $\delta\rho_m$ is strongly concentrated in these coordinates. The first 5 of them carry $98.4\%$ (Ni site) and $96.9\%$ (Al site) of the separable variance, the first 20 carry $99.85\%$ and $99.87\%$, and the first 50 carry $99.99\%$ (Fig.~\ref{fig:importance}). The tensor train therefore needs fine grids and larger ranks only on a few leading cores.

\begin{proposition}[Gaussian-weighted cross]\label{prop:weighted}
Let $\{x^{(k)}_i,w^{(k)}_i\}$ be the Gauss--Hermite nodes and probability weights of core $k$, and let $\mu_{\rm grid}=\bigotimes_kw^{(k)}$. Apply TT cross-interpolation \cite{OseledetsTyrtyshnikov2010,Savostyanov2011} not to $f$ but to the weighted tensor
\begin{equation}\label{eq:weighted}
g(i_1,\dots,i_d)=f\big(x^{(1)}_{i_1},\dots,x^{(d)}_{i_d}\big)\prod_{k=1}^{d}\sqrt{w^{(k)}_{i_k}/w^{(k)}_{\max}},
\end{equation}
and divide the cores back by the same factors. Then
\begin{enumerate}[label=(\alph*)]
\item the Frobenius error of $g$ (the square root of the sum of squared entrywise errors) equals the $L^2(\mu_{\rm grid})$ error of $f$, i.e.\ its root-mean-square error weighted by the Gaussian grid weights, up to the constant $\prod_kw^{(k)}_{\max}$;
\item for every Hermite moment of an order $|\alpha|$ low enough that the grid integrates $\mathrm{He}_\alpha^2$ exactly,
\begin{equation}\label{eq:momerr}
\big|\avg{\mathrm{He}_\alpha(f-\tilde f)}_{\mu_{\rm grid}}\big|\le\sqrt{\alpha!}\;\|f-\tilde f\|_{L^2(\mu_{\rm grid})};
\end{equation}
\item maximum-volume pivots of $g$ lie where the Gaussian probability mass is. The unweighted cross puts them where $|f|$ is largest on the grid, which for a polynomially growing $f$ means the outermost nodes, whose weight is $10^{-2}$--$10^{-5}$ on the grids used.
\end{enumerate}
\end{proposition}

\begin{proof}
(a) $\sum_i|g_i-\tilde g_i|^2=\big(\prod_kw^{(k)}_{\max}\big)^{-1}\sum_i\big(\prod_kw^{(k)}_{i_k}\big)|f_i-\tilde f_i|^2$. (b) The Cauchy--Schwarz inequality $|\langle a,b\rangle|\le\|a\|\,\|b\|$, with $\avg{\mathrm{He}_\alpha^2}=\alpha!$ on a grid exact to degree $2|\alpha|$. (c) is a heuristic statement, not a theorem. Maxvol maximises $|\det|$ of the pivot block \cite{Goreinov1997}; at rank one it selects the largest entry, and in general it favours rows with large entries. The weights move those entries into the bulk of the Gaussian. \end{proof}

\noindent In a test without the weights (NiAl, 54 atoms, unrelaxed $a=2.882$\,\AA, 300\,K), the cross returned $\avg{\tilde F}=6.8\times10^{-4}$\,eV for a remainder whose true mean is $0.90\times10^{-4}$\,eV (Monte Carlo, $\pm0.02$). With the weights, rank 3 already gives $0.88\times10^{-4}$\,eV.

\paragraph{Implementation of the cross.} The tensor train is constructed by a two-site cross \cite{Savostyanov2011}, which sweeps along the train from left to right and back and updates two neighbouring cores at a time; the scheme is borrowed from the density-matrix renormalisation group (DMRG) of quantum many-body physics, and working on two cores at once lets the ranks adapt to the function. At each bond, i.e.\ at the link between cores $k$ and $k+1$, it evaluates the supercore, the array of function values
\[
g(I_{<k},i_k,i_{k+1},J_{>k+1})
\]
for the current left pivots $I_{<k}$, all nodes $i_k$, $i_{k+1}$ of the two cores, and the current right pivots $J_{>k+1}$. It truncates the singular values of this array at relative tolerance $10^{-6}$ (maximum rank $r_{\max}$), and chooses new pivots with maxvol \cite{Goreinov1997,teneva}. After two sweeps, a final pass over single cores builds the cores $\mathcal G_k=\mathcal C_k\,\mathcal C_k[\mathrm{piv}]^{-1}$, where $\mathcal C_k$ is the array of sampled fibres and $\mathcal C_k[\mathrm{piv}]$ its pivot rows; these cores are well conditioned by the maxvol property. The function has the form $\tilde F_m(\delta\rho_m(M\eta))$, with $M$ the $3n_{\rm nb}\times d$ matrix of relative displacements, so each cross point is evaluated as $L_\alpha+M_kx_{i}+M_{k+1}x_{j}+R_\beta$ from pivot sums prepared in advance. An evaluation then costs $O(n_{\rm nb})$ instead of $O(n_{\rm nb}d)$. The first $10$ cores use 9 Gauss--Hermite nodes and the remaining ones 5.

\begin{lemma}[Exact Hermite contraction of a tensor train]\label{lem:contract}
Let $\tilde f=\mathcal G_1\cdots\mathcal G_d$ be given on nodes $x^{(k)}$ with weights $w^{(k)}$. Define the $r_{k-1}\times r_k$ transfer matrices (the cores averaged against the weights, with or without a Hermite factor)
\begin{align*}
E_k&=\sum_iw^{(k)}_i\,\mathcal G_k(i), &
A_k&=\sum_iw^{(k)}_i\,\mathrm{He}_1\big(x^{(k)}_i\big)\,\mathcal G_k(i),\\
B_k&=\sum_i w^{(k)}_i\,\mathrm{He}_2\big(x^{(k)}_i\big)\,\mathcal G_k(i), &
D_k&=\sum_i w^{(k)}_i\,\mathrm{He}_3\big(x^{(k)}_i\big)\,\mathcal G_k(i),
\end{align*}
the left and right environments (the averaged products of all cores to the left or to the right of core $k$) $\mathcal L_k=E_1\cdots E_{k-1}$ and $\mathcal R_k=E_{k+1}\cdots E_d$, and $E_{k:l}=E_k\cdots E_l$. Then, for $k<l<m$,
\begin{align*}
\avg{\tilde f}&=E_1\cdots E_d, &
\avg{\partial_k\tilde f}&=\mathcal L_kA_k\mathcal R_k,\\
\avg{\partial_k\partial_l\tilde f}&=\mathcal L_kA_kE_{k+1:l-1}A_l\mathcal R_l, &
\avg{\partial_k^2\tilde f}&=\mathcal L_kB_k\mathcal R_k,\\
\avg{\partial_k\partial_l\partial_m\tilde f}&=\mathcal L_kA_kE_{k+1:l-1}A_lE_{l+1:m-1}A_m\mathcal R_m, &
\avg{\partial_k^3\tilde f}&=\mathcal L_kD_k\mathcal R_k,
\end{align*}
with $B$ in place of one $A$ for a repeated index. All $d^2$ second moments cost $O(d^2r^2)$ and all $d^3/6$ third moments cost $O(d^3r)$, by propagating row vectors from the left and column blocks from the right.
\end{lemma}

\begin{proof}
The average of a TT over a product measure factorises core by core. Inserting $\mathrm{He}_{\alpha_k}(x^{(k)})$ in core $k$ gives the discrete Hermite moment, which by Lemma~\ref{lem:smooth} (on the grid) is the smoothed derivative. For the cost, define the column blocks and row blocks
\[
Y_l=\big[E_{l+1:m-1}A_m\mathcal R_m\big]_{m>l},\qquad W_l=\big[\mathcal L_kA_kE_{k+1:l-1}\big]_{k<l}.
\]
They satisfy the recursions
\[
Y_l=\big[A_{l+1}\mathcal R_{l+1},\;E_{l+1}Y_{l+1}\big],\qquad W_{l+1}=\begin{bmatrix}W_lE_l\\ \mathcal L_lA_l\end{bmatrix},
\]
and the third moments for fixed $l$ are the single matrix product $(W_lA_l)\,Y_l$. \end{proof}

\noindent The moments are finally rotated back: $\avg{\nabla^2_u\tilde F_m}=(S^{+\top}V)\,\mathcal H_2\,(S^{+\top}V)^{\top}$, and likewise with three factors for $\mathcal H_3$. They are then summed over the translation orbit \eqref{eq:assemble}. The volume derivative of the remainder needs one more tensor train, for the integrand $\big(F_m'(\rho_m)-F_m'(\bar\rho_m)\big)\,\partial_V\rho_m$. This integrand is a product of two near-separable sums, and it needs a somewhat higher rank ($r_{\max}=10$ versus $6$; Table~\ref{tab:rank}).

\subsection{The complete deterministic scheme and its error}

\noindent The complete scheme is summarised in \hyperref[alg:main]{Algorithm~1}; its error is bounded in Theorem~\ref{thm:budget}.

\begin{keybox}[title={\textbf{Algorithm 1: TT-SCHA with the TT cubic correction}}]
\phantomsection\label{alg:main}\small
\emph{Purpose:} the complete deterministic calculation at one volume and temperature, from the trial Gaussian to the pressure, the energy and the bubble correction; the right margin names the statement on which each step rests.\\[2pt]
\textbf{Input:} lattice $x_0(V)$, potential tables, $T$. \quad \textbf{Start:} $\Phi\leftarrow$ point Hessian (or the converged $\Phi$ of the previous $T$).
\begin{enumerate}[leftmargin=14pt]
\item Eigen-decompose $\Phi=Q\,\mathrm{diag}(\phi)\,Q^{\top}$. Set $\sigma_k=(\beta\phi_k)^{-1/2}$ and $S=Q\,\mathrm{diag}(\sigma)$ \hfill (rank-one Gaussian)
\item \emph{Pair part.} For the 336 (NiAl) or 268 (TiAl) representative pairs, apply the 3-D rule \eqref{eq:pair2}--\eqref{eq:pairV} to the density $f_j$ and to $\phi_{ij}$. This gives $\bar\rho_i$ and, by linearity, all averages of $\psi_{ij}$ \hfill (Prop.~\ref{prop:pairrule}, Lemmas~\ref{lem:split}, \ref{lem:orbit})
\item \emph{Remainder.} For each orbit representative $m$: build the density-adapted coordinates, run the weighted TT-cross of $\tilde F_m$ (\hyperref[alg:C1]{Algorithm~C.1}), and contract its Hermite moments of orders 0--2 (0--3 for the bubble; \hyperref[alg:C2]{Algorithm~C.2}) \hfill (Props.~\ref{prop:adapted}, \ref{prop:weighted}, Lemma~\ref{lem:contract})
\item Assemble $\Phi^{*}$ \eqref{eq:assemble}, symmetrise, and set $\Phi\leftarrow\Phi^{*}$. Repeat steps 1--2 with the remainder frozen until $\|\Delta\Phi\|/\|\Phi\|<10^{-7}$; refresh the remainder (step 3) until its change is $<10^{-4}\|\Phi\|$ (at most six times); finish with $n_g=24$
\item \emph{Observables at $\Phi^{\star}$:} $\avg U$, $P_{\rm HF}=-\avg{\partial_VU|_u}$, $F_{GB}$ \hfill (Prop.~\ref{propB:HF}, Thm.~\ref{thmB:GB})
\item \emph{Bubble:} assemble $\Psi^{*}$ and compute $F_b=-\tfrac{\beta}{12}\avg{\Psi^{*},C^{\otimes3}\Psi^{*}}$ \hfill (Lemma~\ref{lem:wick})
\item Repeat at $V(1\pm0.01)$: $P_b=-\Delta F_b/\Delta V$, and check $-\Delta F_{GB}/\Delta V$ against $P_{\rm HF}$; the two must agree at a stationary point \hfill (stationarity test)
\end{enumerate}
\end{keybox}

\begin{theorem}[Deterministic error budget]\label{thm:budget}
Let $\mathcal G(\Phi)=\mathcal S[\Phi^{*}(\Phi)]$ be the exact SCHA map and $\tilde{\mathcal G}$ the computed one. Suppose $\mathcal G$ is a contraction with constant $L<1$ in a neighbourhood of its fixed point $\Phi^{\star}$, i.e.\ it reduces the distance between any two matrices there by at least the factor $L$, and $\|\tilde{\mathcal G}(\Phi)-\mathcal G(\Phi)\|\le\varepsilon_{\rm quad}+\varepsilon_{\rm TT}$ there. Let $\tilde\Phi$ be the returned matrix and $\delta=\|\tilde{\mathcal G}(\tilde\Phi)-\tilde\Phi\|$ its residual. Then:
\begin{enumerate}[label=(\alph*)]
\item the returned matrix satisfies
\begin{equation}\label{eq:fperr}
\|\tilde\Phi-\Phi^{\star}\|\le\frac{\varepsilon_{\rm quad}+\varepsilon_{\rm TT}+\delta}{1-L};
\end{equation}
\item the error of every observable in \eqref{eq:list} is a sum of three terms: a pair-quadrature term, which vanishes as $n_g\to\infty$; the TT interpolation error of the remainder, bounded by \eqref{eq:momerr}; and the Gauss--Hermite error of the remainder grids. The last two tend to zero as $r_{\max}$ and the node counts grow;
\item \emph{there is no statistical error}. For fixed pivots and grids, repeated runs give identical numbers, finite differences in $V$ or $T$ need no common random numbers, and the $d^3$ ``noise bias'' of the Monte Carlo bubble estimator does not arise.
\end{enumerate}
\end{theorem}

\begin{proof}
(a) $\|\tilde\Phi-\Phi^{\star}\|\le\|\tilde\Phi-\tilde{\mathcal G}(\tilde\Phi)\|+\|\tilde{\mathcal G}(\tilde\Phi)-\mathcal G(\tilde\Phi)\|+\|\mathcal G(\tilde\Phi)-\mathcal G(\Phi^{\star})\|\le\delta+\varepsilon_{\rm quad}+\varepsilon_{\rm TT}+L\|\tilde\Phi-\Phi^{\star}\|$. Rearrange. (b) Each observable is a finite sum of pair averages (Proposition~\ref{prop:pairrule}) and remainder moments (Lemma~\ref{lem:contract}); bound each by its own error. (c) Every step is a deterministic function of $\Phi$. \end{proof}

\noindent In practice the plain iteration $\Phi\leftarrow\mathcal G(\Phi)$ reduces the residual by a factor of about $0.087$ per step (Fig.~\ref{fig:conv}), so $L\approx0.087$ (a measured asymptotic ratio for the pair part, not a proven bound on the factor by which the full map can stretch distances), and the factor $1/(1-L)\approx1.1$ in \eqref{eq:fperr} makes the error of the fixed point essentially equal to the error of a single evaluation of the map. The TT approximation of the remainder is, however, rebuilt at every refresh, and its pivots move slightly with $\Phi$. This sets a floor on the change of the remainder Hessian between refreshes, and hence on $\delta$. The floor is about $5\times10^{-5}\,\|\Phi\|$ for NiAl at 300\,K (about $1\%$ of the remainder contribution) and rises to $1$--$4\times10^{-4}\,\|\Phi\|$ for NiAl at 800\,K and for TiAl at 400--800\,K, where the refresh does not reach its tolerance (Section~\ref{sec:results}).


\section{Protocol, validation and results for NiAl and $\gamma$-TiAl}\label{sec:results}

\subsection{Protocol}
The systems are listed in Table~\ref{tab:systems}; the numerical settings are as follows.
\begin{table}[!htbp]\centering\small
\renewcommand{\arraystretch}{1.1}
\caption{Systems studied. For each crystal structure: the interatomic potential (embedded-atom model), the supercell in units of the cubic (B2) or tetragonal (L1$_0$) unit cell, the number of atoms, the supercell volume at which all calculations are performed, and the purpose of the calculations. The 54-atom NiAl cell is used for the comparison with Monte Carlo sampling and for the convergence tests, the two larger cells for the comparison with molecular dynamics.}\label{tab:systems}
\begin{tabular}{@{}lllrrl@{}}
\toprule
System & Potential & Supercell & Atoms & $V$ (\AA$^3$) & used for\\
\midrule
B2 NiAl & Purja Pun \& Mishin 2009 \cite{PurjaPunMishin2009} & $3^3$ & 54 & 613.25 & validation, convergence\\
B2 NiAl & same & $4^3$ & 128 & 1453.63 & results\\
$\gamma$-TiAl (L1$_0$) & Zope \& Mishin 2003 \cite{ZopeMishin2003} & $3^3$ & 108 & 1806.73 & results\\
\bottomrule
\end{tabular}
\end{table}
\begin{itemize}
\item \textbf{Potential engine.} A vectorised Python re-implementation, based on the array library NumPy, of the LAMMPS routine for embedded-atom potentials of alloys (\texttt{pair\_style eam/alloy}), which evaluates many configurations at once. It uses the same tabulated functions (\texttt{setfl} files, the LAMMPS file format for tabulated embedded-atom potentials), the same piecewise-cubic interpolation of the tables and the same treatment of arguments at the ends of the tables. Energies agree with LAMMPS to $2\times10^{-12}$\,eV and forces to $5\times10^{-14}$\,eV/\AA{} for random displacements of $0.12$\,\AA. The fixed-$u$ volume derivative agrees with a LAMMPS finite difference to all printed digits. The re-implementation serves only to accelerate the evaluations; within these tolerances all energies, forces and volume derivatives coincide with those of LAMMPS.
\item \textbf{Reference.} LAMMPS \cite{LAMMPS} NVT molecular dynamics, i.e.\ at constant number of atoms, volume and temperature, at the same $N$ and $V$: 20\,ps equilibration, 100\,ps production, and block errors, i.e.\ statistical errors estimated from the scatter of averages over consecutive blocks of the trajectory.
\item \textbf{Other references.} MD: LAMMPS NVT, 20\,ps equilibration, 100\,ps production, block errors. Monte Carlo of the exact distribution: HMC, with $P=-\avg{\partial U/\partial V|_u}$. Free-energy integration: $F-F_{GB}=\int_0^1\avg{U-U_{\Phi^\star}}_\lambda d\lambda$ with $U_\lambda=(1-\lambda)U_{\Phi^\star}+\lambda U$, Gauss--Legendre nodes (6 for large cells, 8 for 16 atoms), HMC at each node. Frenkel--Ladd: the same integral from an Einstein crystal (springs of equal stiffness $k$ on all internal coordinates, $F_E=-\tfrac d2\kB T\ln(2\pi\kB T/k)$), 8 nodes, 256 chains. The pressures of the free energies use the secant over $V(1\pm1\%)$. The Monte Carlo columns of Tables~\ref{tab:NiAl} and~\ref{tab:TiAl} use antithetic pairs $\pm\xi$, for which the odd part of the integrand cancels in the even weights, the Stein estimator $\widehat{\Phi^{*}}=-\beta\,\Phi\,\widehat{\avg{uf^{\top}}}$ with $f=-\nabla U$ \cite{Errea2014}, and, for the bubble, the split-half estimator: with independent halves $A$ and $B$ the product $\sum\hat X^A\hat X^B$ is unbiased for $\sum X^2$, whereas $\sum\hat X^2$ is biased by the variance of $\hat X$.
\item \textbf{TT-SCHA.} Pair rule with $n_g=12$ nodes per axis during the iteration and $n_g=24$ for the final iterations and all observables. Remainder: $r_{\max}=6$ for $\tilde F$ (third moments included) and $r_{\max}=10$ for its volume derivative (for TiAl and NiAl at 100\,K the energy of the remainder was also evaluated with $r_{\max}=10$); 9 nodes on the 10 leading density-adapted cores and 5 on the rest; two cross sweeps (passes along the train). Iteration tolerance $10^{-7}$ for the pair part; the remainder is refreshed until its change is below $10^{-4}\|\Phi\|$, at most six times. Bubble energy: $E_b=\partial(\beta F_b)/\partial\beta=-F_b\,\mathrm d\ln\sum X^2/\mathrm d\ln T$, evaluated on the temperature grid. This reduces to $-F_b$ for temperature-independent cubic constants; for NiAl the exponent is $0.95\to0.85$. Volumes $V$ and $V(1\pm0.01)$, the latter for $P_b$ and for the stationarity test (Section~\ref{sec:stationarity}).
\end{itemize}

\subsection{Validation at a fixed Gaussian}\label{sec:validation}
Table~\ref{tab:val} compares the TT integration with a $4\times10^4$-sample antithetic Monte Carlo estimate over the same Gaussian. Energy and pressure agree within a fraction of the Monte Carlo standard error. The TT and Monte Carlo averaged Hessians differ by $1.55\%$ in Frobenius norm (the root of the sum of squared entries), and the difference of the two Monte Carlo halves predicts $1.51\%$ from noise alone. The TT result is therefore consistent with the exact average, and its own error is far below the resolution of $4\times10^4$ random samples. For $\gamma$-TiAl (108 atoms, $2.4\times10^4$ samples) the same comparison gives $2.07\%$ against a noise estimate of $2.00\%$, and energy and pressure again agree within $0.4\sigma$.

\begin{figure}[!tbp]\centering
\includegraphics[width=\textwidth]{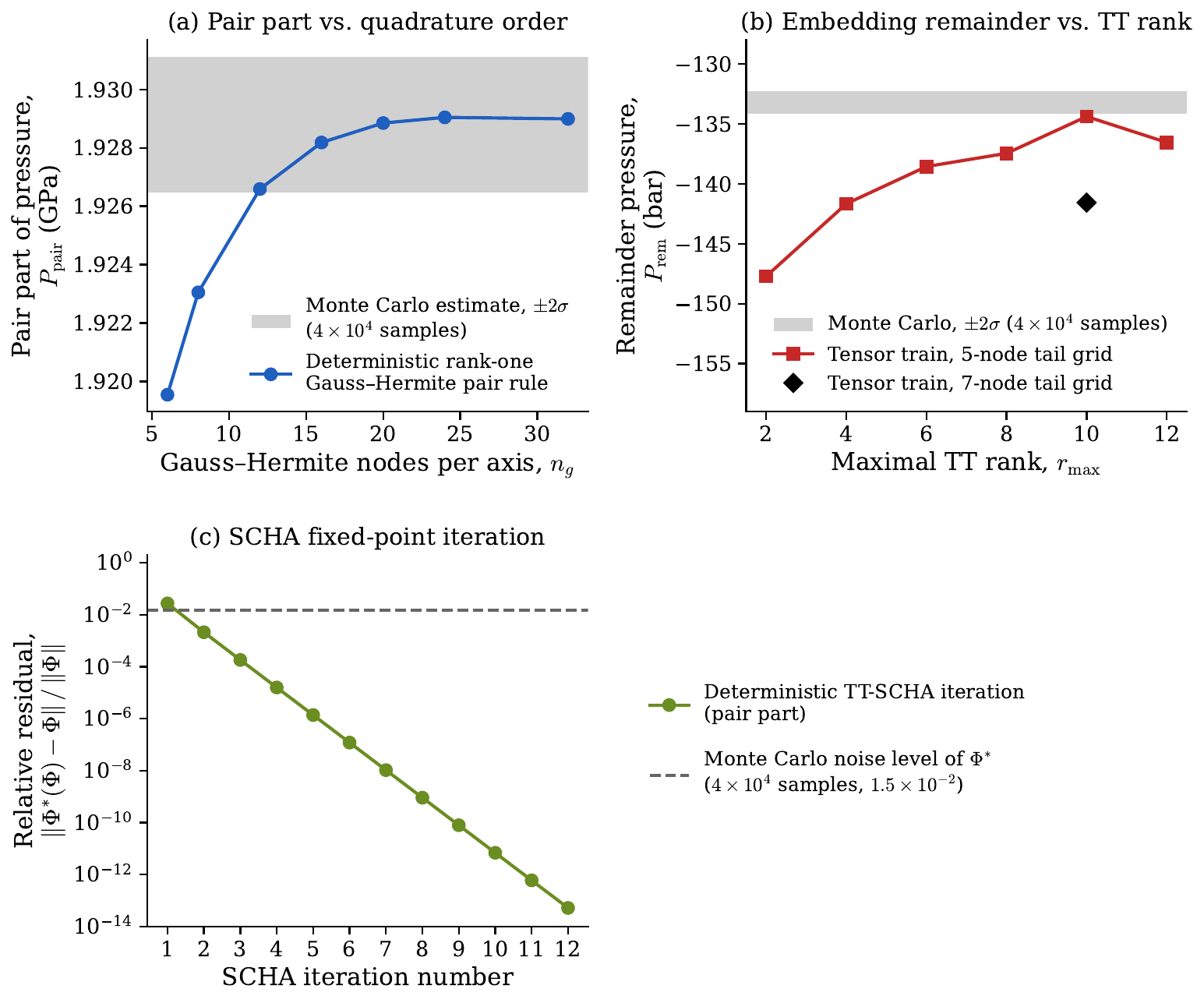}
\caption{Convergence of the deterministic scheme for 54-atom B2 NiAl at 300\,K, evaluated at one fixed Gaussian (the TT-SCHA $\Phis$ of Table~\ref{tab:val}). (a) Pair part of the Hellmann--Feynman pressure, $P_{\rm pair}$ in GPa, against the number $n_g$ of Gauss--Hermite nodes per axis of the three-dimensional pair rule (Proposition~\ref{prop:pairrule}; $n_g=6$--$32$, Table~\ref{tab:ng}; blue circles). The grey band is the Monte Carlo estimate of the same quantity ($\pm2\sigma$, $4\times10^{4}$ antithetic samples), obtained as the Monte Carlo total pressure minus the TT remainder contribution. The quadrature converges algebraically and changes by $0.5$\,bar between $n_g=24$ and $32$, well inside the Monte Carlo band. (b) Contribution of the embedding remainder to the pressure, $P_{\rm rem}$ in bar, against the maximal TT rank $r_{\max}=2$--$12$ (red squares: 9 Gauss--Hermite nodes on the 10 leading density-adapted cores and 5 on the others; Table~\ref{tab:rank}). The black diamond is $r_{\max}=10$ with 7 nodes on the tail cores. The grey band is the Monte Carlo estimate of the same quantity at the same Gaussian and the same $\bar\rho$ ($\pm2\sigma$, $4\times10^{4}$ samples; Section~\ref{sec:validation}). For $r_{\max}\ge6$ the values vary within a range of $4$\,bar, compared with a total pressure of $1.9\times10^{4}$\,bar; the finer tail grid shifts $P_{\rm rem}$ by $7$\,bar. (c) Relative residual $\|\Phi^{*}(\Phi)-\Phi\|/\|\Phi\|$ of the SCHA fixed-point iteration (pair part, started from the point Hessian) against the iteration number, on a logarithmic scale. It decreases by a factor of about $0.087$ per iteration and reaches $5\times10^{-14}$ after 12 iterations. The dashed line marks the statistical noise of a Monte Carlo estimate of $\Phi^{*}$ with $4\times10^{4}$ samples ($1.5\times10^{-2}$, Table~\ref{tab:val}), below which a stochastic iteration cannot converge.}
\label{fig:conv}
\end{figure}

Table~\ref{tab:ng} and Fig.~\ref{fig:conv}(a) show the convergence of the pair rule. $\Phi^{*}$ is converged to $10^{-4}$ already at $n_g=8$. The pressure converges algebraically, because the potential has third-derivative jumps at its spline knots (Remark after Proposition~\ref{prop:pairrule}). Its change from $n_g=24$ to $32$ is $0.5$\,bar.

\begin{figure}[!tbp]\centering
\includegraphics[width=0.55\textwidth]{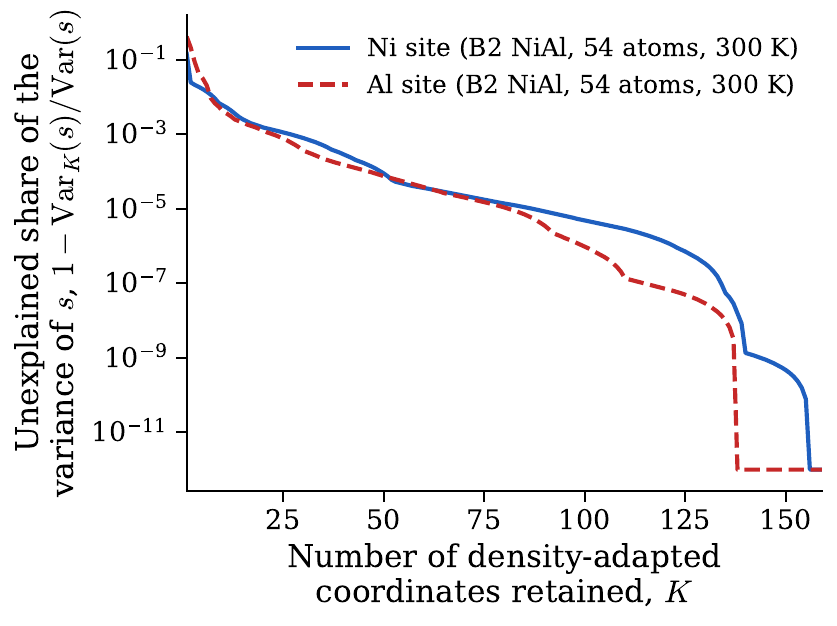}
\caption{Concentration of the density fluctuation in the density-adapted coordinates of Proposition~\ref{prop:adapted}, for 54-atom B2 NiAl at 300\,K (same Gaussian as Table~\ref{tab:val}). Horizontal axis: number $K$ of leading density-adapted coordinates, ordered by their contribution $\tilde c_k^2+\tfrac12b_k^2$ to the variance, up to $K=d=159$. Vertical axis (logarithmic): share of the variance of the separable part $s$ of $\delta\rho_m$ that is \emph{not} carried by the first $K$ coordinates, for the Ni site (blue) and the Al site (red). The first 5 coordinates carry $98.4\%$ (Ni) and $96.9\%$ (Al) of the separable variance, the first 20 carry $99.85\%$ and $99.87\%$, and the first 50 carry $99.99\%$. Fine grids and larger TT ranks are therefore required only on a few leading cores of the tensor train.}
\label{fig:importance}
\end{figure}

Table~\ref{tab:rank} and Fig.~\ref{fig:conv}(b) show the remainder. For $r_{\max}\ge6$ its energy stays within $0.3\%$, and the norm of its Hessian within $0.2\%$, of the values at $r_{\max}=12$, and the remainder carries only $0.54\%$ of $\Phi^{*}$. For $r_{\max}\ge6$ its pressure contribution varies within a range of $4$\,bar, out of a total pressure of $1.9\times10^{4}$\,bar. A separate Monte Carlo estimate at the same Gaussian and the same $\bar\rho$ ($4\times10^{4}$ samples) gives $\sum_m|\mathcal O_m|\avg{\tilde F_m}=5.839\pm0.009$\,meV and $P_{\rm rem}=-133.2\pm0.5$\,bar. The TT values are therefore biased low by $0.4$--$0.7\%$ in energy and high by $1$--$8$\,bar in magnitude in pressure. The size is consistent with the non-separable part $R$ of Proposition~\ref{prop:adapted}. By orthogonality of the Hermite chaos components (Lemma~\ref{lemB:chaos}), $\avg{\tilde F(s+R)}-\avg{\tilde F(s)}\approx\tfrac12F''\avg{R^2}$, which with $\|R\|^2\approx2\%$ of $\operatorname{Var}\rho_m$ is about $2\%$ of the remainder. The finite Gauss--Hermite grids of the train contribute as well: changing the tail grid from 5 to 7 nodes moves $P_{\rm rem}$ by $7$\,bar. In absolute terms the bias is $3\times10^{-5}$\,eV in $\avg U$ and at most $0.05\%$ of $P$, below the MD error bars. It could be removed by treating the quadratic term $\tfrac12F''(\bar\rho)\,\delta\rho^2$ exactly as a sum of two-neighbour integrals, leaving only the cubic and higher terms of $\tilde F$ to the tensor train (a control variate: a part of the integrand with known average is subtracted and added back exactly). The density-adapted coordinates work as Proposition~\ref{prop:adapted} predicts. The separable part $s$ carries $98.0\%$ (Ni site) and $98.5\%$ (Al site) of the variance of $\rho_m$. The variance itself is a Monte Carlo estimate with a relative error of about $1\%$, which is why the Ti value in Table~\ref{tab:val} exceeds one. Within $s$, the variance is concentrated on a handful of coordinates (Fig.~\ref{fig:importance}).

\begin{table}[t]\centering\small
\caption{Validation of the deterministic TT integration against Monte Carlo sampling at one fixed Gaussian, the converged TT-SCHA $\Phis$ at 300\,K, for 54-atom B2 NiAl ($4\times10^{4}$ samples) and 108-atom $\gamma$-TiAl ($2.4\times10^{4}$ samples). TT: pair rule with $n_g=24$ nodes per axis and remainder with $r_{\max}=10$. Monte Carlo (MC) errors are $1\sigma$, estimated from antithetic pairs, and the last column gives the TT$-$MC difference in units of this error. The ``MC noise'' of $\Phi^{*}$ is half the norm of the difference between the estimates from two independent halves of the sample, i.e.\ the discrepancy expected from sampling noise alone. The remainder share is the relative norm of the embedding-remainder contribution to $\Phi^{*}$, and $\operatorname{Var}(s)/\operatorname{Var}(\rho_m)$ is the fraction of the variance of the site density carried by the separable part of Proposition~\ref{prop:adapted} on the two sublattices (the variance itself is a Monte Carlo estimate with a relative error of about $1\%$, so that values slightly above one can occur). Energies and pressures agree within $0.4\sigma$, and the TT$-$MC difference of $\Phi^{*}$ coincides with the Monte Carlo noise level: the error of the TT integration is not resolved by the sampling.}\label{tab:val}
\begin{tabular}{@{}lrrr@{}}\toprule
quantity & TT (deterministic) & Monte Carlo & (TT$-$MC)$/\sigma_{\rm MC}$\\\midrule
\multicolumn{4}{@{}l}{\emph{B2 NiAl, 54 atoms}, 40000 Monte Carlo samples}\\
$\avg{U}$ [eV] & -241.55756 & -241.55753 $\pm$ 0.00116 & -0.0\\
$P_{\rm HF}$ [GPa] & 1.91560 & 1.91536 $\pm$ 0.00116 & +0.2\\
\quad remainder part of $P$ [bar] & -134.4 & & \\
$\|\Phi^{*}_{\rm TT}-\Phi^{*}_{\rm MC}\|/\|\Phi^{*}\|$ & 1.55e-02 & MC noise 1.51e-02 & \\
$\|\Phi^{*}_{\rm rem}\|/\|\Phi^{*}\|$ (remainder share) & 5.39e-03 & & \\
$\operatorname{Var}(s)/\operatorname{Var}(\rho_m)$, two sublattices & 0.980, 0.985 & & \\[3pt]
\multicolumn{4}{@{}l}{\emph{$\gamma$-TiAl, 108 atoms}, 24000 Monte Carlo samples}\\
$\avg{U}$ [eV] & -482.77377 & -482.77345 $\pm$ 0.00213 & -0.1\\
$P_{\rm HF}$ [GPa] & 1.47246 & 1.47218 $\pm$ 0.00077 & +0.4\\
\quad remainder part of $P$ [bar] & -22.4 & & \\
$\|\Phi^{*}_{\rm TT}-\Phi^{*}_{\rm MC}\|/\|\Phi^{*}\|$ & 2.07e-02 & MC noise 2.00e-02 & \\
$\|\Phi^{*}_{\rm rem}\|/\|\Phi^{*}\|$ (remainder share) & 8.15e-03 & & \\
$\operatorname{Var}(s)/\operatorname{Var}(\rho_m)$, two sublattices & 1.003, 0.987 & & \\[3pt]
\bottomrule
\end{tabular}
\end{table}

\begin{table}[t]\centering\small
\caption{Convergence of the pair part with the number $n_g$ of Gauss--Hermite nodes per axis of the three-dimensional rank-one rule (Proposition~\ref{prop:pairrule}), for 54-atom B2 NiAl at 300\,K and the same Gaussian as Table~\ref{tab:val}. Columns: pair contribution to the average potential energy, pair contribution to the Hellmann--Feynman pressure, and wall-clock time on one core for the averages of all representative pairs. The pair Hessian changes by $\|\Phi_8-\Phi_{32}\|/\|\Phi_{32}\|=9.3e-05$ between $n_g=8$ and $n_g=32$. The pressure converges algebraically, because the tabulated potential has third-derivative jumps at its spline knots; between $n_g=24$ and $n_g=32$ it changes by $0.5$\,bar.}\label{tab:ng}
\begin{tabular}{@{}rrrr@{}}\toprule
$n_g$ & $\avg{U}_{\rm pair}$ [eV] & $P_{\rm pair}$ [bar] & time [s]\\\midrule
6 & -241.56347 & 19195.5 & 0.06\\
8 & -241.56329 & 19230.5 & 0.08\\
12 & -241.56327 & 19265.9 & 0.31\\
16 & -241.56331 & 19281.8 & 0.86\\
20 & -241.56334 & 19288.5 & 2.32\\
24 & -241.56337 & 19290.4 & 3.09\\
32 & -241.56339 & 19289.9 & 7.03\\
\bottomrule
\end{tabular}
\end{table}

\begin{table}[t]\centering\small
\caption{Convergence of the embedding remainder with the maximal TT rank $r_{\max}$ for 54-atom B2 NiAl at 300\,K (same Gaussian as Table~\ref{tab:val}; two orbit sites, Ni and Al). The weighted TT-cross uses 9 Gauss--Hermite nodes on the 10 leading density-adapted cores and $n_t$ nodes on the remaining cores. Columns: remainder contribution to the average energy, $\sum_m|\mathcal O_m|\avg{\tilde F_m}$; remainder contribution to the pressure, $P_{\rm rem}$; norm of the remainder contribution to $\Phi^{*}$; and number of evaluations of the site function (``evals'') for both sites and both integrands, $\tilde F_m$ and its volume derivative. For $r_{\max}\ge6$ the energy of the remainder stays within $0.3\%$, and the norm of its Hessian within $0.2\%$, of the values at $r_{\max}=12$, and $P_{\rm rem}$ varies within a range of $4$\,bar, i.e.\ $0.02\%$ of the total pressure of $1.9\times10^{4}$\,bar. The last row shows the effect of a finer grid on the tail cores, which moves $P_{\rm rem}$ by $7$\,bar ($0.04\%$ of the total pressure). A Monte Carlo reference for the remainder with the same $\bar\rho$ is discussed in Section~\ref{sec:validation}.}\label{tab:rank}
\begin{tabular}{@{}rrrrrr@{}}\toprule
$r_{\max}$ & $n_t$ & $\sum_m|\mathcal O_m|\avg{\tilde F_m}$ [meV] & $P_{\rm rem}$ [bar] & $\|\Phi^{*}_{\rm rem}\|$ [eV/\AA$^2$] & evals\\\midrule
2 & 5 & 6.214 & -147.71 & 0.6325 & 2.49e+05\\
4 & 5 & 5.846 & -141.66 & 0.7417 & 7.59e+05\\
6 & 5 & 5.816 & -138.56 & 0.7422 & 1.49e+06\\
8 & 5 & 5.806 & -137.46 & 0.7405 & 2.11e+06\\
10 & 5 & 5.809 & -134.39 & 0.7434 & 3.01e+06\\
12 & 5 & 5.801 & -136.55 & 0.7419 & 3.81e+06\\
10 & 7 & 5.818 & -141.57 & 0.7438 & 5.17e+06\\
\bottomrule
\end{tabular}
\end{table}

\begin{table}[t]\centering\small
\caption{Pressure and potential energy of 128-atom B2 NiAl at fixed volume $V=1453.63$\,\AA$^3$ for $T=100$--$800$\,K. LAMMPS MD: NVT molecular dynamics at the same $N$ and $V$, with the $1\sigma$ block error of the pressure given in parentheses in units of the last digit. TT-SCHA + TT bubble: Hellmann--Feynman pressure $P_{\rm HF}$ of the self-consistent Gaussian, bubble pressure $P_b=-\Delta F_b/\Delta V$ from the TT third moments, total pressure $P=P_{\rm HF}+P_b$ (all in GPa), and potential energy per atom $\avg U_{\rm SCHA}+E_b$, where $E_b=\partial(\beta F_b)/\partial\beta=-F_b\,\mathrm d\ln\sum X^2/\mathrm d\ln T$. MC: total pressure from a Monte Carlo evaluation of the same SCHA and bubble averages (Appendix~\ref{app:B}). The last two columns give the deviation of the TT-SCHA + TT bubble pressure (in \%) and potential energy per atom (in meV) from MD.}\label{tab:NiAl}
\setlength{\tabcolsep}{4pt}
\begin{tabular}{@{}r|rr|rrr r|r|rr@{}}\toprule
 & \multicolumn{2}{c|}{LAMMPS MD} & \multicolumn{4}{c|}{TT-SCHA + TT bubble (this work)} & MC & \multicolumn{2}{c}{$\Delta$ vs MD}\\
$T$ [K] & $P$ [GPa] & $\avg U/N$ [eV] & $P_{\rm HF}$ & $P_b$ & $P$ [GPa] & $\avg U/N$ [eV] & $P$ [GPa] & $P$ [\%] & $U$ [meV]\\\midrule
100 & 0.6712(34) & -4.49803 & 0.6654 & +0.0028 & 0.6682 & -4.49810 & 0.6685 & -0.45 & -0.07\\
200 & 1.3170(56) & -4.48537 & 1.3028 & +0.0121 & 1.3149 & -4.48535 & 1.3140 & -0.16 & +0.02\\
300 & 1.9666(63) & -4.47236 & 1.9253 & +0.0269 & 1.9523 & -4.47258 & 1.9533 & -0.73 & -0.22\\
400 & 2.6027(77) & -4.45973 & 2.5348 & +0.0468 & 2.5817 & -4.45977 & 2.5847 & -0.81 & -0.05\\
600 & 3.8726(169) & -4.43391 & 3.7280 & +0.0983 & 3.8263 & -4.43404 & 3.8310 & -1.19 & -0.13\\
800 & 5.0955(125) & -4.40768 & 4.8847 & +0.1617 & 5.0464 & -4.40815 & 5.0419 & -0.97 & -0.47\\
\bottomrule
\end{tabular}
\end{table}

\begin{table}[t]\centering\small
\caption{Pressure and potential energy of 108-atom $\gamma$-TiAl (L1$_0$) at fixed volume $V=1806.73$\,\AA$^3$ for $T=100$--$800$\,K. Columns and conventions as in Table~\ref{tab:NiAl}.}\label{tab:TiAl}
\setlength{\tabcolsep}{4pt}
\begin{tabular}{@{}r|rr|rrr r|r|rr@{}}\toprule
 & \multicolumn{2}{c|}{LAMMPS MD} & \multicolumn{4}{c|}{TT-SCHA + TT bubble (this work)} & MC & \multicolumn{2}{c}{$\Delta$ vs MD}\\
$T$ [K] & $P$ [GPa] & $\avg U/N$ [eV] & $P_{\rm HF}$ & $P_b$ & $P$ [GPa] & $\avg U/N$ [eV] & $P$ [GPa] & $P$ [\%] & $U$ [meV]\\\midrule
100 & 0.4963(26) & -4.49570 & 0.4985 & +0.0030 & 0.5016 & -4.49570 & 0.5023 & +1.06 & +0.00\\
200 & 0.9977(47) & -4.48265 & 0.9882 & +0.0117 & 0.9999 & -4.48275 & 1.0009 & +0.22 & -0.11\\
300 & 1.5034(70) & -4.46972 & 1.4725 & +0.0255 & 1.4979 & -4.46971 & 1.4996 & -0.37 & +0.00\\
400 & 2.0081(71) & -4.45637 & 1.9512 & +0.0439 & 1.9952 & -4.45659 & 1.9964 & -0.65 & -0.22\\
600 & 3.0063(76) & -4.42987 & 2.8960 & +0.0926 & 2.9886 & -4.43018 & 2.9674 & -0.59 & -0.31\\
800 & 4.0091(152) & -4.40316 & 3.8209 & +0.1558 & 3.9767 & -4.40359 & 3.9706 & -0.81 & -0.43\\
\bottomrule
\end{tabular}
\end{table}

\begin{table}[t]\centering\small
\caption{Internal consistency and computational cost of the TT-SCHA + TT bubble calculations for 128-atom B2 NiAl and 108-atom $\gamma$-TiAl. $-\Delta F_{GB}/\Delta V$ is the secant pressure (GPa) over $V(1\pm0.01)$, obtained from the Gibbs--Bogoliubov free energy alone, i.e.\ from $\avg U$ and $\ln\det\Phis$ without use of $\partial_VU$. $P_{\rm Simpson}=(P_{\rm HF}^{+}+4P_{\rm HF}+P_{\rm HF}^{-})/6$ is the Simpson average of the Hellmann--Feynman pressure over the same interval, i.e.\ its mean by Simpson's rule, the three-point quadrature with weights $1/6$, $4/6$, $1/6$, $P_{\rm HF}^{\pm}$ denoting its values at $V(1\pm0.01)$. The two agree if and only if the computed $\Phis$ is stationary (Proposition~\ref{propB:HF}); their difference is given in MPa. $\sum X^2$ is the dimensionless bubble sum of Theorem~\ref{thmB:bubble}, given in total and for the embedding-remainder part alone. Time: wall-clock seconds per temperature on one core for all three volumes, including the bubble.}\label{tab:diag}
\setlength{\tabcolsep}{4pt}
\begin{tabular}{@{}lr|rrr|rr|r@{}}\toprule
system & $T$ & $-\Delta F_{GB}/\Delta V$ & $P_{\rm Simpson}$ & diff. [MPa] & $\sum X^2$ & remainder & time [s]\\\midrule
NiAl & 100 & 0.68009 & 0.68006 & +0.0 & 7.023 & 7.8e-04 & 1429\\
NiAl & 200 & 1.31761 & 1.31733 & +0.3 & 13.565 & 1.5e-03 & 1117\\
NiAl & 300 & 1.93974 & 1.93967 & +0.1 & 19.716 & 2.2e-03 & 983\\
NiAl & 400 & 2.55032 & 2.54977 & +0.6 & 25.557 & 2.8e-03 & 941\\
NiAl & 600 & 3.74280 & 3.74196 & +0.8 & 36.523 & 4.0e-03 & 997\\
NiAl & 800 & 4.90440 & 4.89935 & +5.0 & 46.651 & 5.2e-03 & 1218\\
TiAl & 100 & 0.51524 & 0.51515 & +0.1 & 7.093 & 2.4e-04 & 628\\
TiAl & 200 & 1.00476 & 1.00473 & +0.0 & 13.975 & 4.3e-04 & 803\\
TiAl & 300 & 1.48919 & 1.48905 & +0.1 & 20.857 & 6.1e-04 & 853\\
TiAl & 400 & 1.96735 & 1.96804 & -0.7 & 27.796 & 7.4e-04 & 909\\
TiAl & 600 & 2.91151 & 2.91283 & -1.3 & 41.812 & 9.5e-04 & 910\\
TiAl & 800 & 3.83622 & 3.83670 & -0.5 & 55.924 & 1.3e-03 & 946\\
\bottomrule
\end{tabular}
\end{table}


\subsection{Self-consistency and the stationarity test}\label{sec:stationarity}
From the point Hessian, the pair part of the deterministic SCHA iteration converges linearly with ratio $0.087$ (Fig.~\ref{fig:conv}(c)), down to $5\times10^{-14}$. At the fixed point of the 54-atom test the full TT map, including a freshly built remainder, returns its input to $3\times10^{-5}$.

Proposition~\ref{propB:HF} provides a stringent test of the whole construction. If $\Phis$ is stationary, the Hellmann--Feynman pressure, an average of $\partial_VU$ at fixed $\Phi$, must equal minus the volume derivative of the variational free energy $F_{GB}(V)$. The latter is computed from $\avg U$ and $\ln\det\Phis$ at three volumes and does not involve $\partial_VU$ at all. Since $P(V)$ is curved over $V(1\pm0.01)$, the secant $-\Delta F_{GB}/\Delta V$ has to be compared with the average of $P_{\rm HF}$ over the same interval by Simpson's rule. Table~\ref{tab:diag} shows agreement to within $1.3$\,MPa, i.e.\ $5\times10^{-4}$ of the pressure or better, in all cases except NiAl at 800\,K, where the difference is $5$\,MPa ($10^{-3}$). This is not a quadrature effect: the second difference of $P_{\rm HF}$ over the volume interval is nearly the same at all temperatures. For NiAl, 800\,K is the only temperature at which the refresh of the remainder Hessian did not reach its tolerance within six refreshes (next paragraph). For TiAl above 300\,K the refresh stalls as well, at a similar or larger level, but there the stationarity error remains below $5\times10^{-4}$. With Monte Carlo averages this test is limited by the statistical error of $F_{GB}$, about $10^{-3}$\,eV for $10^3$ samples, i.e.\ about $5$\,MPa after division by $\Delta V$.

\paragraph{Self-consistency of the remainder.} The pair part of the fixed point converges to $10^{-7}$ or better at every $(T,V)$. The remainder Hessian is rebuilt by TT-cross at each refresh, and its change between refreshes falls below $10^{-4}\,\|\Phi\|$ within two refreshes for NiAl up to 600\,K and for TiAl up to 300\,K. At higher temperatures it stalls at $1$--$4\times10^{-4}\,\|\Phi\|$. That is up to about $5\%$ of the remainder contribution and $4\times10^{-4}$ of $\Phi$, and it sets the level of residual self-consistency there. The total number of fixed-point iterations per $(T,V)$, including all refreshes, is 10--19, rising to 26--36 where the refresh loop runs to its limit.

\subsection{Pressure and energy of NiAl and $\gamma$-TiAl}
\begin{figure}[!tbp]\centering
\includegraphics[width=\textwidth]{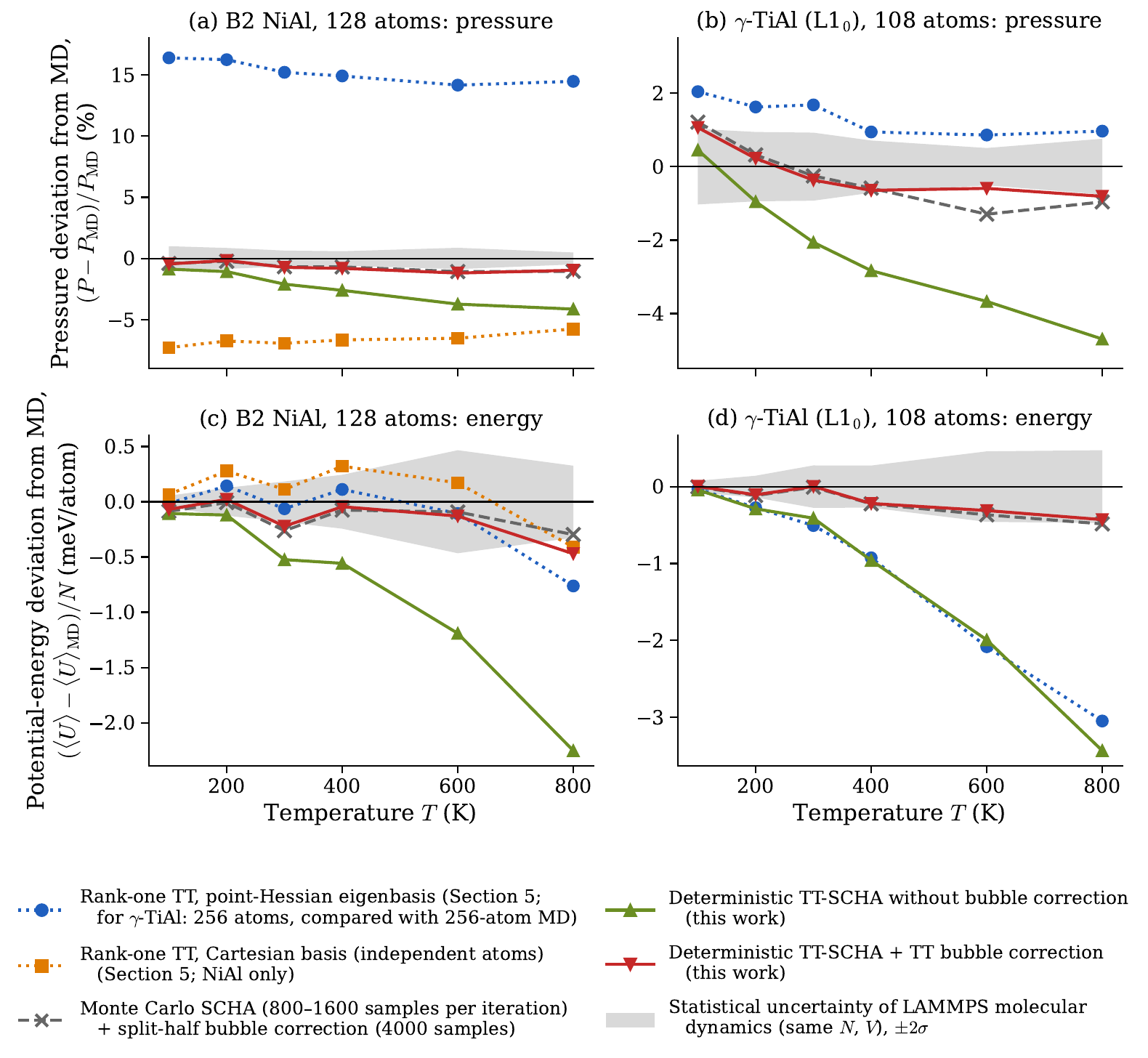}
\caption{Deviation from LAMMPS NVT molecular dynamics at the same $N$ and $V$, against temperature ($100$--$800$\,K). Top row: relative pressure deviation $(P-P_{\rm MD})/P_{\rm MD}$ in \%; bottom row: deviation of the potential energy per atom, $\avg U/N-\avg U_{\rm MD}/N$ in meV. Left column: 128-atom B2 NiAl; right column: 108-atom $\gamma$-TiAl. The grey band is the MD $\pm2\sigma$ block error. Grey crosses (dashed): SCHA + bubble with all Gaussian averages evaluated by Monte Carlo sampling (Appendix~\ref{app:B}). Green upward triangles: deterministic TT-SCHA without the cubic correction ($P_{\rm HF}$, $\avg U_{\rm SCHA}$). Red downward triangles: deterministic TT-SCHA + TT bubble of this work ($P_{\rm HF}+P_b$, $\avg U_{\rm SCHA}+E_b$). The rank-one eigenbasis result overestimates the NiAl pressure by $14$--$16\%$ and the Cartesian (independent-atom) result underestimates it by about $6$--$7\%$; the TT-SCHA + TT bubble pressure lies within $1.2\%$ of MD for both alloys, and its energy within $0.5$\,meV per atom.}
\label{fig:dev}
\end{figure}

\begin{figure}[!tbp]\centering
\includegraphics[width=\textwidth]{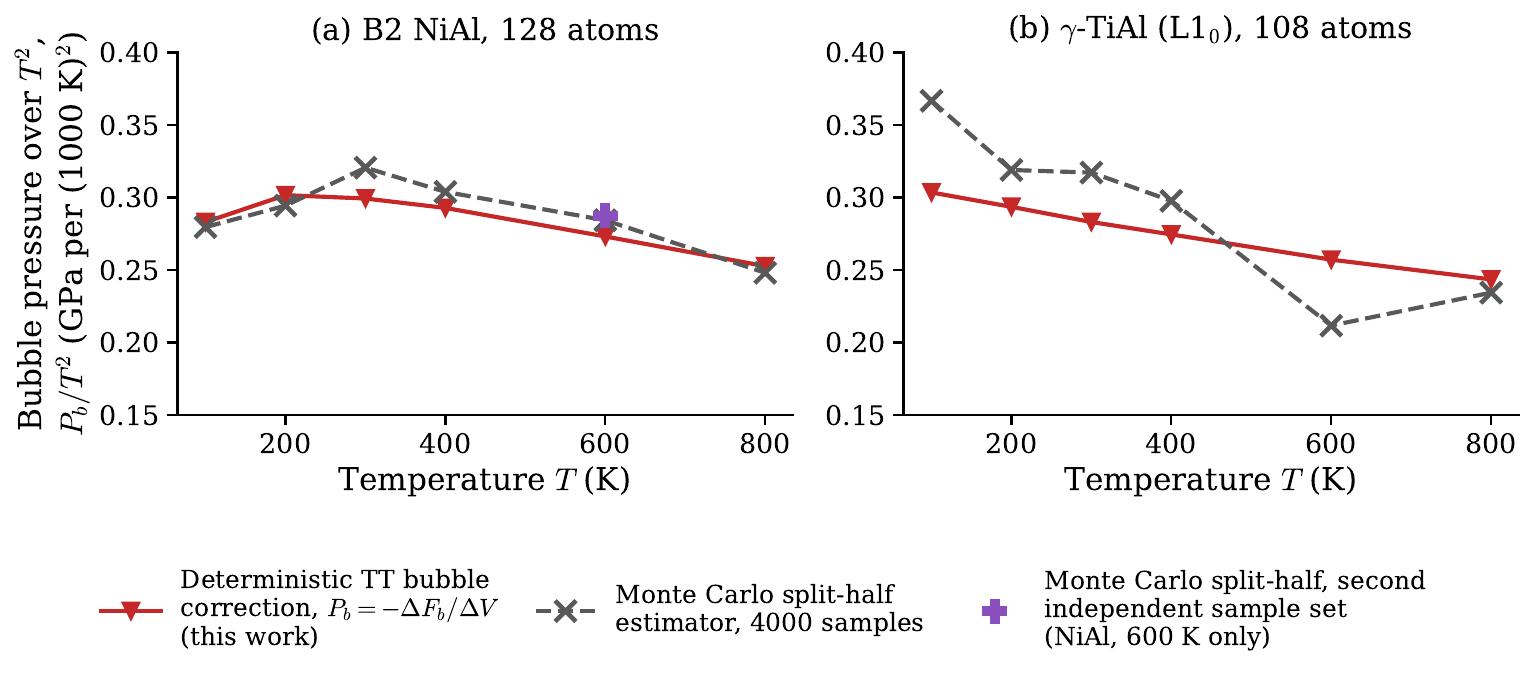}
\caption{Bubble (cubic) contribution to the pressure divided by the square of the temperature, $P_b/T^2$ in GPa per $(1000\,{\rm K})^2$, against temperature at fixed volume for 128-atom B2 NiAl (left) and 108-atom $\gamma$-TiAl (right), $T=100$--$800$\,K. This ratio would be independent of temperature if the cubic force constants and their volume derivatives did not depend on $T$. Red downward triangles: deterministic TT result, $P_b=-\Delta F_b/\Delta V$ over $V(1\pm0.01)$, with $F_b$ from the Wick contraction (Lemma~\ref{lem:wick}) of the TT third moments. Grey crosses (dashed): Monte Carlo split-half estimator with $4000$ antithetic Gaussian samples; for NiAl an estimate from a second, independent sample set at 600\,K is shown as a plus sign. The TT curve is smooth; for NiAl it rises by $7\%$ between 100 and 200\,K and then decreases by $16\%$ up to 800\,K, and for TiAl it decreases monotonically by $20\%$ between 100 and 800\,K. This temperature dependence reflects the change with temperature of the thermally averaged cubic constants and of their volume dependence. The Monte Carlo estimates scatter about the TT curve by up to $20\%$.}
\label{fig:bubble}
\end{figure}

Tables~\ref{tab:NiAl} and \ref{tab:TiAl} and Figs.~\ref{fig:dev}--\ref{fig:bubble} give the results.

\paragraph{Agreement with molecular dynamics.} For B2 NiAl (128 atoms) the deterministic TT-SCHA + TT-bubble pressure lies $0.2$--$1.2\%$ below LAMMPS, and the potential energy agrees within $0.25$\,meV per atom up to 600\,K and $0.5$\,meV at 800\,K. For $\gamma$-TiAl (108 atoms) the pressure deviation ranges from $+1.1\%$ at 100\,K, a $2\sigma$ MD fluctuation, to $-0.8\%$ at 800\,K, and the energy agrees within $0.5$\,meV per atom. The deviations are shown in Fig.~\ref{fig:dev}.

\paragraph{Comparison with a Monte Carlo evaluation of the same Gaussian averages.} The TT results and a Monte Carlo (MC) evaluation of the same SCHA and bubble averages (column ``MC'') differ by at most $0.15\%$ in pressure, with one exception: TiAl at 600\,K, where the Monte Carlo bubble pressure is $0.021$\,GPa too low. The TT bubble pressure is a smooth function of temperature. It rises from $0.4\%$ (100\,K) to $3.2\%$ (800\,K) of the total pressure for NiAl, and from $0.6\%$ to $3.9\%$ for TiAl (Fig.~\ref{fig:bubble}). For TiAl, $\sum X^2/T$ is constant to $2\%$ over the whole range (Table~\ref{tab:diag}), as Theorem~\ref{thmB:bubble} predicts for temperature-independent cubic constants. For NiAl it falls by $17\%$ between 100 and 800\,K. This is the thermal smoothing of the knot structure of the Mishin potential (Lemma~\ref{lem:smooth}), here acting on the third derivatives, and the TT result resolves it without statistical noise. That the structure is real, and not an artefact of finite differences, was checked with analytic derivatives of the potential. Replacing the tabulated pair function by a version smoothed over $0.03$\,\AA{} removes $91\%$ of the anomaly in the thermal pressure per kelvin (which falls from $81$ to $67$\,bar/K between low and high temperature), while smoothing the embedding function or the density function removes none of it. Two shells of the pair function, the Al--Al shell at $4.005$\,\AA{} and the Ni--Al shell at $4.696$\,\AA, together account for $89\%$. These are knots of the tabulated file; the knots of the original published spline have not been compared. $P_b/T^2$ is not constant either, because the volume derivative of $\sum X^2$ changes with temperature as well: for NiAl it rises by $7\%$ between 100 and 200\,K and then decreases by $16\%$ up to 800\,K, and for TiAl it decreases monotonically, by $20\%$ between 100 and 800\,K. The Monte Carlo estimates of $P_b$ are differences of two noisy $F_b$ values at $V(1\pm0.01)$ and differ from the TT values by up to $20\%$. For TiAl below 600\,K they lie systematically above the TT values; statistical errors are not available for these estimates.

\paragraph{Origin of the remaining deviation.} The TT errors of Theorem~\ref{thm:budget} are below $0.05\%$ of $P$, far smaller than the remaining deviation from MD, which is therefore not caused by the numerical integration. Section~\ref{sec:T3} shows that it is the sum of the terms of order $T^3$ in the free energy, which the bubble truncation omits: the quartic loop $\|W_4\|^2$, the product $\avg{W_3^2W_4}$ and the fourth cumulant of the cubic component (Proposition~\ref{propB:T3}). They are evaluated there by TT-cross, and they remove most of the deficit at 600 and 800\,K.

\subsection{Next-order terms of the free energy by tensor-train cross}\label{sec:T3}
\paragraph{What is computed.} Beyond the cubic bubble, the free energy contains three terms of order $T^3$ (Proposition~\ref{propB:T3}): $F_{4a}$, the second-order contribution of the average fourth derivatives of the potential; $F_{4b}$, the product of the square of the cubic component with the quartic component; and $F_{4c}$, the fourth cumulant (a measure of the non-Gaussian shape) of the cubic component. They are built from the averaged third-derivative tensor $D^{(3)}$, which is the one already used for the bubble, and from the averaged fourth-derivative tensor $D^{(4)}$ in the whitened coordinates.

\paragraph{Next order.} By the power counting of Appendix~\ref{app:B} ($\|W_n\|^2\propto T^n$) the terms of order $T^4$ contain derivatives up to the sixth order and higher cumulants of the cubic component, for example $\beta\|W_5\|^2$ and $\beta^2\langle W_3W_3W_6\rangle$. They have the same local structure, so in principle the same machinery applies, but they need the fifth- and sixth-derivative tensors and larger contractions than the $O(d^5)$ tetrahedron sum used here. We have not derived the complete list or computed any of them; the rough estimate of Table~\ref{tab:pnext} is not a substitute.

\paragraph{How.} The fourth derivatives follow the same splitting of the embedded-atom energy as the third ones. The pair part is evaluated with the three-dimensional Gauss--Hermite rule of Proposition~\ref{prop:pairrule}, using Stein's identity with the Hermite polynomial of order four ($16$ nodes per axis). For the many-body remainder the weighted TT-cross of Appendix~\ref{app:C} supplies the train, and the Hermite moments of order four are contracted exactly from its cores by a left-to-right sweep that carries one row vector per way of distributing the four derivatives over the cores (the extension of Lemma~\ref{lem:contract} to order four; tested against the full tensor to $2\times10^{-16}$). For cells with up to 16 atoms ($d=45$) the tensor $D^{(4)}$ is stored in full. For the large cells it is never formed: $\|W_4\|^2$ and $\avg{W_3^2W_4}$ are computed as sums over pairs of atom pairs, in which two pair separations are linked by their $3\times3$ covariance block, and the tetrahedron sum $K_4$ is evaluated in blocks with $O(d^5)$ operations. These scalable formulas reproduce the full-tensor values of the 16-atom cell to better than $10^{-8}$ relative. In the 16-atom cell the many-body remainder changes the sum of the three terms by only $0.02\%$ (the remainder's own contribution to $\|D^{(4)}\|^2$ is $6\times10^{-4}$ of the pair part's); in the large cells $D^{(4)}$ therefore contains the pair part only, while $D^{(3)}$ contains both parts as before. The cost per temperature and volume for 128-atom NiAl on four cores is about $2.5$ minutes for $D^{(3)}$ (the same as the bubble), $1$ minute for the pair parts and $2$ minutes for the tetrahedron sum.

\paragraph{Reference for the free energy.} To test the terms, the free energy of the true potential relative to the Gaussian was obtained by direct numerical integration along $U_\lambda=(1-\lambda)U_{\Phi^*}+\lambda U$ ($F_{\rm exact}-F_{GB}=\int_0^1\avg{U-U_{\Phi^*}}_\lambda\,\mathrm d\lambda$; thermodynamic integration, 6 Gauss--Legendre nodes for the large cells and 8 for the 16-atom cell), the averages at each node being obtained by Hamiltonian Monte Carlo (sampling with fictitious momenta) in the whitened coordinates. This reference has a statistical error, given in Tables~\ref{tab:t3free} and \ref{tab:t3small}; it starts from the same Gaussian as the method. A separate check that shares nothing with the tensor-train method was therefore made on the 16-atom cell at 800\,K: a Frenkel--Ladd integration, i.e.\ thermodynamic integration from an Einstein crystal (every internal coordinate held by an identical spring of stiffness $k$, whose free energy is known in closed form, $F_E=-\tfrac d2\kB T\ln(2\pi\kB T/k)$) to the true potential, $F=F_E+\int_0^1\avg{U-U_E}_\lambda\,\mathrm d\lambda$, with 8 Gauss--Legendre nodes and Hamiltonian Monte Carlo at each node (256 independent chains, 150 trajectories each). It gives $F_{\rm exact}-F_{GB}=-19.62\pm0.74$\,meV, against $-20.31\pm0.70$\,meV from the integration starting at the Gaussian, and against $-20.77$\,meV for bubble plus $T^3$ terms; the difference to the latter is $+1.14\pm0.74$\,meV, i.e.\ $1.5$ standard errors, while the bubble alone misses by $-1.95\pm0.74$\,meV. For the 128- and 108-atom cells this check has not been made, because the noise of the Einstein-crystal path grows roughly in proportion to the number of atoms times the cost of one sample.

\paragraph{Free energy.} In the 16-atom cell (Table~\ref{tab:t3small}) the direct integration at 800\,K gives $-20.31\pm0.70$\,meV, the bubble $-17.68$\,meV, and the three terms of order $T^3$ add $-3.09$\,meV, which leaves $+0.46\pm0.70$\,meV, compatible with zero; before adding them, $-2.63\pm0.70$\,meV was missing. At 400\,K the missing part is $-0.19\pm0.25$\,meV before and $+0.41\pm0.25$\,meV after, so that the test cannot resolve the terms at that temperature. For the large cells (Tables~\ref{tab:t3free} and~\ref{tab:t3}) the part missing after the bubble at 800\,K, $-31.5\pm2.9$\,meV for NiAl and $-23.8\pm3.3$\,meV for TiAl, falls to $-6.3$ and $-4.1$\,meV ($2.2$ and $1.3$ standard errors); at 600\,K to $+0.3$ and $-2.3$\,meV ($0.2$ and $1.1$ standard errors); at 400\,K to $+2.1$ and $-1.2$\,meV. The three terms have different signs ($F_{4b}$ is positive, the other two negative) and are individually larger than their sum, up to $60$\,meV for TiAl at 800\,K; the cancellation must therefore be computed with all three.

\paragraph{Pressure.} The pressure contribution $P_{T^3}=-\Delta F_{T^3}/\Delta V$ is $+0.029$\,GPa (NiAl) and $+0.019$\,GPa (TiAl) at 800\,K, and $+0.014$ and $+0.009$\,GPa at 600\,K (Table~\ref{tab:t3p}). Relative to Monte Carlo sampling of the exact Boltzmann distribution at the same volume, the pressure of the TT-SCHA + bubble was low by $0.047$ (NiAl) and $0.020$\,GPa (TiAl) at 800\,K; with the $T^3$ terms the deviations are $-0.018\pm0.008$ and $-0.001\pm0.006$\,GPa. At 600\,K they are $-0.001\pm0.005$ and $+0.004\pm0.005$\,GPa. Relative to molecular dynamics the pressure deviation at 600 and 800\,K changes from $-1.2\%$ and $-1.0\%$ to $-0.8\%$ and $-0.4\%$ for NiAl, and from $-0.6\%$ and $-0.8\%$ to $-0.3\%$ and $-0.3\%$ for TiAl. The $T^3$ pressure is a difference of free energies at two volumes and carries the secant bias discussed for the stationarity test; its size is of the order of $0.01$\,GPa. In the 16-atom cell (Monte Carlo with two independent runs, combined $4.874\pm0.007$\,GPa at 800\,K) the deficit of the TT-SCHA + bubble pressure is $0.026$\,GPa and the $T^3$ terms add $0.037$\,GPa, which gives $+0.011\pm0.007$\,GPa.

\begin{table}[!htbp]\centering\small
\caption{\textbf{Free-energy terms of order $T^3$.} Free energy of the whole supercell, in meV, at fixed volume. $F_b$: cubic (bubble) term. $\sum T^3$: sum of the three next-order terms $F_{4a}+F_{4b}+F_{4c}$. ``Exact$-$GB'': free energy of the true potential minus that of the self-consistent Gaussian, from thermodynamic integration with Hamiltonian Monte Carlo ($\pm1\sigma$). ``After $F_b$'' and ``after $T^3$'': the difference of Exact$-$GB from $F_b$, and from $F_b+\sum T^3$; the last column gives the latter in units of the statistical error. Individual terms are in Table~\ref{tab:t3free}.}\label{tab:t3}
\setlength{\tabcolsep}{5pt}
\begin{tabular}{@{}lr|rr|r|rr|r@{}}\toprule
 & $T$ [K] & $F_b$ & $\sum T^3$ & Exact$-$GB & after $F_b$ & after $T^3$ & [$\sigma$]\\\midrule
NiAl & 400 & $-73.4$ & $-4.09$ & $-75.4\pm1.1$ & $-2.0$ & $+2.1$ & 1.9\\
128 atoms & 600 & $-157.4$ & $-11.86$ & $-168.9\pm2.0$ & $-11.5$ & $+0.3$ & 0.2\\
 & 800 & $-268.0$ & $-25.13$ & $-299.5\pm2.9$ & $-31.5$ & $-6.3$ & 2.2\\\midrule
TiAl & 400 & $-79.8$ & $-2.85$ & $-83.9\pm1.1$ & $-4.1$ & $-1.2$ & 1.1\\
108 atoms & 600 & $-180.2$ & $-8.93$ & $-191.4\pm2.1$ & $-11.2$ & $-2.3$ & 1.1\\
 & 800 & $-321.3$ & $-19.69$ & $-345.1\pm3.3$ & $-23.8$ & $-4.1$ & 1.3\\
\bottomrule
\end{tabular}
\end{table}
\begin{figure}[!htbp]\centering
\includegraphics[width=\textwidth]{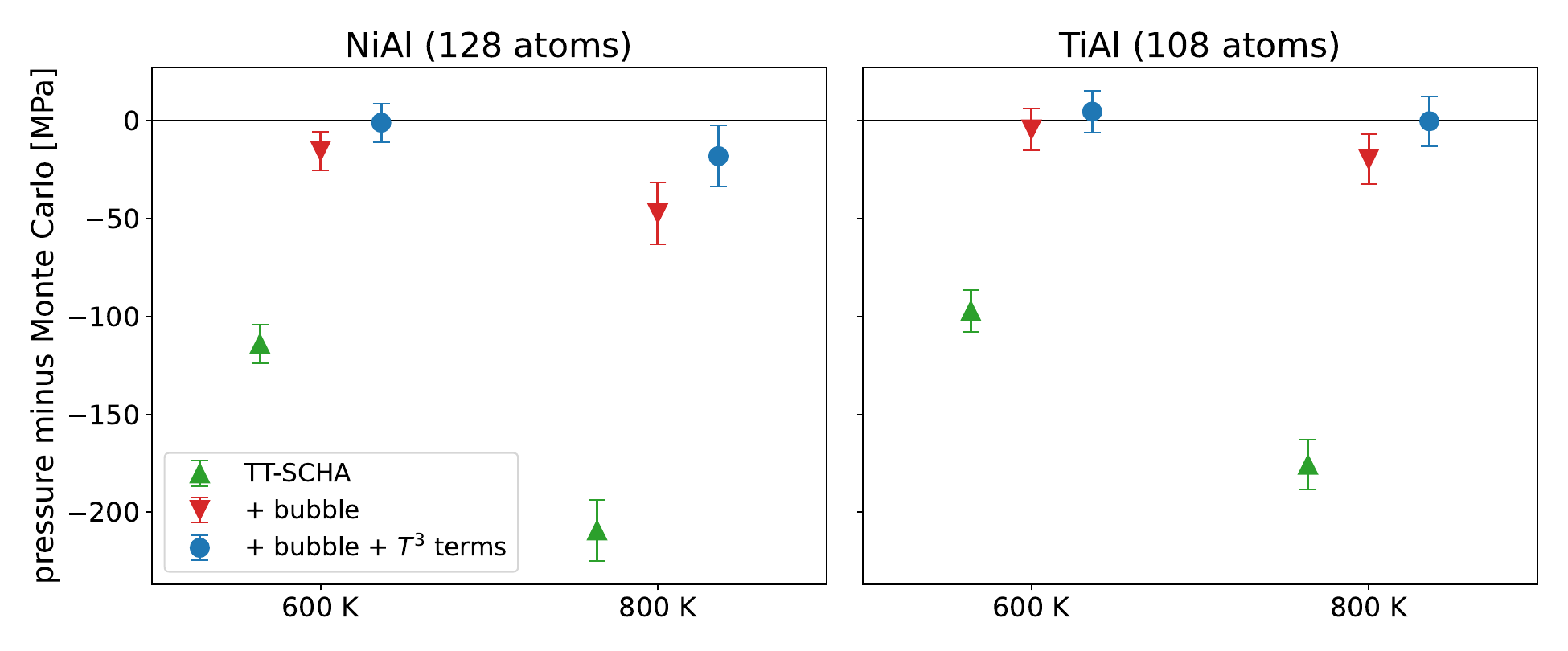}
\caption{\textbf{Effect of the next-order terms on the pressure.} Difference from Monte Carlo sampling of the exact Boltzmann distribution of the same potential, in MPa, at 600 and 800\,K, for 128-atom NiAl (left) and 108-atom TiAl (right). Green: self-consistent reference alone. Red: with the cubic correction. Blue: with the cubic and the three $T^3$ terms. Bars: $\pm2\sigma$ error of the Monte Carlo reference. Zero is exact agreement. The blue points lie on zero within the error bars, except NiAl at 800\,K ($-18$\,MPa, $2.3\sigma$).}\label{fig:t3}
\end{figure}
\begin{table}[!htbp]\centering\small
\caption{\textbf{Pressure with the next-order terms.} Pressure in GPa at fixed volume. MD: molecular dynamics; MC: Hamiltonian Monte Carlo of the exact distribution ($1\sigma$ in parentheses); TT$+b$: self-consistent reference plus cubic term; $+T^3$: also with the three terms of order $T^3$. Last column: $+T^3$ relative to MD in \%.}\label{tab:t3p}
\setlength{\tabcolsep}{6pt}
\begin{tabular}{@{}lr|rrrr|r@{}}\toprule
 & $T$ [K] & MD & MC & TT$+b$ & $+T^3$ & $\Delta$ vs MD [\%]\\\midrule
NiAl & 600 & 3.8726(169) & 3.8420(49) & 3.8263 & 3.8407 & $-0.8$\\
 & 800 & 5.0955(125) & 5.0939(78) & 5.0464 & 5.0756 & $-0.4$\\\midrule
TiAl & 600 & 3.0063(76) & 2.9933(53) & 2.9886 & 2.9976 & $-0.3$\\
 & 800 & 4.0091(152) & 3.9966(64) & 3.9767 & 3.9961 & $-0.3$\\
\bottomrule
\end{tabular}
\end{table}

\paragraph{Guard for the range of validity.} The series is asymptotic and no radius of convergence is available. As a practical guard we use the ratio of the $T^3$ terms to the cubic term, in the free energy and in the pressure (Table~\ref{tab:ratio}); it grows with temperature, and we make no claim where it is not small. For 128-atom NiAl at 800\,K the cubic term is $-268.0$\,meV per cell and the three $T^3$ terms together are $-25.1$\,meV, a ratio of 0.09. A rough estimate of the next order is given in Table~\ref{tab:pnext}.

\begin{table}[!htbp]\centering\small
\caption{\textbf{Ratio of the $T^3$ terms to the cubic (bubble) terms, 600 to 1000\,K.} $|F_4|/|F_b|$: free energy; $P_{T^3}/P_b$: pressure, as used for $P_{\rm next}$ in Table~\ref{tab:pnext}. Three-run means; the standard deviation over the starting pivots is at most 1.4\,MPa in the pressure terms. A growing ratio means that the next order grows faster than the last included one; we make no claim where it is no longer small.}\label{tab:ratio}
\footnotesize\setlength{\tabcolsep}{4pt}
\begin{tabular}{@{}rrrcrr@{}}\toprule
 & \multicolumn{2}{c}{NiAl} & & \multicolumn{2}{c}{$\gamma$-TiAl}\\\cmidrule(lr){2-3}\cmidrule(l){5-6}
$T$ [K] & $|F_4|/|F_b|$ & $P_{T^3}/P_b$ & & $|F_4|/|F_b|$ & $P_{T^3}/P_b$\\\midrule
600 & 0.084 & 0.159 & & 0.061 & 0.120\\
700 & 0.095 & 0.179 & & 0.068 & 0.136\\
800 & 0.105 & 0.198 & & 0.075 & 0.150\\
900 & 0.115 & 0.215 & & 0.082 & 0.164\\
1000 & 0.125 & 0.232 & & 0.087 & 0.177\\
\bottomrule
\end{tabular}
\end{table}

\begin{table}[t]\centering\small
\caption{Next-order ($T^3$) terms of the free energy, evaluated with tensor-train integration, compared with the free energy obtained by direct numerical integration from the Gaussian to the true potential (thermodynamic integration along $U_\lambda=(1-\lambda)U_{\Phi^*}+\lambda U$ with 6 Gauss--Legendre nodes and Hamiltonian Monte Carlo at each node; Section~\ref{sec:T3}). All free energies are for the whole supercell, in meV, at fixed volume. $F_b$: cubic (bubble) term. $F_{4a}=-\frac\beta2\|W_4\|^2$, $F_{4b}=\frac{\beta^2}{2}\langle W_3^2W_4\rangle$, $F_{4c}=-\frac{\beta^3}{24}\kappa_4(W_3)$: the three terms of order $T^3$ (defined in Theorem~\ref{thmB:bubble}(iii)), and their sum. ``exact$-$GB'' is the free energy of the true potential minus the Gibbs--Bogoliubov free energy of the self-consistent Gaussian, $\pm1\sigma$ statistical error of the integration. The last three columns give what remains after the bubble, what remains after the bubble and the three $T^3$ terms, and the latter in units of the statistical error $\sigma$.}\label{tab:t3free}
\setlength{\tabcolsep}{3.5pt}
\begin{tabular}{@{}r|rrrrr|r|rrr@{}}\toprule
$T$ [K] & $F_b$ & $F_{4a}$ & $F_{4b}$ & $F_{4c}$ & $\sum T^3$ & exact$-$GB & after $F_b$ & after $T^3$ & [$\sigma$]\\\midrule
\multicolumn{10}{@{}l}{\emph{NiAl (128 atoms)}}\\
400 & -73.4 & -2.91 & +3.68 & -4.86 & -4.09 & $-75.4\pm1.1$ & $-2.0$ & $+2.1$ & 1.9\\
600 & -157.4 & -7.54 & +10.38 & -14.71 & -11.86 & $-168.9\pm2.0$ & $-11.5$ & $+0.3$ & 0.2\\
800 & -268.0 & -14.81 & +21.38 & -31.70 & -25.13 & $-299.5\pm2.9$ & $-31.5$ & $-6.3$ & 2.2\\
\multicolumn{10}{@{}l}{\emph{TiAl (108 atoms)}}\\
400 & -79.8 & -2.69 & +6.99 & -7.16 & -2.85 & $-83.9\pm1.1$ & $-4.1$ & $-1.2$ & 1.1\\
600 & -180.2 & -9.67 & +25.17 & -24.43 & -8.93 & $-191.4\pm2.1$ & $-11.2$ & $-2.3$ & 1.1\\
800 & -321.3 & -23.90 & +62.79 & -58.58 & -19.69 & $-345.1\pm3.3$ & $-23.8$ & $-4.1$ & 1.3\\
\bottomrule
\end{tabular}
\end{table}

\begin{table}[t]\centering\small
\caption{Check of the $T^3$ terms on a small cell: B2 NiAl with 16 atoms ($2^3$ unit cells; $d=45$), where the fourth-order tensor can be stored in full. Columns and conventions as in Table~\ref{tab:t3free}; free energies for the whole cell in meV. The statistical error of the direct integration is smaller here than for the large cells.}\label{tab:t3small}
\setlength{\tabcolsep}{3.5pt}
\begin{tabular}{@{}r|rrrrr|r|rr@{}}\toprule
$T$ [K] & $F_b$ & $F_{4a}$ & $F_{4b}$ & $F_{4c}$ & $\sum T^3$ & exact$-$GB & after $F_b$ & after $T^3$\\\midrule
400 & -4.62 & -0.55 & +0.12 & -0.17 & -0.60 & $-4.81\pm0.25$ & $-0.19$ & $+0.41$\\
800 & -17.68 & -2.68 & +0.83 & -1.25 & -3.09 & $-20.31\pm0.70$ & $-2.63$ & $+0.46$\\
\bottomrule
\end{tabular}
\end{table}


\subsection{Spectrum of the self-consistent Hessian}\label{sec:spectrum}
\begin{figure}[!htbp]\centering
\includegraphics[width=\textwidth]{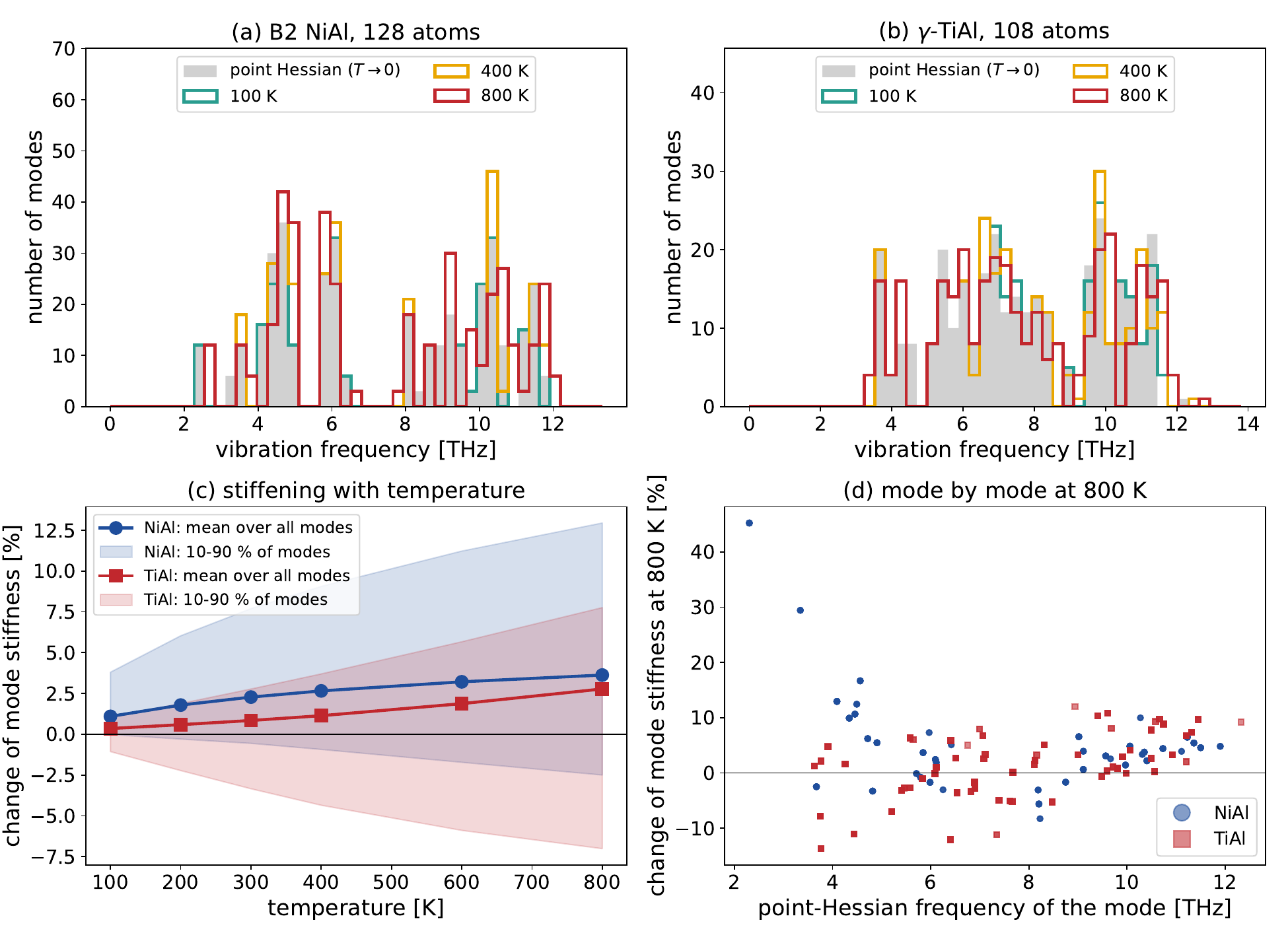}
\caption{\textbf{Vibration spectrum of the self-consistent Gaussian against temperature.} (a), (b) Number of vibration modes per frequency interval (the three translations of the crystal excluded) for 128-atom NiAl and 108-atom $\gamma$-TiAl: grey, the point Hessian (curvature of the potential at the lattice sites); coloured lines, the eigenvalues of the self-consistent Hessian $\Phi^\star(T)$ computed by tensor-train integration at 100, 400 and 800\,K, converted to frequencies with the atomic masses. (c) Change of the stiffness of the vibration modes relative to the point Hessian against temperature: lines, mean over all modes; shaded bands, range between the 10th and 90th percentile of the change of individual modes (the change of a mode is the diagonal element of $\Phi^\star(T)$ in the point-Hessian mode divided by the point value, minus one). (d) The same change for each mode at 800\,K against the frequency of the mode in the point Hessian. All at constant volume. Full data in Table~\ref{tab:eig}.}\label{fig:eig}
\end{figure}

The Gaussian that minimizes the free-energy bound of Gibbs and Bogoliubov is fixed by its stiffness matrix $\Phi^\star(T)$, a $384\times384$ ($3N$ with $N=128$) matrix for NiAl and $324\times324$ for TiAl that holds the stiffness between every pair of atomic coordinates. We compute it by the TT integration of Section~\ref{sec:tt}, for every temperature and without sampling. Its eigenvalues, after division by the atomic masses, are the squared frequencies of the collective vibration modes of the crystal at that temperature, so the TT calculation gives the whole temperature-dependent vibration spectrum at once. Figure~\ref{fig:eig} and Table~\ref{tab:eig} compare it with the spectrum of the ordinary (point) Hessian, the curvature of the potential at the lattice sites, which is what the harmonic approximation uses. The three translations of the whole crystal remain exactly free (their eigenvalues are zero to $10^{-16}$ in all cases), as they must. At constant volume the modes become stiffer with temperature: the mean stiffness is 1.1\,\% (NiAl) and 0.35\,\% (TiAl) above the point value at 100\,K and 3.6\,\% and 2.8\,\% above it at 800\,K. This is not uniform. At 800\,K the change of the stiffness of individual modes spans $-7.0$ to $+7.8$\,\% (10th to 90th percentile) for TiAl and $-2.5$ to $+12.9$\,\% for NiAl, the strongest stiffening occurring in the softest modes of NiAl (the four modes at 2.30\,THz in the point Hessian stiffen by 45\,\%); the lowest vibration frequency of NiAl rises from 2.30 to 2.77\,THz, whereas that of TiAl falls from 3.63 to 3.49\,THz above 400\,K. The modes of the point Hessian are also no longer exact eigenmodes: the share of the stiffness matrix outside their diagonal grows with temperature to 1.7\,\% (NiAl) and 2.9\,\% (TiAl) at 800\,K. The whole temperature-dependent spectrum comes from the same calculation, with no extra cost.

\begin{table}[!htbp]\centering\small
\caption{\textbf{Eigenvalues of the self-consistent Hessian against temperature.} For each temperature: the mean, lowest and highest vibration frequency in THz (three translations excluded; the row ``0 (point)'' is the point Hessian at the lattice sites); the change of the mean stiffness (mean eigenvalue) relative to the point Hessian in \%; the 10th and 90th percentile of the change of the stiffness of the individual point-Hessian modes in \% (these changes are defined in the caption of Fig.~\ref{fig:eig}); and the share of the norm of $\Phi^\star(T)$ that lies outside the diagonal in the basis of the point-Hessian modes, in \%.}\label{tab:eig}
\setlength{\tabcolsep}{4pt}
\begin{tabular}{@{}r|rrr|r|rr|r@{}}\toprule
$T$ [K] & mean & lowest & highest & mean & 10th & 90th & off-diag.\\
 & \multicolumn{3}{c|}{frequency [THz]} & \multicolumn{3}{c|}{stiffness change [\%]} & share [\%]\\\midrule
\multicolumn{8}{@{}l}{\emph{B2 NiAl (128 atoms)}}\\
0 (point) & 7.412 & 2.303 & 11.899 & 0 & 0 & 0 & --\\
100 & 7.459 & 2.429 & 11.971 & 1.09 & -0.02 & 3.80 & 0.4\\
200 & 7.487 & 2.511 & 12.024 & 1.79 & -0.30 & 6.03 & 0.7\\
300 & 7.507 & 2.572 & 12.063 & 2.28 & -0.56 & 7.70 & 1.0\\
400 & 7.522 & 2.623 & 12.095 & 2.66 & -0.93 & 9.06 & 1.2\\
600 & 7.544 & 2.704 & 12.145 & 3.21 & -1.72 & 11.22 & 1.5\\
800 & 7.560 & 2.769 & 12.184 & 3.63 & -2.50 & 12.94 & 1.7\\
\midrule
\multicolumn{8}{@{}l}{\emph{$\gamma$-TiAl (108 atoms)}}\\
0 (point) & 7.715 & 3.630 & 12.324 & 0 & 0 & 0 & --\\
100 & 7.724 & 3.632 & 12.395 & 0.35 & -1.07 & 1.01 & 0.4\\
200 & 7.728 & 3.631 & 12.457 & 0.59 & -2.23 & 1.88 & 0.7\\
300 & 7.734 & 3.631 & 12.521 & 0.84 & -3.34 & 2.77 & 1.1\\
400 & 7.740 & 3.625 & 12.588 & 1.14 & -4.35 & 3.70 & 1.4\\
600 & 7.759 & 3.557 & 12.730 & 1.87 & -5.89 & 5.67 & 2.2\\
800 & 7.785 & 3.494 & 12.879 & 2.78 & -7.01 & 7.76 & 2.9\\
\bottomrule
\end{tabular}
\end{table}

\paragraph{Computation.} The matrix $\Phi^\star(T)$ is the fixed point of $\Phi=\avg{\nabla^2U}_\Phi$, computed with the pair rule and the TT remainder at each temperature and volume; the iteration is repeated until the change is below $10^{-7}$ for the pair part. For the spectrum, $\Phi^\star$ is divided by the square roots of the atomic masses (Ni 58.6934, Al 26.9815, Ti 47.867\,u), $D=M^{-1/2}\Phi^\star M^{-1/2}$, and its eigenvalues $\lambda$ give frequencies $\nu=\sqrt{\lambda}/2\pi$ (with $\lambda$ in eV\,\AA$^{-2}$\,u$^{-1}$, $1\,\mathrm{eV}\,\text{\AA}^{-2}\,\mathrm{u}^{-1}=9.64853\times10^{27}\,\mathrm s^{-2}$). The point Hessian is the central finite difference (step $2\times10^{-3}$\,\AA) of the LAMMPS forces at the relaxed lattice, with the acoustic sum rule imposed. For the spectrum alone the calculation can be shortened: only one volume is needed, the cubic term is not, and the iteration can start from the point Hessian and stop at a relative change of $10^{-5}$ in $\Phi$. This takes 17\,s (NiAl, 400\,K) to 56\,s (NiAl, 800\,K) and 42\,s (TiAl, 800\,K) on one workstation, instead of 10--24 minutes for the full calculation, and the eigenvalues differ from those of the fully converged run by at most $0.04$--$0.06$\,\% (NiAl) and $0.4$\,\% (TiAl). The change of the stiffness of a point-Hessian mode $k$ is $v_k^{\mathsf T}Dv_k/\lambda^{(0)}_k-1$ with $v_k,\lambda^{(0)}_k$ its eigenvector and eigenvalue.


\section{Thermodynamic-integration chains, cell-size dependence and the local formulation}\label{sec:ti}
This section describes the free-energy chains used as references, the cell-size study with the dense and with the local formulation of the self-consistent reference, the local cubic and $T^3$ terms, and the reuse of cross pivots.

\subsection{Frenkel--Ladd chains}
\paragraph{Method.} The Helmholtz free energy is obtained by thermodynamic integration from an Einstein crystal (Frenkel--Ladd): $H_\lambda=\lambda U+(1-\lambda)\frac k2\sum_i|\mathbf u_i-\bar{\mathbf u}|^2$, where $\mathbf u_i$ is the displacement of atom $i$ from its lattice site, $\bar{\mathbf u}$ the mean displacement (so that the translation of the whole crystal is free) and $k=6.221$\,eV\,\AA$^{-2}$ the spring stiffness (chosen as the harmonic mean of the eigenvalues of the self-consistent Hessian; any $k$ is exact). With $d=3N-3=381$ internal coordinates, $F=U_0+F_E+\int_0^1\langle U-U_E\rangle_\lambda\,\mathrm d\lambda$ with $F_E=-\frac d2\kB T\ln(2\pi\kB T/k)=34.994$\,eV and $U_0=-577.3915$\,eV. The integral is done by 6-point Gauss--Legendre quadrature, $\lambda=0.0338,0.1694,0.3807,0.6193,0.8306,0.9662$. The average at each node is obtained in two ways. (a) Hamiltonian Monte Carlo (HMC) in the whitened internal coordinates with our NumPy implementation of the embedded-atom energy; chains started from the Gaussian of the self-consistent reference (a start from the Einstein distribution is not equilibrated at the nodes $\lambda\ge0.83$: at $\lambda=0.966$ it gave $-0.32$ instead of $-0.65$\,eV in 40 trajectories, which would have shifted the free energy by 67\,meV). (b) LAMMPS Langevin molecular dynamics (damping time 0.1\,ps, 20\,\% of each run discarded), with the interaction scaled by $\lambda$ (pair style \texttt{hybrid/scaled}) and the spring force added with \texttt{fix addforce}. The time step matters: at 1\,fs the chain gave $-283.3\pm3.3$\,meV and at 0.5\,fs $-284.8\pm2.3$\,meV (18 runs per node); the node $\lambda=0.38$ changed from $1.9068$ to $1.8933$ ($\pm0.005$) between 1 and 0.5\,fs.

\begin{table}[!htbp]\centering\small
\caption{Average of $U-U_E$ (eV, whole cell) at the six integration nodes for 128-atom NiAl at 800\,K from the two chains. HMC: 13 batches of 64 chains of 44 trajectories; MD: 18 runs of 100\,ps per node, 0.5\,fs time step. The error of the MD values is the standard error of the 18 run means. The last column is the MD$-$HMC difference in units of the combined error.}\label{tab:tinodes}
\begin{tabular}{@{}rrrrrr@{}}\toprule
node & $\lambda$ & weight & HMC & MD & (MD$-$HMC)/$\sigma$\\\midrule
1 & 0.0338 & 0.0857 & $6.5808\pm0.0061$ & $6.5951\pm0.0064$ & 1.6\\
2 & 0.1694 & 0.1804 & $5.0314\pm0.0058$ & $5.0620\pm0.0043$ & 4.2\\
3 & 0.3807 & 0.2340 & $3.3740\pm0.0058$ & $3.3729\pm0.0060$ & $-0.1$\\
4 & 0.6193 & 0.2340 & $1.8866\pm0.0069$ & $1.8999\pm0.0040$ & 1.7\\
5 & 0.8306 & 0.1804 & $0.5295\pm0.0083$ & $0.5472\pm0.0059$ & 1.7\\
6 & 0.9662 & 0.0857 & $-0.6489\pm0.0099$ & $-0.6435\pm0.0076$ & 0.4\\
\bottomrule
\end{tabular}
\end{table}

\begin{table}[!htbp]\centering\small
\caption{Free energy of the 128-atom NiAl cell at 800\,K at fixed volume, relative to the Gibbs--Bogoliubov free energy $F_{GB}=-539.3579$\,eV of the self-consistent Gaussian, in meV, and the cost of each route (one workstation with 4 cores; core-hours are wall-clock time times 4 for the chains and for the tensor-train $T^3$ step, and wall-clock time on one core for the rest). Route B: integration from the Gaussian to the true potential (Table~\ref{tab:t3free}). Predicted: $F_b+\sum T^3=-268.0-25.1$\,meV.}\label{tab:ticost}
\resizebox{\textwidth}{!}{%
\begin{tabular}{@{}lrrr@{}}\toprule
route & $F-F_{GB}$ [meV] & error [meV] & cost\\\midrule
tensor train: bubble & $-268.0$ & none (truncation) & 7\,min\\
tensor train: bubble $+\,T^3$ & $-293.1$ & none (truncation) & 9\,min, about 0.3 core-h\\
HMC chain from the Gaussian (route B) & $-299.5$ & 2.9 & not recorded\\
HMC chain from the Einstein crystal & $-298.1$ & 3.0 & 2.2 core-h (33\,min wall)\\
LAMMPS MD chain, 1\,fs & $-283.3$ & 3.3 & 0.3 core-h\\
LAMMPS MD chain, 0.5\,fs & $-284.8$ & 2.3 & 0.9 core-h (0.5 scaled to 3\,meV)\\
\bottomrule
\end{tabular}%
}
\end{table}
\paragraph{Findings.} (i) The two original chains disagree by $15.1\pm3.4$\,meV per cell ($4.4\sigma$, about 0.12\,meV per atom; all 36 molecular-dynamics seeds pooled against the HMC chain), with the MD values higher at all six nodes. A repeat of the MD chain with 18 new seeds reproduced the earlier result ($+3.6$\,meV per cell from the old seeds, $+1.1\sigma$), the NumPy and LAMMPS embedded-atom codes agree to $2\times10^{-8}$\,meV per cell, and neither the time step (1 and 0.5\,fs gave the same chain value within error) nor the constants ($F_E$ and $U_0$ are the same in both) explains the gap. A rerun of the HMC chain with a much longer burn-in did not move its node values (new minus old $-2.7$\,meV, $-0.6\sigma$; integrated shift $-2.9\pm7.2$\,meV). The tests on a harmonic crystal below trace the gap to two sampling artifacts of the chains. (ii) With the HMC chain, bubble plus $T^3$ is $5.0\pm3.0$\,meV below the integration; with the MD chain it is $8.3\pm2.3$\,meV above it. (iii) The cost of a chain at $\pm3$\,meV is about 0.5 core-hours (LAMMPS) or 2.2 core-hours (HMC in NumPy), against about 0.3 core-hours for the tensor-train free energy: the difference is a factor of about two compared with molecular dynamics and is not a speed advantage of the tensor-train method. Chain costs grow with the cell size at fixed absolute error, because the variance of $U-U_E$ is proportional to $N$; the cost of the tensor-train calculation at larger $N$ was not measured. Scripts: \texttt{fl\_node2.py}, \texttt{lmp\_ti.py}, \texttt{lmp\_ti\_combine.py}.

\paragraph{Resolution of the discrepancy.} We ran both chains on a purely harmonic version of the same 128-atom NiAl cell at 800\,K, for which the integral is known exactly (Table~\ref{tab:harm}). The production LAMMPS chain reads $+16.0\pm2.4$\,meV per cell too high, at every node; the same value results with the momentum of the cell zeroed after each step, with a 1\,ps damping time and three times the length. The cause is the option \texttt{zero yes} of the Langevin thermostat: it subtracts the same net random force from every atom whatever its mass, which in a Ni/Al crystal unbalances the random kicks against the friction. With \texttt{zero no} the chain reads $+2.1\pm2.4$\,meV and passes at every node (within $1.2\sigma$); giving both elements the same mass also removes the bias, and halving the time step does not. We checked that LAMMPS samples the intended Hamiltonian: its $U_\lambda$ equals $\lambda U$ plus the spring term to $2\times10^{-11}$\,eV in energy and $9\times10^{-12}$\,eV/\AA\ in the forces. The HMC chain with a fixed trajectory length reads $-21.9\pm6.5$\,meV ($-3.4\sigma$) on the harmonic crystal, because modes whose phase advance per trajectory is near a multiple of $\pi$ barely mix (42 such modes at the largest $\lambda$); started from the Einstein crystal it is off by $+74.0\pm5.1$\,meV. With a randomly chosen trajectory length it reads $-0.7\pm3.5$ and $+1.4\pm3.3$\,meV for the two starts. A harmonic crystal with exactly resonant modes is the worst case for a fixed trajectory length; on the embedded-atom potential the original HMC chain is not measurably biased (Table~\ref{tab:tirerun}).
\begin{table}[!htbp]\centering\small
\caption{Each chain minus the exact value of the integral for the harmonic version of 128-atom NiAl at 800\,K, in meV per cell (sum over the six nodes). Pass: within $3\sigma$ at every node.}\label{tab:harm}
\begin{tabular}{@{}lrl@{}}\toprule
Chain & integral $-$ exact & verdict\\\midrule
LAMMPS Langevin, production (\texttt{zero yes}) & $+16.0\pm2.4$ & fail\\
LAMMPS, momentum zeroed & $+16.0\pm2.4$ & fail\\
LAMMPS, 1\,ps damping, three times longer & $+15.9\pm4.1$ & fail\\
LAMMPS, \texttt{zero no} & $+2.1\pm2.4$ & pass\\
HMC, fixed trajectory length (S6 protocol) & $-21.9\pm6.5$ & integral off\\
HMC, fixed length, Einstein-crystal start & $+74.0\pm5.1$ & fail\\
HMC, random trajectory length, two starts & $-0.7\pm3.5$, $+1.4\pm3.3$ & pass\\\bottomrule
\end{tabular}
\end{table}
We then reran the chains on the embedded-atom potential with the two corrections (rule fixed beforehand; Table~\ref{tab:tirerun}). LAMMPS with \texttt{zero no} (36 seeds) and HMC with random trajectory length started from the Gaussian agree: $+1.8\pm3.8$\,meV per cell ($+0.5\sigma$; limit $3\sigma$), which passes. HMC started from the Einstein crystal differs from the Gaussian start by $-7.9\pm4.8$\,meV ($-1.7\sigma$). The corrected LAMMPS chain lies $13.1\pm2.4$\,meV below the original one ($5.6\sigma$, lower at all six nodes); the corrected HMC chain agrees with the original ($+0.2\pm4.5$\,meV). The gap of 15\,meV per cell was therefore the \texttt{zero yes} bias of the molecular-dynamics chain. The common value of the free energy is $-539.65429$\,eV, against $-539.65099$\,eV from the tensor-train terms (Gaussian $+$ bubble $+$ $T^3$) in the same cell: $F_{\rm TI}-F_{\rm TT}=-3.3\pm1.5$\,meV per cell ($-0.026$\,meV per atom, $2.2\sigma$), six times below our limit for $\Delta F$ (0.15\,meV per atom). Two causes are of the right size and we have not separated them. (i) The terms beyond $T^3$: the geometric estimate used for the pressure, the $T^3$ terms ($-25.1$\,meV per cell) times their ratio to the bubble ($25.1/268.0=0.094$), gives about $-2.4$\,meV per cell, which would reduce the difference to about $-1\pm1.5$\,meV; this is an estimate of the same kind as Table~\ref{tab:pnext}, not a calculation. (ii) The scatter of the TT-cross with its starting pivots: at most 0.013\,meV per atom in $F_{\rm GB}$ (Table~\ref{tab:pivot}), i.e.\ 1.7\,meV for the 128 atoms. The three chains cost 7\,038, 9\,788 and 9\,358 core-seconds.
\begin{table}[!htbp]\centering\small
\caption{Integral $I=\int_0^1\langle U-U_E\rangle_\lambda\,d\lambda$ (eV per cell, $\pm1\sigma$) for 128-atom NiAl at 800\,K from the corrected chains and the original ones. A: LAMMPS, \texttt{zero no}, 36 seeds. B, C: HMC with random trajectory length, started from the Gaussian and from the Einstein crystal. Differences in meV per cell.}\label{tab:tirerun}
\begin{tabular}{@{}lr@{}}\toprule
Chain & $I$ [eV per cell]\\\midrule
A: LAMMPS, \texttt{zero no} & $2.74397\pm0.00166$\\
B: HMC, random length, Gaussian start & $2.74213\pm0.00336$\\
C: HMC, random length, Einstein start & $2.75006\pm0.00338$\\
original LAMMPS, \texttt{zero yes} & $2.75705\pm0.00167$\\
original HMC, fixed length & $2.74195\pm0.00302$\\\midrule
A $-$ B & $+1.8\pm3.8$ meV (pass)\\
B $-$ C & $-7.9\pm4.8$ meV\\
A $-$ original LAMMPS & $-13.1\pm2.4$ meV\\
B $-$ original HMC & $+0.2\pm4.5$ meV\\\bottomrule
\end{tabular}
\end{table}

\subsection{Cell-size dependence of the dense tensor-train reference}\label{sec:sizedense}
LAMMPS NVT molecular dynamics of B2 NiAl at 800\,K at the same density as the 128-atom cell (relaxed lattice constant), 20\,ps equilibration and 100\,ps production, time step 1\,fs, the same thermostat and settings as the reference values of the Letter but a different random seed. Wall-clock times are for one workstation. The tensor-train rows give the Hellmann--Feynman pressure of the self-consistent harmonic reference (no cubic term; iteration from the point Hessian to $10^{-5}$), so they are not comparable in absolute value with the molecular-dynamics rows, only in their change with size.
\begin{table}[!htbp]\centering\small
\caption{Pressure of B2 NiAl at 800\,K against the number of atoms. MD: molecular dynamics, pressure in GPa with $1\sigma$ block error, potential energy per atom in eV, and wall-clock time for 120\,ps. TT: tensor-train self-consistent reference, pressure $P_{\rm HF}$ in GPa and wall-clock time of the calculation at one volume on one core (all fixed-point iterations; the tensor-train code was run in saved pieces for 432 and 1024 atoms). The earlier value 4.8834\,GPa for 128 atoms came from an earlier run of the same code that we could not reproduce; the present value is from the code distributed with this work, which reproduces the 54- and 250-atom results of the earlier runs exactly.}\label{tab:mdsize}
\begin{tabular}{@{}rr|rrrr|rr@{}}\toprule
 & & \multicolumn{4}{c|}{MD} & \multicolumn{2}{c}{TT, $P_{\rm HF}$}\\
cells & atoms & $P$ [GPa] & $\sigma$ & $U/N$ [eV] & time [s] & $P_{\rm HF}$ [GPa] & time [s]\\\midrule
$3^3$ & 54 & 5.0085 & 0.0364 & $-4.40921$ & 6 & 4.8542 & 44\\
$4^3$ & 128 & 5.0731 & 0.0211 & $-4.40767$ & 13 & 4.8890 & 67\\
$5^3$ & 250 & 5.1001 & 0.0113 & $-4.40734$ & 26 & 4.9027 & 103\\
$6^3$ & 432 & 5.1073 & 0.0093 & $-4.40725$ & 48 & 4.9101 & 230\\
$8^3$ & 1024 & 5.1254 & 0.0055 & $-4.40694$ & 128 & 4.9207 & 1183\\
\bottomrule
\end{tabular}
\end{table}
The molecular-dynamics pressure rises by $2.3\,\%$ from 54 to 1024 atoms and the 128-atom value is $1.0\,\%$ ($2.4\sigma$ of the combined error) below the 1024-atom value; the potential energy per atom changes by $0.7$\,meV. The tensor-train reference for the same cells is discussed next.

\begin{figure}[!htbp]\centering
\includegraphics[width=\textwidth]{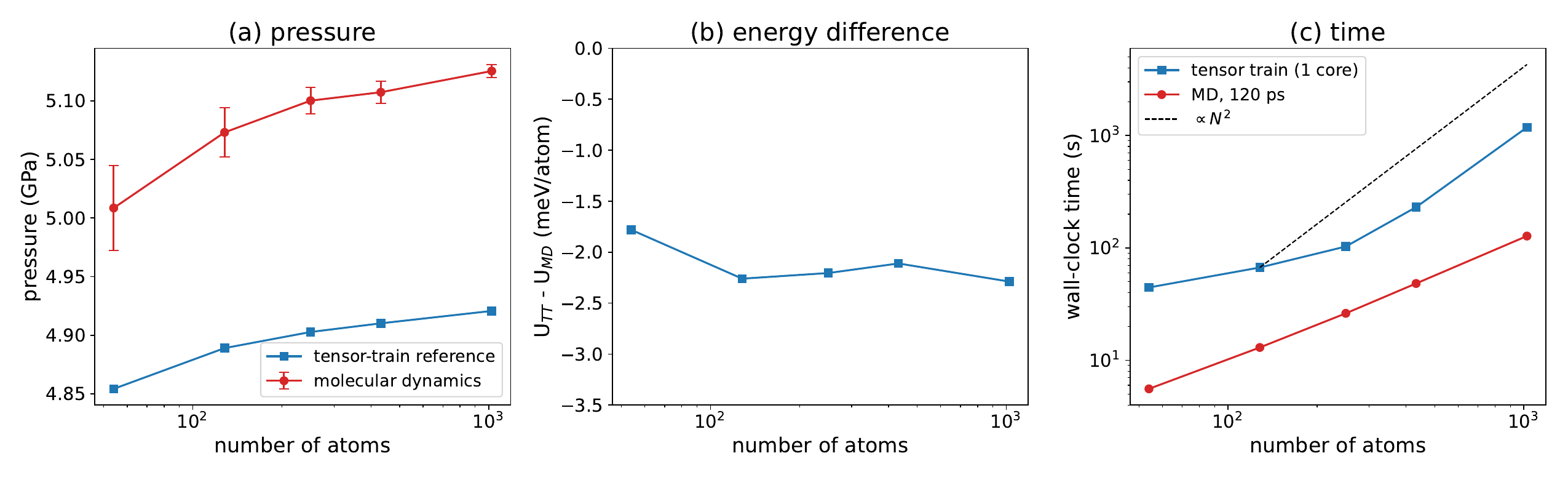}
\caption{\textbf{Cell-size dependence of the tensor-train reference for B2 NiAl at 800\,K.} (a) Pressure from molecular dynamics (red, with $1\sigma$ error bars) and from the self-consistent tensor-train Gaussian (blue) against the number of atoms. The gap between them is the anharmonic part. (b) Difference between the mean potential energy per atom of the Gaussian and the molecular-dynamics value, in meV per atom. (c) Wall-clock time against the number of atoms on logarithmic axes: tensor train on one core (blue), molecular dynamics for 120\,ps (red), and the line proportional to $N^2$ (dashed). The values are those of Table~\ref{tab:mdsize}.}\label{fig:ttsize}
\end{figure}

\paragraph{Tensor-train reference.} The self-consistent tensor-train Gaussian was computed for the same five cells (54 to 1024 atoms, $d=159$ to $3069$ internal coordinates) at 800\,K and the same density, on one processor core, with the settings used everywhere else (24 Gauss--Hermite nodes for the pair part, tensor-train rank at most 6). The cubic and $T^3$ terms were not computed for the larger cells, because the dense third-derivative tensor of $d$ coordinates needs $8d^3$ bytes (3.3\,GB for 250 atoms, which does not fit the 3\,GB workstation used here). Table~\ref{tab:mdsize} and Figure~\ref{fig:ttsize} give the results. (i) \emph{Accuracy.} The difference between the molecular-dynamics pressure and the pressure of the Gaussian reference, which is the anharmonic part that the cubic and quartic terms must supply, is $0.154$, $0.184$, $0.197$, $0.197$ and $0.205$\,GPa for 54, 128, 250, 432 and 1024 atoms. It therefore rises by 0.02\,GPa between 128 and 1024 atoms and is flat between 250 and 432 atoms, while the pressure of the reference itself rises by $0.032$\,GPa (0.65\,\%). The average potential energy per atom of the Gaussian differs from the molecular-dynamics value by $-1.8$, $-2.3$, $-2.2$, $-2.1$ and $-2.3$\,meV per atom: it does not change with the cell size beyond the statistical error of the molecular-dynamics energy. (ii) \emph{Cost.} The time grows from 44\,s (54 atoms) to 67, 103, 230 and 1183\,s (1024 atoms, 20\,min). The local exponent of the time against the number of atoms $N$ is 1.1 between 54 and 128 atoms, 1.7 between 250 and 1024 atoms and 1.9 between 432 and 1024 atoms, i.e.\ it approaches $N^2$, the cost of the dense $d\times d$ matrix operations of the iteration; the pair part and the embedding train, which are the parts that exploit the tensor-train structure, grow more slowly. The molecular-dynamics time grows as $N^{1.1}$, so for 1024 atoms molecular dynamics (128\,s for 120\,ps) is 9 times faster than this dense code for the pressure. (iii) \emph{Limits of the dense code.} The cubic term, the $T^3$ terms and the free energy were computed with it for 128 atoms only. The growth of the time is a property of this dense code; the local formulation of the next subsection removes it.

\subsection{Local formulation: the cost does not depend on the number of atoms}\label{sec:cellsize}
The growth of the time in Table~\ref{tab:mdsize} comes from the way the first code was organised, not from the tensor-train integrals. That code builds the full $d\times d$ matrix of the Gaussian (every atom with every other atom) and puts every site integral in the space of all $d$ modes of the cell. But each integral that the method needs involves one atom and its neighbours only: the pair integral of an atom $m$ with a neighbour $j$ depends on the relative displacement $u_j-u_m$, and the embedding function of atom $m$ depends on the relative displacements of the atoms within the cutoff radius plus a skin of 1.2\,\AA{} (168 neighbours for B2 NiAl). In a perfect crystal all atoms of one sublattice give the same integral, so there are only two site integrals (one per sublattice), whatever the size of the sample. We therefore rewrote the calculation in terms of these local objects (the file \texttt{tt\_local.py}). The matrix of force constants is stored as a short list $\Phi(s,s',\Delta)$, the $3\times3$ block between an atom of sublattice $s$ in one unit cell and an atom of sublattice $s'$ in the cell displaced by the integer vector $\Delta$ ($\Delta$ up to $\pm5$ cells). The only place where the size of the sample enters is the covariance of the Gaussian, which is obtained in a few seconds from a Fourier transform over $n\times n\times n$ wave vectors (each wave vector needs the inverse of one $6\times6$ matrix). The covariance of the 169 atoms of one neighbourhood is read from this table, the 504 relative-displacement coordinates are whitened, the embedding function is approximated as a tensor train of rank 10 by TT-cross exactly as before, and the resulting $3\times3$ blocks are added to the list of force constants. With $n$ wave vectors per direction the result is the one of a periodic cell of $2n^3$ atoms; letting $n\to\infty$ gives the bulk crystal.

We tested the new code against the old one in three ways. (i) The covariance read from the table agrees with the covariance of the dense 128-atom calculation to $3\times10^{-17}$\,\AA$^2$. (ii) For one evaluation of the averages at the same trial matrix, the pressure agrees with the dense code to a relative $4\times10^{-13}$, the energy per atom to $10^{-15}$ relative, and the new force constants to $10^{-15}$ (pair part) and $4\times10^{-11}$ (embedding part). (iii) The converged self-consistent pressures agree with the dense ones to within the noise of the tensor-train approximation of the embedding function (Table~\ref{tab:local}). This noise comes from the sweeps of TT-cross and is not removed by iterating: in the 128-atom cell two runs that differ only in the stopping rule gave 4.8846 and 4.8890\,GPa (relative difference 0.09\,\%), and over the five sizes where both codes were run the root-mean-square difference is 0.0025\,GPa. We take 0.003\,GPa as the numerical uncertainty of every tensor-train pressure in Table~\ref{tab:local}.

\begin{table}[!htbp]\centering\small
\caption{\textbf{Self-consistent tensor-train reference for B2 NiAl at 800\,K from the local formulation, for cells of 54 to 524\,288 atoms.} Columns: cells per direction, number of atoms, pressure of the Gaussian reference in GPa (numerical uncertainty 0.003\,GPa), energy per atom in eV, number of whitened coordinates of the embedding integral, wall-clock time in seconds on one processor core for the whole self-consistent calculation, and, for comparison, pressure and time of the dense code of Table~\ref{tab:mdsize} where it was run. The number of coordinates stops growing at 504 once the cell is larger than twice the neighbourhood radius; the time then stays at 70 to 81\,s.}\label{tab:local}
\begin{tabular}{@{}rrrrrr|rr@{}}\toprule
cells & atoms & $P$ [GPa] & $U/N$ [eV] & coordinates & time [s] & dense $P$ & dense time [s]\\\midrule
$3^3$ & 54 & 4.8543 & $-4.41099$ & 159 & 28 & 4.8542 & 44 \\
$4^3$ & 128 & 4.8846 & $-4.40993$ & 351 & 37 & 4.8890 & 67 \\
$5^3$ & 250 & 4.9003 & $-4.40954$ & 468 & 69 & 4.9027 & 103 \\
$6^3$ & 432 & 4.9135 & $-4.40936$ & 504 & 69 & 4.9101 & 230 \\
$8^3$ & 1\,024 & 4.9199 & $-4.40923$ & 504 & 70 & 4.9207 & 1183 \\
$12^3$ & 3\,456 & 4.9185 & $-4.40916$ & 504 & 71 & -- & -- \\
$16^3$ & 8\,192 & 4.9211 & $-4.40915$ & 504 & 70 & -- & -- \\
$24^3$ & 27\,648 & 4.9209 & $-4.40913$ & 504 & 72 & -- & -- \\
$32^3$ & 65\,536 & 4.9234 & $-4.40914$ & 504 & 72 & -- & -- \\
$64^3$ & 524\,288 & 4.9216 & $-4.40913$ & 504 & 81 & -- & -- \\
\bottomrule
\end{tabular}
\end{table}
\begin{figure}[!htbp]\centering
\includegraphics[width=\textwidth]{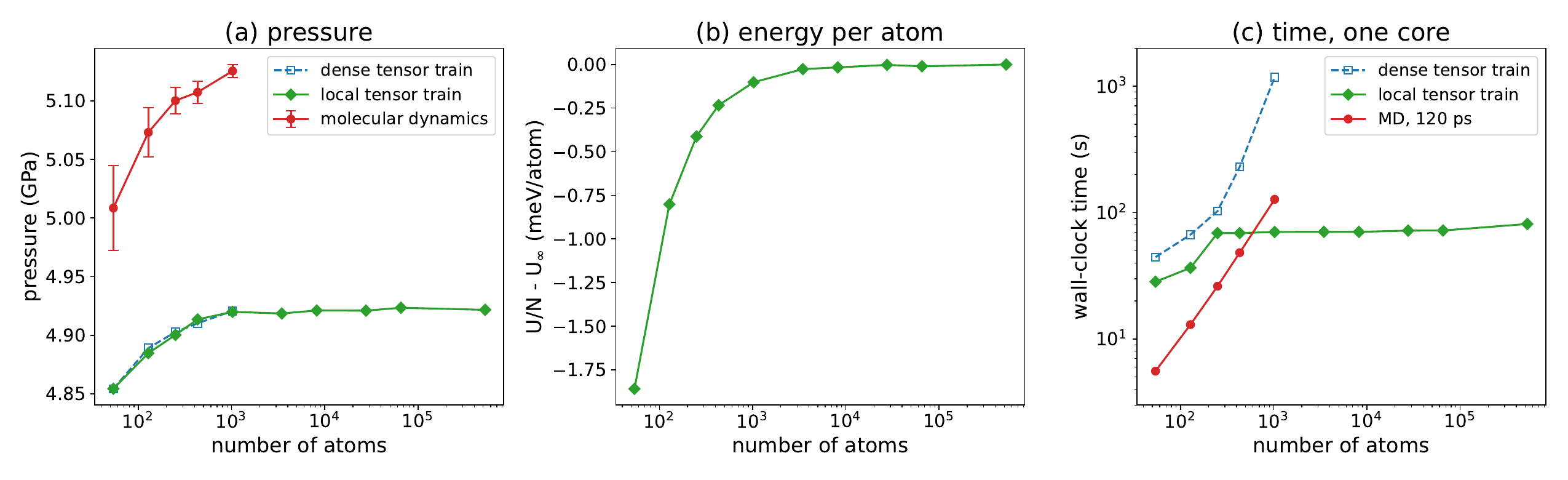}
\caption{\textbf{The local tensor-train formulation: pressure, energy and time against the number of atoms, B2 NiAl at 800\,K.} (a) Pressure from molecular dynamics (red, $1\sigma$ error bars), from the dense tensor-train code (blue, open squares) and from the local code (green diamonds, up to 524\,288 atoms). (b) Energy per atom of the local reference, relative to its value for the largest cell. (c) Wall-clock time on one core against the number of atoms on logarithmic axes: dense code (blue), local code (green), molecular dynamics for 120\,ps (red). The local time is constant, 70 to 81\,s, from 432 to 524\,288 atoms.}\label{fig:ttlocal}
\end{figure}
\paragraph{Findings.} (i) The time of the local calculation does not depend on the number of atoms once the cell is larger than the neighbourhood: 69\,s for 250 and 432 atoms and 70, 71, 70, 72, 72 and 81\,s for 1\,024, 3\,456, 8\,192, 27\,648, 65\,536 and 524\,288 atoms. The small increase at the end is the Fourier transform over $64^3$ wave vectors. The dense code needed 1\,183\,s for 1\,024 atoms. (ii) The pressure and the energy converge with the size of the cell: for 1\,024 atoms and more the pressure is $4.9209\pm0.0015$\,GPa (standard deviation over the six cells, inside the numerical uncertainty of 0.003\,GPa) and the energy per atom is $-4.40913$\,eV, to within $0.1$\,meV for 3\,456 atoms and more. The pressure of the dense reference rose with the size up to the 1\,024-atom cell and the local calculation shows that it has stopped rising there. (iii) The 65\,536-atom cell, comparable to the largest molecular-dynamics cell reported for NiAl, takes 72\,s. (iv) The cubic and $T^3$ terms are made local in the next subsection.

\subsection{Cubic and $T^3$ terms in the local formulation}\label{sec:loc3}
The sums over atoms in the cubic and $T^3$ terms were also rewritten so that no tensor of the size of the cell is formed. In the local formulation the third derivative of the energy is the sum over cells of a local vertex $K^{(3)}_s$ of sublattice $s$ (pair part, dense, and embedding part, the third Hermite moment of the tensor train of that sublattice). Expanding the displacements in the modes of the Gaussian, with six modes at each of the $n^3$ wave vectors, the third-derivative tensor becomes a vertex $V(\mathbf k_1j_1,\mathbf k_2j_2,\mathbf k_3j_3)$ with $\mathbf k_1+\mathbf k_2+\mathbf k_3=0$, obtained by contracting $K^{(3)}_s$ with the mode vectors projected onto the neighbourhood of one atom. The closed diagrams are then sums over wave vectors: $F_b$ is the sunset sum $N_c^{-1}\sum|V|^2$, $\operatorname{tr}M^2$ a sum over one wave vector, the cross term of $F_{4b}$ a sum over three free wave vectors with the pair part of the fourth derivative factorised over pairs, $F_{4a}$ a sum over pairs of pairs in real space, and $K_4$ the tetrahedron, a sum over three free wave vectors of a contraction of four vertices. The embedding part of $V$ never needs the $d\times d\times d$ tensor: by the sweep of Methods, one left-to-right pass over the cores of the train, carrying which of the three mode vectors have been placed (8 states), gives $V$ for given mode vectors; with rank 6 and $d=351$ one vertex costs 15\,ms (both sublattices) and agrees with the dense contraction to $10^{-14}$ (code \texttt{tt\_vertex.py}).

\emph{Validation.} For $4^3$ cells (128 atoms) the local code reproduces the dense code: $F_b=-268.007$\,meV, $F_{4a}=-14.813$, $F_{4b}=+21.381$, $F_{4c}=-31.696$ and the $T^3$ sum $-25.129$\,meV, equal to $10^{-12}$ relative for the terms and the sum. Table~\ref{tab:loc3} gives the per-atom values for 54 to 432 atoms, each evaluated at the self-consistent force constants of its own cell. All terms converge with the cell size, with steps that shrink: for $F_b$ the changes between successive cells are $-0.110$, $-0.034$ and $-0.016$\,meV per atom, for the $T^3$ sum $-0.0076$, $-0.0018$ and $-0.0014$\,meV per atom. A fit $y_\infty+c/N$ to the four cells gives $F_b/N=-2.168$ and $\sum T^3/N=-0.2010$\,meV per atom (rms of the fit 0.002 and 0.0003); a fit in $1/N^{2/3}$ gives $-2.205$ and $-0.2034$. We take the spread as the extrapolation uncertainty, about 0.04\,meV per atom ($2\%$) for $F_b$ and 0.003\,meV per atom ($1.5\%$) for the $T^3$ sum. The $T^3$ sum is 9\,\% of $F_b$.

\emph{The tetrahedron.} $K_4$ is a sum over $N_k^3$ triples of wave vectors; with the exact sum its time grows as $N_k^3$ (about 30 minutes on four cores for $6^3$ cells). We replaced it by an average over random triples, whose cost does not depend on the cell: $2\times10^4$ triples (4\,s) give $K_4$ to $0.4\,\%$ and $10^5$ triples (20\,s) to $0.15\,\%$, unbiased against the exact sums for $4^3$, $5^3$ and $6^3$ cells ($3.604\pm0.014$ against $3.592$, $7.517\pm0.012$ against $7.505$, and $13.312\pm0.021$ against $13.317$, all $\times10^{-5}$\,eV). Because $K_4$ is about 20\,\% of the $T^3$ sum, the sampling error of the sum is $0.03\,\%$. We also evaluated $K_4$ for continuous wave vectors, with the Gaussian built from the force constants of the $24^3$ cell and the vertex contracted on demand, so that no table is stored and the cost is independent of the cell (code \texttt{tt\_kint.py}; it reproduces the table values to $2\times10^{-15}$). A rank-one lattice rule whose generating vector is the powers of one integer converged poorly (in units of $10^{-4}$\,eV for a $6^3$ cell, $1.301$, $1.332$, $1.342$, $1.365$ and $1.373$ after $2^8,\dots,2^{12}$ points); we attribute this to the choice of generating vector and not to the acoustic singularity at $\mathbf k=0$, because the integrand is bounded (its standard deviation is 0.43 of its mean). An unscrambled Sobol net with one fixed shift, which is deterministic, gives $1.342$, $1.365$, $1.346$, $1.353$, $1.354$ and $1.363$ after $2^8,\dots,2^{13}$ points, i.e.\ $1.357\pm0.01$ (mean and spread of the last three). This is 2\,\% above the exact value of the $6^3$ cell ($1.332$) and agrees with the geometric extrapolation of the exact values of the $3^3$ to $6^3$ cells ($0.863$, $1.212$, $1.297$, $1.332$, extrapolated $1.356$); the effect on the $T^3$ sum is $0.4\,\%$. The integrand is rough: its value at a point and at the nearest point of a $6^3$ mesh are correlated only to 0.6, so quasi-Monte-Carlo is not much better than sampling here, and we claim no deterministic accuracy better than about $1\,\%$ for $K_4$, which is $0.2\,\%$ of the $T^3$ sum. A tensor-train cross over the nine mesh indices of the three wave vectors was not useful: the momentum-conserving couplings (for instance $\mathbf k_a-\mathbf k_d$) keep the rank of the tensor at the size of the mesh, and more than $10^4$ evaluations were needed at a $6^3$ mesh without convergence.

\emph{What is not yet independent of the cell.} The real-space sum for $F_{4a}$ and the cross term still take times that grow with the number of cells (for $3^3$ to $6^3$ cells: 6.8, 15.9, 33.5 and 53\,s and 0.9, 3.6, 12.7 and 36.6\,s), since they run over all pairs of pairs of the cell instead of one reference pair; this has not been removed. The pressure from the local $T^3$ terms (by volume differencing) has not been computed. For 432 atoms the vertex build took about 12 minutes on four cores (measured in resumed pieces, not timed separately), the real-space terms 1.5 minutes and the sampled tetrahedron 20\,s, so the calculation is not faster than molecular dynamics, which takes 26\,s for 250 atoms.

\begin{table}[!htbp]\centering\small
\caption{\textbf{Cubic and $T^3$ free-energy terms per atom from the local formulation, B2 NiAl at 800\,K.} meV per atom, each cell at its own self-consistent force constants, $K_4$ by the exact sum. The last two rows are fits of $y_\infty+c/N$ (all four cells) and of $y_\infty+c/N^{2/3}$.}\label{tab:loc3}
\begin{tabular}{@{}rrrrrrr@{}}\toprule
cells & atoms & $F_b/N$ & $F_{4a}/N$ & $F_{4b}/N$ & $F_{4c}/N$ & $\sum T^3/N$\\\midrule
$3^3$ & 54 & $-1.9833$ & $-0.1190$ & $+0.1463$ & $-0.2160$ & $-0.18873$\\
$4^3$ & 128 & $-2.0939$ & $-0.1158$ & $+0.1671$ & $-0.2476$ & $-0.19633$\\
$5^3$ & 250 & $-2.1277$ & $-0.1151$ & $+0.1730$ & $-0.2559$ & $-0.19810$\\
$6^3$ & 432 & $-2.1434$ & $-0.1154$ & $+0.1755$ & $-0.2596$ & $-0.19946$\\
fit $1/N$ & $\infty$ & $-2.168$ & & & & $-0.2010$\\
fit $1/N^{2/3}$ & $\infty$ & $-2.205$ & & & & $-0.2034$\\
\bottomrule
\end{tabular}
\end{table}

\subsection{Reuse of cross pivots and the full pressure of the 128-atom cell}\label{sec:fast}
The profile of the first implementation (Table~\ref{tab:mdsize}; stored timings of every run) shows that the cubic stage, not the self-consistent iteration, takes most of the time: per volume, at 100--800\,K, the self-consistent iteration took 37--132\,s and the cubic stage (embedding trains with third moments, assembly of the $d\times d\times d$ tensor over the translations and the $d^4$ contraction) 262--427\,s, i.e.\ 85\,\% of the 940--1430\,s per temperature, which is three volumes. We replaced the cubic stage by the local vertex of the previous section, with $\sum X^2$ evaluated as the sunset sum over wave vectors, in which only one representative of each class of triples $(\mathbf k_1,\mathbf k_2,\mathbf k_3)$ under permutation and inversion is evaluated (379 of 4096 pairs for $4^3$ cells), and the self-consistent iteration by the local code at each volume, warm-started from the converged force constants of the central volume. We also stop the outer iteration, which rebuilds the embedding trains, when their Hessian changes by less than $3\times10^{-4}$ instead of $10^{-4}$, since the change stagnates at $1.4\times10^{-4}$, the noise of the tensor-train cross; this halves the number of tensor-train builds and changes the pressure of the reference by 0.002\,GPa. The three volumes are independent and are run on three cores.

Table~\ref{tab:fast} compares the two for B2 NiAl, 128 atoms. The pressures $P=P_{\rm HF}+P_b$ differ by $2\times10^{-4}$\,GPa at 300\,K and 0.003\,GPa at 800\,K (0.01\,\% and 0.06\,\%), which is the noise of the tensor-train cross (0.003\,GPa); the bubble pressures agree to 0.0005\,GPa and the bubble free energies to 0.01\,meV. The wall-clock time for one temperature falls from 983 and 1218\,s to 38 and 45\,s on three cores, i.e.\ the work from 16 and 20 core-minutes to about 2 core-minutes. \emph{Reuse of cross pivots.} The tensor-train cross of the embedding train is the largest remaining cost: each build sweeps the 351 cores twice, with $\sim10^4$ evaluations of the site energy per pass, and the self-consistent iteration rebuilds the trains at every outer iteration and once more at the end. The pivots, that is the sets of index paths $(i_1,\dots,i_{k})$ and $(i_{k+1},\dots,i_d)$ that define the interpolation, change very little between iterations, volumes and temperatures, because they depend only on where in the Gaussian the integrand varies. We therefore store the index paths of the final cross, restart every later build from them with no sweep (only the one-site interpolation pass that fills the cores, of cost $\sum_k r^2 n$ evaluations), and use the pivots of the train of $F$ also for the train of $\partial F/\partial V$ and for the third-moment trains of the cubic stage. The pivots stored from one run (800\,K, central volume, file \texttt{piv\_n4.pkl}, 14\,MB) are used for the three volumes at both 300 and 800\,K. The pressure changes by $-4\times10^{-4}$\,GPa at 300\,K and $-2.5\times10^{-3}$\,GPa at 800\,K relative to the first implementation, again the noise of the cross, and the bubble pressures by 0.0001 and 0.0008\,GPa. The wall-clock time falls to 16 and 22\,s (1.4 and 1.9 core-minutes in total, per volume 7 and 12\,s of self-consistent iteration, 2.4\,s of vertex build and 5.7\,s of sunset sum). The first build of a new system, without stored pivots, still takes the longer time. (Only B2 NiAl at 300 and 800\,K has been checked; $\gamma$-TiAl is not implemented in the local code. The timings are from one workstation; each volume used one thread.)

\begin{table}[!htbp]\centering\small
\caption{\textbf{Full pressure of 128-atom B2 NiAl, first implementation and local implementation.} $P_{\rm HF}$: pressure of the self-consistent Gaussian, $P_b$: bubble pressure from the free energies at $V(1\pm1\%)$, $P$: their sum, in GPa; $F_b$ in meV for the whole cell at $V$. Times are per volume, in s on one core: self-consistent iteration (SCHA), cubic stage (bubble); total: wall-clock time per temperature (first implementation: three volumes in sequence on one core; local: three volumes in parallel on three cores). ``Pivots reused'': the tensor-train cross starts from the interpolation sets of an earlier run (see text).}\label{tab:fast}
\begin{tabular}{@{}rl|rrrr|rrr@{}}\toprule
$T$ [K] & code & $P_{\rm HF}$ & $P_b$ & $P$ & $F_b$ & SCHA & bubble & total [s]\\\midrule
300 & first & 1.92531 & 0.02694 & 1.95225 & $-42.476$ & 37 & 282 & 983\\
300 & local & 1.92514 & 0.02692 & 1.95206 & $-42.471$ & 24--27 & 11 & 38\\
300 & local, pivots reused & 1.92483 & 0.02698 & 1.95181 & $-42.471$ & 7 & 8 & 16\\
800 & first & 4.88468 & 0.16168 & 5.04636 & $-268.007$ & 120 & 278 & 1218\\
800 & local & 4.88791 & 0.16162 & 5.04952 & $-268.014$ & 30--31 & 13 & 45\\
800 & local, pivots reused & 4.88299 & 0.16089 & 5.04388 & $-267.981$ & 11--13 & 8 & 22\\
\bottomrule
\end{tabular}
\end{table}

\section{Comparison with molecular dynamics and Frenkel--Ladd free energies for 65\,536 atoms}\label{sec:md65k}
\paragraph{Reference calculations.} For each alloy the free energy, energy, pressure and entropy were obtained for a cell of 65\,536 atoms at one fixed volume per alloy (Table~\ref{tab:md65k}), chosen so that the pressure is close to zero at 600\,K. The absolute free energy at $T_0=600$\,K comes from a Frenkel--Ladd integration from an Einstein crystal (ten nodes in the coupling parameter, statistical error 0.004\,meV per atom); at the other temperatures from 600 to 1000\,K it follows by integrating the molecular-dynamics energy along the temperature, $\partial(F/T)/\partial T=U/T^2$ for the potential energy $U$. The energy $U$ and the pressure $P$ are NVT averages, the entropy is $S=(U-F)/T$. The comparison is therefore a comparison with sampling at a size at which the finite-size error of the cell is negligible; it is not a comparison at the cell sizes of the other tables, and the volumes differ from those of the comparison with 1\,024 atoms in the Letter.

\paragraph{Tensor-train calculation.} The thermodynamic-limit calculation of the Letter (local formulation, Brillouin-zone integrals for the bubble and $T^3$ terms, $8^3$-cell covariance grid, three volumes $V(1\pm1\%)$) was run at the volume of the molecular-dynamics cell for the temperatures of the scan. The free energy per atom is the absolute classical free energy of the Gaussian, $F_{GB}=\langle U\rangle_{\Phi^\star}+3\kB T\,\overline{\ln(\hbar\omega/\kB T)}-\tfrac32\kB T$, where the bar is the mean over the six modes on a shifted $24^3$ mesh of wave vectors, plus the bubble and $T^3$ terms per atom, $F=F_{GB}+(F_b+F_{4a}+F_{4b}+F_{4c})/2$. The pressure is $P_{\rm HF}+P_b+P_{T^3}$ as in the Letter. The entropy $S=-\partial F/\partial T$ is obtained from a third-order polynomial fit in $T$ of $F+3\kB T\ln(T/T_0)$ through the values of the five temperatures of the scan, and the energy from $U=F+TS$; the fit introduces no information from molecular dynamics.

\begin{table}[!htbp]\centering\scriptsize
\caption{\textbf{Tensor-train (TT) calculation against molecular dynamics with Frenkel--Ladd free energy for 65\,536 atoms, 600 to 1000\,K, at fixed volume.} TT values are the means of three runs from different starting pivots (Table~\ref{tab:seedmean}), for the infinite crystal, with no sampling. $F$: free energy per atom in eV (TT: absolute, from the Gaussian plus the bubble and $T^3$ terms; MD: Frenkel--Ladd at 600\,K and integration of the energy), $\Delta F$ TT$-$MD in meV per atom. $P$: pressure in GPa (MD: $1\sigma$ block error in units of the last digit), $\Delta P$ in MPa. The entropy and energy differences are in Table~\ref{tab:seedmean}.}\label{tab:md65k}
\setlength{\tabcolsep}{4pt}
\begin{tabular}{@{}r|rrr|rrr@{}}\toprule
$T$ [K] & $F_{\rm TT}$ & $F_{\rm MD}$ & $\Delta F$ & $P_{\rm TT}$ & $P_{\rm MD}$ & $\Delta P$\\\midrule
\multicolumn{7}{@{}l}{\emph{B2 NiAl, $V=11.6484$\,\AA$^3$/atom}}\\
600 & -4.60952 & -4.60941 & -0.12 & -0.0050 & -0.0000(3) & -5.0\\
700 & -4.65445 & -4.65433 & -0.12 & 0.6168 & 0.6258(3) & -9.0\\
800 & -4.70311 & -4.70298 & -0.13 & 1.2299 & 1.2455(3) & -15.6\\
900 & -4.75503 & -4.75490 & -0.13 & 1.8442 & 1.8588(3) & -14.6\\
1000 & -4.80985 & -4.80972 & -0.13 & 2.4470 & 2.4670(7) & -20.0\\
\midrule
\multicolumn{7}{@{}l}{\emph{$\gamma$-TiAl, $V=17.1561$\,\AA$^3$/atom}}\\
600 & -4.59934 & -4.59923 & -0.11 & -0.0044 & 0.0002(2) & -4.6\\
700 & -4.64346 & -4.64334 & -0.12 & 0.5353 & 0.5391(2) & -3.8\\
800 & -4.69136 & -4.69124 & -0.13 & 1.0717 & 1.0739(3) & -2.2\\
900 & -4.74258 & -4.74246 & -0.12 & 1.6012 & 1.6033(4) & -2.1\\
1000 & -4.79676 & -4.79663 & -0.12 & 2.1221 & 2.1281(5) & -6.0\\
\bottomrule
\end{tabular}
\end{table}

\begin{figure}[!htbp]\centering
\includegraphics[width=\textwidth]{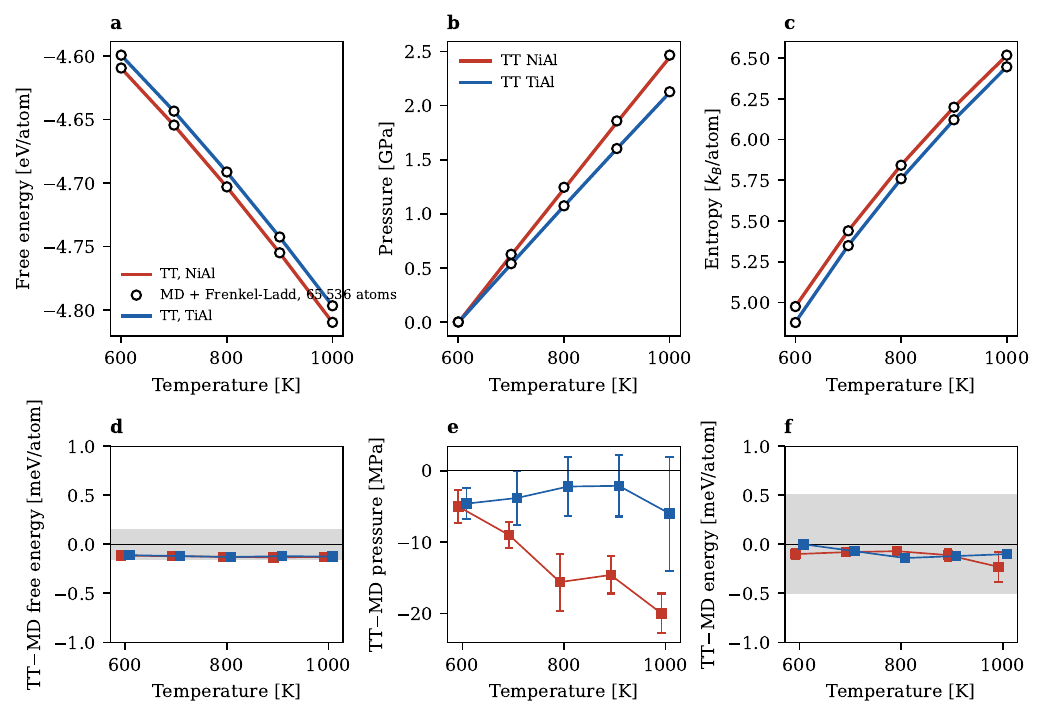}
\caption{\textbf{Tensor-train results against molecular dynamics with Frenkel--Ladd free energy for 65\,536 atoms, 600 to 1000\,K.} (a)--(c) Free energy, pressure and entropy per atom of B2 NiAl (red) and $\gamma$-TiAl (blue) at fixed volume: lines, tensor-train calculation for the infinite crystal (mean of three runs from different starting pivots); circles, molecular dynamics. (d)--(f) Differences TT$-$MD of the free energy, the pressure and the potential energy; error bars, $\pm1$ standard deviation over the three TT runs; grey bands, the pass limits (0.15 and 0.5\,meV/atom in the free energy and the potential energy). Data in Table~\ref{tab:seedmean}.}\label{fig:m65}
\end{figure}

\paragraph{Findings.} Figure~\ref{fig:m65} and Table~\ref{tab:md65k} give the result. The free energy of the three-run means agrees to $0.11$--$0.13$\,meV per atom at all five temperatures and for both alloys, i.e.\ to $3\times10^{-5}$ of its value; at 600\,K, the temperature of the Frenkel--Ladd integration, it is $-0.11$\,meV per atom (the Frenkel--Ladd error is 0.004\,meV per atom), and it hardly changes up to 1000\,K. The entropy agrees to $0.0012\,\kB$ (NiAl) and $0.0021\,\kB$ ($\gamma$-TiAl) and the energy to $0.23$ and $0.14$\,meV per atom (Table~\ref{tab:seedmean}). The pressure agrees to within 6\,MPa for $\gamma$-TiAl and within $5$--$20$\,MPa for NiAl, where the difference grows with temperature, from $-1.4\,\%$ at 700\,K to $-0.8\,\%$ at 1000\,K of the pressure. The tensor-train pressure carries a scatter of 2--4\,MPa from the starting pivots of the TT-cross (up to 9\,MPa at single points; see below) and a possible bias of order 10\,MPa from the finite difference of the $T^3$ term (Letter); at 600\,K, where the pressure of the cell is zero, the TT value is $-5$\,MPa for both alloys, a measure of the offset that is common to all temperatures. The growth of the NiAl difference with temperature may come from the terms of the expansion beyond order $T^3$, which were not computed. A rough estimate of the next order, the $T^3$ pressure times its ratio to the cubic term, reduces the NiAl difference to $-2$ to $-8$\,MPa but does not remove it (Table~\ref{tab:pnext}); for $\gamma$-TiAl it makes the difference positive, up to $+5$\,MPa at 900\,K. The cause is therefore not established. The scatter from the starting pivots (2.4\,MPa typically, at most 7.7\,MPa) is smaller than the difference at 1000\,K (20\,MPa), so it does not explain the growth. The free energy and the entropy are the stricter tests of the expansion, since they are not differences between two volumes.

\paragraph{Dependence on the starting pivots of the TT-cross.} The TT-cross has no random seed, but its result depends on the pivots it starts from. We therefore ran the whole pipeline of \texttt{bulk\_65k.sh} with three random sets of starting pivots (90 runs: two alloys, five temperatures, three volumes, three pivot sets), and all values of this section are means over the three sets. An earlier single run, whose starting pivots were not recorded, is not used. Table~\ref{tab:pivot} gives the scatter: typically 2.4\,MPa (NiAl) and 3.5\,MPa (TiAl) in $P_{\rm HF}$, up to 7.7 and 8.9\,MPa at single points, and at most 0.013\,meV per atom in $F_{\rm GB}$. This is the size of the 3--4\,MPa noise floor quoted in the Letter; its maximum is about twice larger.

\begin{table}[!htbp]\centering\small
\caption{\textbf{Scatter of the TT-cross results with the starting pivots.} 15 points per row (five temperatures from 600 to 1000\,K, three volumes), three random pivot sets. Median and maximum of the standard deviation (SD) of the three runs. The largest SD of the Gaussian free energy $F_{\rm GB}$ is 0.0127\,meV per atom (NiAl) and 0.0117\,meV per atom ($\gamma$-TiAl).}\label{tab:pivot}
\footnotesize\setlength{\tabcolsep}{4pt}
\begin{tabular}{@{}llrr@{}}\toprule
Alloy & Quantity & median SD & max SD\\\midrule
NiAl & $P_{\rm HF}$ (MPa) & 2.43 & 7.67\\
TiAl & $P_{\rm HF}$ (MPa) & 3.50 & 8.90\\
\bottomrule
\end{tabular}
\end{table}

\paragraph{Is one sweep enough?} For screening, yes at the level of the scatter in Table~\ref{tab:pivot}: the Gaussian free energy changes by at most 0.013\,meV per atom (limit for $\Delta F$: 0.15) and the pressure by at most 8.9\,MPa between pivot sets, against differences from MD of up to 20\,MPa. For the entropy and energy, which come from a fit over five temperatures, the run-to-run scatter is at most $0.0019\,\kB$ and $0.15$\,meV (NiAl, 1000\,K), 30 to 40\,\% of the limits of 0.005\,$\kB$ and 0.5\,meV, so we averaged three runs for the comparison with MD and recommend at least three when quoting $S$ and $U$. We did not test a single sweep for $S$ and $U$ separately.

\paragraph{Does the scatter shrink with more sweeps or a higher rank?} We tested this for NiAl at 800\,K, with three random pivot sets per setting (Table~\ref{tab:sweep}). The rule fixed beforehand was that the scatter counts as controllable if the standard deviation of $P_{\rm HF}$ falls by at least a factor of two from the lowest to the highest setting. \emph{It does not meet the rule:} from one to four sweeps the standard deviation falls by a factor of 1.18, and from rank cap 6 to 12 by a factor of 1.13. The scatter of the Gaussian free energy $F_{\rm GB}$ does fall with the number of sweeps (0.0054 to 0.0007\,meV per atom), but it is already small. A standard deviation from three pivot sets is itself uncertain by about 50\,\%, so factors close to one cannot be told apart from noise. We therefore treat the pressure scatter as a noise floor of the method at the present settings, not as something the sweeps or the rank remove, and average three runs where it matters. A fixed, non-random starting set of pivots (the same for every run) is reproducible: its $P_{\rm HF}$ lies between $-0.2$ and $+1.0$ standard deviations from the three-set mean at the three volumes, and its $F_{\rm GB}$ within 0.6 standard deviations at two volumes and $-0.007$\,meV per atom ($-4.2$ standard deviations of a particularly narrow spread) at the third.
\begin{table}[!htbp]\centering\small
\caption{\textbf{Scatter of $P_{\rm HF}$ and $F_{\rm GB}$ with the number of sweeps and the rank cap, NiAl at 800\,K.} Median over the three volumes of the standard deviation over three random pivot sets, and core-seconds per run. The row with two sweeps and rank cap 6 is the production setting.}\label{tab:sweep}
\footnotesize\setlength{\tabcolsep}{4pt}
\begin{tabular}{@{}lrrr@{}}\toprule
Setting & SD of $P_{\rm HF}$ [MPa] & SD of $F_{\rm GB}$ [meV/atom] & core-s per run\\\midrule
$r_{\max}=6$, 1 sweep & 4.54 & 0.0054 & 64\\
$r_{\max}=6$, 2 sweeps (production) & 3.26 & 0.0054 & 71\\
$r_{\max}=6$, 3 sweeps & 2.23 & 0.0022 & 81\\
$r_{\max}=6$, 4 sweeps & 3.86 & 0.0007 & 87\\
$r_{\max}=8$, 2 sweeps & 4.66 & 0.0041 & 95\\
$r_{\max}=12$, 2 sweeps & 2.88 & 0.0023 & 132\\
\bottomrule
\end{tabular}
\end{table}

\paragraph{Is the rank cap harmless?} The production runs cap the rank of the embedding remainder at $r_{\max}=6$ (10 for the pressure integrand), while the singular values in Appendix~\ref{app:C} are shown for $r_{\max}=12$. Table~\ref{tab:rankcap} gives the change of the result when the cap is changed, relative to $r_{\max}=6$, at 600, 800 and 1000\,K for both alloys (three pivot sets each, the same sets for every rank). The limits fixed beforehand were 0.03\,meV per atom for the sum of the bubble and $T^3$ terms, 0.15\,meV per atom for $F$ and 5\,MPa for $P$. \emph{Both alloys pass at every rank and temperature tried.} The largest changes are 0.006\,meV per atom in the sum of the bubble and $T^3$ terms, 0.012\,meV per atom in $F$ and 4.5\,MPa in $P$; every pressure change is at or below the standard deviation over the pivot sets (0.5--12.5\,MPa). The pressure change does not grow with the rank, except for $\gamma$-TiAl at 800\,K (0.8, 3.5 and 4.5\,MPa for $r_{\max}=4,8,12$), which is still below the limit and within its pivot scatter (4.2--5.5\,MPa). The rank cap therefore does not bias the bubble and $T^3$ terms at the level of our limits. The runs cost on average 51, 65, 82 and 121 core-seconds per volume for $r_{\max}=4,6,8,12$.
\begin{table}[!htbp]\centering\small
\caption{\textbf{Effect of the rank cap on the result.} Change relative to $r_{\max}=6$, mean $\pm$ standard deviation over three pivot sets; the tolerance of the TT-cross is $10^{-6}$ except where stated. $\Delta(F_b+F_4)$ and $\Delta F$ in meV per atom, $\Delta P$ in MPa. Limits: 0.03, 0.15 and 5.}\label{tab:rankcap}
\scriptsize\setlength{\tabcolsep}{2.2pt}
\begin{tabular}{@{}rr|rrr|rrr@{}}\toprule
 & & \multicolumn{3}{c|}{NiAl} & \multicolumn{3}{c}{$\gamma$-TiAl}\\
$T$ [K] & $r_{\max}$ & $\Delta(F_b+F_4)$ & $\Delta F$ & $\Delta P$ & $\Delta(F_b+F_4)$ & $\Delta F$ & $\Delta P$\\\midrule
600 & 4 & $+0.0001\pm0.0012$ & $-0.0018\pm0.0017$ & $-0.72\pm4.85$ & $-0.0006\pm0.0013$ & $+0.0037\pm0.0001$ & $+2.64\pm0.52$\\
 & 8 & $+0.0005\pm0.0003$ & $+0.0009\pm0.0032$ & $-3.75\pm4.97$ & $+0.0004\pm0.0001$ & $+0.0002\pm0.0017$ & $-0.15\pm3.07$\\
 & 12 & $-0.0000\pm0.0002$ & $+0.0014\pm0.0008$ & $-0.22\pm0.69$ & $+0.0010\pm0.0006$ & $+0.0021\pm0.0011$ & $+0.83\pm1.65$\\
 & 12 ($10^{-10}$) & $+0.0004\pm0.0005$ & $+0.0017\pm0.0016$ & $-0.07\pm1.85$ & $+0.0015\pm0.0006$ & $+0.0015\pm0.0002$ & $+1.45\pm3.68$\\
\midrule
800 & 4 & $+0.0018\pm0.0015$ & $+0.0033\pm0.0060$ & $-2.89\pm4.86$ & $-0.0008\pm0.0019$ & $-0.0022\pm0.0062$ & $-0.82\pm4.69$\\
 & 8 & $-0.0013\pm0.0012$ & $-0.0033\pm0.0102$ & $+4.09\pm0.86$ & $+0.0018\pm0.0013$ & $+0.0051\pm0.0072$ & $-3.49\pm5.46$\\
 & 12 & $-0.0001\pm0.0005$ & $+0.0010\pm0.0066$ & $+3.54\pm4.00$ & $+0.0045\pm0.0011$ & $+0.0103\pm0.0071$ & $-4.50\pm4.16$\\
 & 12 ($10^{-10}$) & $+0.0000\pm0.0013$ & $+0.0017\pm0.0069$ & $+2.60\pm4.80$ & $+0.0035\pm0.0024$ & $+0.0080\pm0.0013$ & $-2.07\pm3.21$\\
\midrule
1000 & 4 & $+0.0005\pm0.0027$ & $+0.0014\pm0.0114$ & $-3.24\pm3.05$ & $-0.0007\pm0.0018$ & $+0.0030\pm0.0166$ & $-2.30\pm12.51$\\
 & 8 & $-0.0011\pm0.0011$ & $+0.0097\pm0.0211$ & $-1.27\pm3.35$ & $+0.0037\pm0.0026$ & $+0.0116\pm0.0070$ & $-0.02\pm5.90$\\
 & 12 & $-0.0002\pm0.0014$ & $+0.0059\pm0.0186$ & $+0.32\pm5.13$ & $+0.0057\pm0.0012$ & $+0.0084\pm0.0052$ & $-2.36\pm9.93$\\
 & 12 ($10^{-10}$) & $-0.0006\pm0.0006$ & $+0.0020\pm0.0141$ & $+2.05\pm7.02$ & $+0.0059\pm0.0022$ & $+0.0074\pm0.0140$ & $-2.98\pm8.79$\\
\bottomrule
\end{tabular}
\end{table}

\paragraph{Comparison with the means of three starting-pivot runs.} Table~\ref{tab:seedmean} gives the differences TT$-$MD from the 90 runs above, the three-run mean at each temperature with its standard deviation. The entropy is the slope of a degree-3 fit of $F+3\kB T\ln(T/600\,\mathrm{K})$ over the five temperatures of the same run, and $U=F+TS$; runs are never mixed. The limits were fixed before the computation: $|\Delta F|\le0.15$\,meV, $|\Delta S|\le0.005\,\kB$, $|\Delta U|\le0.5$\,meV per atom. \emph{Both alloys pass at all five temperatures.} The largest differences are, for NiAl, $0.133$\,meV in $F$ (at 900\,K, within 0.02\,meV of the limit), $0.0012\,\kB$ in $S$ and $0.23$\,meV in $U$, and for $\gamma$-TiAl $0.129$\,meV, $0.0021\,\kB$ and $0.14$\,meV. A fit of five points with four coefficients is only mildly overdetermined, so the entropy and energy at the ends of the range are less well constrained than in the middle; a degree-2 fit changes $\Delta S$ by at most $0.0004\,\kB$ and $\Delta U$ by at most $0.04$\,meV per atom, and all points still pass.

\begin{table}[!htbp]\centering\small
\caption{\textbf{TT$-$MD for 65\,536 atoms, mean of three runs from different starting pivots $\pm$ their standard deviation, 600 to 1000\,K.} $\Delta F$ and $\Delta U$ in meV per atom, $\Delta S$ in $\kB$ per atom, $\Delta P$ in MPa. Limits (fixed beforehand): $|\Delta F|\le0.15$, $|\Delta S|\le0.005$, $|\Delta U|\le0.5$; the verdict is for these three. $S$ and $U$ come from a degree-3 fit over the five temperatures of each run.}\label{tab:seedmean}
\footnotesize\setlength{\tabcolsep}{2pt}
\resizebox{\textwidth}{!}{%
\begin{tabular}{@{}rrrrrl@{\quad}rrrrl@{}}\toprule
 & \multicolumn{5}{c}{NiAl} & \multicolumn{5}{c}{$\gamma$-TiAl}\\\cmidrule(lr){2-6}\cmidrule(l){7-11}
$T$ [K] & $\Delta F$ & $\Delta S$ & $\Delta U$ & $\Delta P$ & verdict & $\Delta F$ & $\Delta S$ & $\Delta U$ & $\Delta P$ & verdict\\\midrule
600 & $-0.115\pm0.001$ & $+0.0003\pm0.0010$ & $-0.10\pm0.05$ & $-5.0\pm2.3$ & pass & $-0.111\pm0.001$ & $+0.0021\pm0.0002$ & $-0.00\pm0.01$ & $-4.6\pm2.2$ & pass\\
700 & $-0.121\pm0.004$ & $+0.0006\pm0.0001$ & $-0.08\pm0.00$ & $-9.0\pm1.8$ & pass & $-0.120\pm0.004$ & $+0.0008\pm0.0002$ & $-0.07\pm0.01$ & $-3.8\pm3.8$ & pass\\
800 & $-0.128\pm0.006$ & $+0.0008\pm0.0004$ & $-0.07\pm0.03$ & $-15.6\pm4.0$ & pass & $-0.129\pm0.003$ & $-0.0002\pm0.0002$ & $-0.14\pm0.01$ & $-2.2\pm4.1$ & pass\\
900 & $-0.133\pm0.003$ & $+0.0003\pm0.0008$ & $-0.11\pm0.06$ & $-14.6\pm2.6$ & pass & $-0.121\pm0.007$ & $-0.0000\pm0.0003$ & $-0.12\pm0.02$ & $-2.1\pm4.3$ & pass\\
1000 & $-0.129\pm0.013$ & $-0.0012\pm0.0019$ & $-0.23\pm0.15$ & $-20.0\pm2.8$ & pass & $-0.125\pm0.009$ & $+0.0003\pm0.0005$ & $-0.10\pm0.03$ & $-6.0\pm8.0$ & pass\\
\bottomrule
\end{tabular}%
}
\end{table}

\begin{table}[!htbp]\centering\small
\caption{\textbf{Pressure difference TT$-$MD, estimate of the next order, and their sum (MPa), three-run means.} $\Delta P$: difference to MD. $P_{\rm next}=P_{T^3}\,(P_{T^3}/P_b)$: the $T^3$ pressure times its ratio to the cubic (bubble) pressure, a rough geometric extrapolation of the series, not a calculation of the next order. A power-law fit of $\Delta P$ against $T$ is not meaningful for $\gamma$-TiAl, where $\Delta P$ is small and the sum changes sign. The MD error is 0.2--0.7\,MPa.}\label{tab:pnext}
\footnotesize
\begin{tabular}{@{}rrrrcrrr@{}}\toprule
 & \multicolumn{3}{c}{NiAl} & & \multicolumn{3}{c}{$\gamma$-TiAl}\\\cmidrule(lr){2-4}\cmidrule(l){6-8}
$T$ [K] & $\Delta P$ & $P_{\rm next}$ & sum & & $\Delta P$ & $P_{\rm next}$ & sum\\\midrule
600 & $-5.0$ & 2.92 & $-2.1$ & & $-4.6$ & 1.82 & $-2.8$\\
700 & $-9.0$ & 4.78 & $-4.2$ & & $-3.8$ & 3.02 & $-0.8$\\
800 & $-15.6$ & 7.23 & $-8.4$ & & $-2.2$ & 4.62 & $+2.4$\\
900 & $-14.6$ & 10.34 & $-4.3$ & & $-2.1$ & 6.64 & $+4.5$\\
1000 & $-20.0$ & 13.96 & $-6.0$ & & $-6.0$ & 9.06 & $+3.1$\\
\bottomrule
\end{tabular}
\end{table}

\paragraph{Is the embedding remainder needed?} Removing the TT of the embedding remainder from the pipeline changes the free energy by $+0.38$ to $+0.67$\,meV per atom and the pressure by $-66$ to $-120$\,MPa for NiAl, and the free energy by $+0.44$ to $+0.72$\,meV per atom and the pressure by $-28$ to $-53$\,MPa for $\gamma$-TiAl (ranges over 600 to 1000\,K; one pivot set), against our criterion of 0.03\,meV per atom and 5\,MPa, so the remainder is needed; both changes are far above the scatter of Table~\ref{tab:pivot}. Dropping the $T^3$ terms instead changes $F$ by $+0.13$ to $+0.47$\,meV per atom (NiAl) and $+0.12$ to $+0.49$\,meV per atom ($\gamma$-TiAl) and $P$ by $-18$ to $-60$\,MPa and $-15$ to $-51$\,MPa. The free energy then differs from MD by $+0.010$ to $+0.336$\,meV per atom (NiAl) and $+0.013$ to $+0.363$\,meV per atom ($\gamma$-TiAl), against $-0.11$ to $-0.13$\,meV per atom with them.

\paragraph{Thermal quantities at zero pressure.} The free energy at several volumes also gives the thermal expansion, the bulk modulus and the heat capacities of the two crystals, which can be compared with published calculations. For each temperature we added the volumes $V_0(1+2\%)$, $V_0(1+3\%)$ and $V_0(1+4\%)$ to the three of the comparison above (90 new runs, 2\,946 core-seconds for NiAl and 2\,465 for $\gamma$-TiAl, three pivot sets each), and repeated the procedure at 300, 400 and 500\,K with a bracket of volumes around the zero-pressure volume (3\,074 and 2\,295 core-seconds; $P=0$ fell inside the bracket at every temperature), fitted the pressure $P(V)$ with a quadratic through the six volumes, and took the volume $V_{\rm eq}(T)$ at which $P=0$ (the fits are over 300--1000\,K); a cubic fit changes $V_{\rm eq}$ by at most 0.001\,\AA$^3$ per atom, $\alpha_V$ by at most $0.06\times10^{-5}$\,K$^{-1}$ and $B_T$ by at most 0.7\,GPa. The volume expansion coefficient is $\alpha_V=V^{-1}dV_{\rm eq}/dT$ (degree-3 fit of $V_{\rm eq}(T)$), the isothermal bulk modulus $B_T=-V\,\partial P/\partial V$ at $P=0$, $C_V=-T\,\partial^2F/\partial T^2$ at $V_{\rm eq}$, and $C_P=C_V+TV\alpha_V^2B_T$. Everything is classical: there is no zero-point motion and no electronic contribution. Tables~\ref{tab:thermNiAl} and \ref{tab:thermTiAl} give the results.
\begin{table}[!htbp]\centering\small
\caption{\textbf{B2 NiAl at zero pressure, 300 to 1000\,K.} Mean $\pm$ standard deviation over three pivot sets. $V_{\rm eq}$ in \AA$^3$ per atom, $a_{\rm eq}=(2V_{\rm eq})^{1/3}$ in \AA, linear expansion relative to 300\,K, $\alpha_V$ in $10^{-5}$\,K$^{-1}$, $B_T$ in GPa, heat capacities in $\kB$ per atom. Classical, no electronic or quantum terms.}\label{tab:thermNiAl}
\footnotesize\setlength{\tabcolsep}{3pt}
\begin{tabular}{@{}rrrrrrrr@{}}\toprule
$T$ [K] & $V_{\rm eq}$ & $a_{\rm eq}$ & expansion [\%] & $\alpha_V$ & $B_T$ & $C_V$ & $C_P$\\\midrule
300 & 11.4999 & 2.84386 & 0 & $4.164\pm0.016$ & $152.75\pm0.03$ & $3.003\pm0.000$ & $3.069\pm0.001$\\
400 & 11.5483 & 2.84784 & 0.140 & $4.239\pm0.005$ & $150.05\pm0.04$ & $3.005\pm0.001$ & $3.095\pm0.001$\\
500 & 11.5976 & 2.85189 & 0.282 & $4.298\pm0.003$ & $147.18\pm0.07$ & $3.008\pm0.001$ & $3.122\pm0.000$\\
600 & 11.6480 & 2.85602 & 0.427 & $4.34\pm0.04$ & $144.44\pm0.19$ & $3.013\pm0.002$ & $3.151\pm0.003$\\
700 & 11.6988 & 2.86016 & 0.573 & $4.37\pm0.01$ & $141.84\pm0.12$ & $3.014\pm0.003$ & $3.174\pm0.003$\\
800 & 11.7503 & 2.86435 & 0.721 & $4.38\pm0.02$ & $139.27\pm0.14$ & $3.015\pm0.003$ & $3.197\pm0.005$\\
900 & 11.8018 & 2.86853 & 0.868 & $4.39\pm0.00$ & $136.91\pm0.19$ & $3.014\pm0.004$ & $3.217\pm0.004$\\
1000 & 11.8538 & 2.87273 & 1.015 & $4.39\pm0.06$ & $134.52\pm0.03$ & $3.012\pm0.005$ & $3.235\pm0.003$\\
\bottomrule
\end{tabular}
\end{table}
\begin{table}[!htbp]\centering\small
\caption{\textbf{$\gamma$-TiAl at zero pressure, 300 to 1000\,K.} As Table~\ref{tab:thermNiAl}; the ratio $c/a=1.0485$ of the cell is kept fixed, and the linear expansion is $(V_{\rm eq}/V_{\rm eq}(300\,{\rm K}))^{1/3}-1$.}\label{tab:thermTiAl}
\footnotesize\setlength{\tabcolsep}{3pt}
\begin{tabular}{@{}rrrrrrr@{}}\toprule
$T$ [K] & $V_{\rm eq}$ & expansion [\%] & $\alpha_V$ & $B_T$ & $C_V$ & $C_P$\\\midrule
300 & 16.9226 & 0 & $4.146\pm0.017$ & $126.15\pm0.02$ & $3.039\pm0.003$ & $3.118\pm0.002$\\
400 & 16.9955 & 0.143 & $4.437\pm0.002$ & $120.72\pm0.03$ & $3.047\pm0.002$ & $3.164\pm0.002$\\
500 & 17.0729 & 0.295 & $4.709\pm0.011$ & $115.10\pm0.04$ & $3.054\pm0.000$ & $3.212\pm0.001$\\
600 & 17.1563 & 0.458 & $5.01\pm0.02$ & $109.24\pm0.09$ & $3.061\pm0.004$ & $3.265\pm0.003$\\
700 & 17.2438 & 0.629 & $5.18\pm0.01$ & $104.65\pm0.10$ & $3.063\pm0.002$ & $3.309\pm0.001$\\
800 & 17.3352 & 0.806 & $5.38\pm0.01$ & $100.04\pm0.21$ & $3.063\pm0.003$ & $3.353\pm0.004$\\
900 & 17.4304 & 0.990 & $5.60\pm0.04$ & $95.79\pm0.02$ & $3.059\pm0.009$ & $3.400\pm0.011$\\
1000 & 17.5304 & 1.183 & $5.84\pm0.08$ & $91.38\pm0.21$ & $3.052\pm0.019$ & $3.448\pm0.025$\\
\bottomrule
\end{tabular}
\end{table}
\emph{Comparison with a published calculation.} Zope and Mishin~\cite{ZopeMishin2003} give the linear thermal expansion of $\gamma$-TiAl with this potential, from the quasiharmonic approximation and from Monte Carlo sampling (for $\gamma$-TiAl they state only ``by the Monte Carlo method''; the ensemble and the cell size are not given). Their table does not state its reference temperature, so we take differences of its entries relative to 400\,K, which is valid if the entries are expansions from one common reference. Their Monte Carlo values are then 0.32, 0.67 and 1.05\,\% at 600, 800 and 1000\,K and their quasiharmonic values 0.30, 0.67 and 1.18\,\%. Ours relative to 400\,K, from Table~\ref{tab:thermTiAl}, are 0.314, 0.662 and 1.038\,\%: they agree with the Monte Carlo values to 0.012 percentage points, and at 1000\,K they differ from the quasiharmonic value by 0.14 points, which is what one expects of a classical calculation with the full anharmonicity. This compares two calculations with one potential. The potential gives $c/a=1.047$ at equilibrium; our cell keeps $c/a=1.0485$ fixed, so the $\gamma$-TiAl expansion and modulus here describe this potential at a fixed ratio. The classical $C_P$ of 3.07--3.24\,$\kB$ (NiAl) and 3.12--3.45\,$\kB$ ($\gamma$-TiAl) per atom over 300--1000\,K omits the electronic contribution and zero-point effects of real crystals.

\paragraph{Elastic constants.} The isothermal elastic constants follow from the second derivative of the free energy per volume with respect to a homogeneous Lagrangian strain, taken at the zero-pressure volume $V_{\rm eq}(T)$ of Tables~\ref{tab:thermNiAl} and \ref{tab:thermTiAl}. For B2 NiAl we used stretching along $x$ ($C_{11}$), equal scaling of the three axes (the bulk modulus $B=(C_{11}+2C_{12})/3$, from which $C_{12}$ follows), the shear $yz$ ($C_{44}$) and $x$ stretched with $y$ compressed ($C_{11}-C_{12}$, a second route to $C_{12}$). For $\gamma$-TiAl we used stretching along $x$, $x$ stretched with $y$ compressed, stretching along $z$ ($C_{33}$), $x$ and $z$ stretched together (the curvature is $C_{11}+C_{33}+2C_{13}$), the shears $yz$ ($C_{44}$) and $xy$ ($C_{66}$), and equal scaling. The strains are applied in the conventional cell, 45$^\circ$ from the axes of the primitive cell used by the code, and the lattice sites of the neighbour clusters are kept at their unstrained positions so that the TT-cross pivots can be reused in every strained run. The amplitude, 1\,\%, was chosen at 300\,K from one pivot set by the rule written before the runs: the curvatures from 1 and from 2\,\% strain differ by at most 1.9\,GPa (limit 3\,GPa), and $F(+e)-F(-e)$ of the shears is at most 0.0012\,meV per atom (limit 0.013\,meV). All runs use the three pivot sets of the zero-pressure study; there are 216 production runs.

\emph{Check of the strained cells at zero temperature.} The paper of Zope and Mishin does not state the method or the temperature of its elastic constants. Without the tensor train, our static (zero-temperature) constants of their potential at its own relaxed cell ($a=3.99789$\,\AA, $c/a=1.04721$) reproduce the published values to within 1.0\,GPa (Table~\ref{tab:elstat}); the sublattice forces under strain are at most $4\times10^{-12}$\,eV/\AA{}, so there is no internal relaxation. For NiAl we computed the static constants but have no published values to compare them with.
\begin{table}[!htbp]\centering\small
\caption{\textbf{Static elastic constants at zero temperature (GPa).} $\gamma$-TiAl at the relaxed cell of the potential; the published values are those of Zope and Mishin~\cite{ZopeMishin2003}. B2 NiAl at $a=2.83199$\,\AA{}: $C_{11}=190.7$, $C_{12}=142.9$, $C_{44}=121.5$, $B=158.8$\,GPa (no published value used).}\label{tab:elstat}
\footnotesize\setlength{\tabcolsep}{4pt}
\begin{tabular}{@{}lrrrrrr@{}}\toprule
 & $C_{11}$ & $C_{12}$ & $C_{13}$ & $C_{33}$ & $C_{44}$ & $C_{66}$\\\midrule
this work & 196.0 & 107.0 & 113.6 & 213.3 & 92.1 & 84.7\\
published & 195 & 107 & 113 & 213 & 92 & 84\\
\bottomrule
\end{tabular}
\end{table}
\emph{Results.} Tables~\ref{tab:elNiAl} and \ref{tab:elTiAl} give the constants with the standard deviation over the pivot sets. At 300 and 600\,K every standard deviation of the three sets is at most 0.9\,GPa. At 1000\,K the scatter between the first three pivot sets was 4--6\,GPa, and two of the rules written before the runs failed there: the standard deviation of a single run was above 3\,GPa for several constants, and the two routes to $C_{11}-C_{12}$ of NiAl differed by 5.4\,GPa (53.4 and 58.8\,GPa) against a limit of 3\,GPa. We therefore ran six more pivot sets at 1000\,K (144 runs, 8\,946 core-seconds; the seeds were written down before the runs, and all six sets are used). The single-run scatter over nine sets is still above 3\,GPa for NiAl $C_{11}$ (3.8), $C_{44}$ (4.2) and for $C_{11}-C_{12}$ from $C_{11}$ and $B$ (5.2\,GPa), and the 3\,GPa rule for the two routes still fails (they differ by 3.3\,GPa), because these rules measure the scatter of one run, which more sets cannot reduce. Three further rules, written before the extra runs, concern the mean of the nine sets and all pass: the standard error of the nine-set mean is at most 1.7\,GPa for every constant (limit 2.5\,GPa); the two routes to $C_{11}-C_{12}$ differ by $-3.3$\,GPa, within twice their combined standard error (3.9\,GPa), with $53.5\pm0.9$ and $56.9\pm1.7$\,GPa; and the equal-scaling bulk modulus differs from the zero-pressure study by at most 0.14\,GPa. All $\gamma$-TiAl constants have a single-run scatter of at most 3.0\,GPa over the nine sets. Two single values lie more than 2.5 standard deviations from the mean of the other eight sets and are kept: NiAl $B$ in set 2 ($+0.9$\,GPa) and $\gamma$-TiAl $C_{12}$ in set 3 ($-6.0$\,GPa). The 1000\,K entries of Tables~\ref{tab:elNiAl} and \ref{tab:elTiAl} are the nine-set means; a single run at 1000\,K is uncertain by up to about 5\,GPa. The bulk modulus from equal scaling agrees with that of the zero-pressure study to within 0.5\,GPa at every temperature for both alloys (Table~\ref{tab:elB}), as required.
\begin{table}[!htbp]\centering\small
\caption{\textbf{B2 NiAl, isothermal elastic constants (GPa)} at the zero-pressure volume; mean $\pm$ standard deviation over three pivot sets (300 and 600\,K) or nine ($^{*}$1000\,K). The last two columns are the two routes to $C_{11}-C_{12}$ (tetragonal shear; from $C_{11}$ and $B$).}\label{tab:elNiAl}
\footnotesize\setlength{\tabcolsep}{3pt}
\begin{tabular}{@{}rrrrrrr@{}}\toprule
$T$ [K] & $C_{11}$ & $C_{12}$ & $C_{44}$ & $B$ & $C_{11}-C_{12}$ & $C_{11}-C_{12}$\\\midrule
300 & $188.5\pm0.1$ & $134.9\pm0.0$ & $111.3\pm0.2$ & $152.8\pm0.0$ & $53.7\pm0.0$ & $53.6\pm0.1$\\
600 & $180.3\pm0.6$ & $126.5\pm0.3$ & $100.8\pm0.4$ & $144.5\pm0.1$ & $54.1\pm0.3$ & $53.8\pm0.9$\\
1000$^{*}$ & $172.4\pm3.8$ & $115.5\pm1.4$ & $88.8\pm4.2$ & $134.4\pm0.4$ & $53.5\pm2.7$ & $56.9\pm5.2$\\
\bottomrule
\end{tabular}
\end{table}
\begin{table}[!htbp]\centering\small
\caption{\textbf{$\gamma$-TiAl, isothermal elastic constants (GPa)} at the zero-pressure volume (cell with $c/a=1.0485$); mean $\pm$ standard deviation over three pivot sets (300 and 600\,K) or nine ($^{*}$1000\,K). $C_{13}$ follows from the curvature $C_{11}+C_{33}+2C_{13}$, and $B$ is the equal-scaling value.}\label{tab:elTiAl}
\footnotesize\setlength{\tabcolsep}{3pt}
\begin{tabular}{@{}rrrrrrrr@{}}\toprule
$T$ [K] & $C_{11}$ & $C_{12}$ & $C_{13}$ & $C_{33}$ & $C_{44}$ & $C_{66}$ & $B$\\\midrule
300 & $170.6\pm0.1$ & $93.8\pm0.1$ & $102.1\pm0.0$ & $197.7\pm0.1$ & $80.8\pm0.2$ & $73.2\pm0.1$ & $126.1\pm0.0$\\
600 & $144.1\pm0.2$ & $79.4\pm0.1$ & $89.9\pm0.1$ & $180.1\pm0.3$ & $69.0\pm0.1$ & $60.9\pm0.3$ & $109.7\pm0.0$\\
1000$^{*}$ & $115.1\pm3.0$ & $64.5\pm2.7$ & $76.5\pm1.9$ & $155.5\pm2.8$ & $56.3\pm1.6$ & $46.6\pm1.7$ & $91.5\pm0.2$\\
\bottomrule
\end{tabular}
\end{table}
\begin{table}[!htbp]\centering\small
\caption{\textbf{Bulk modulus: equal-scaling strain against the zero-pressure study (GPa).} Mean of three pivot sets at 300 and 600\,K, of nine at 1000\,K.}\label{tab:elB}
\footnotesize\setlength{\tabcolsep}{3pt}
\begin{tabular}{@{}lrrrrrr@{}}\toprule
 & \multicolumn{3}{c}{NiAl} & \multicolumn{3}{c}{$\gamma$-TiAl}\\\cmidrule(lr){2-4}\cmidrule(lr){5-7}
$T$ [K] & 300 & 600 & 1000 & 300 & 600 & 1000\\\midrule
equal-scaling strain & 152.76 & 144.47 & 134.45 & 126.13 & 109.74 & 91.52\\
zero-pressure study & 152.75 & 144.44 & 134.52 & 126.15 & 109.24 & 91.38\\
\bottomrule
\end{tabular}
\end{table}
\emph{Comparison with the potential.} Table~\ref{tab:elpot} compares the $\gamma$-TiAl constants at 300\,K with the zero-temperature constants of the potential (Table~\ref{tab:elstat}). Ours lie 7--13\,\% below them, as expected from thermal softening, and the Voigt bulk modulus $(2C_{11}+2C_{12}+C_{33}+4C_{13})/9$ of the published constants is 141.0\,GPa, against our static 141.5\,GPa and our 126.1\,GPa at 300\,K.
\begin{table}[!htbp]\centering\small
\caption{\textbf{$\gamma$-TiAl elastic constants at 300\,K and the zero-temperature constants of the potential (GPa).} ``Potential'' is the published zero-temperature value of Table~\ref{tab:elstat}. Percentages are (TT $-$ potential)/potential.}\label{tab:elpot}
\footnotesize\setlength{\tabcolsep}{4pt}
\begin{tabular}{@{}lrrr@{}}\toprule
 & TT, 300\,K & potential, 0\,K & TT$-$potential\\\midrule
$C_{11}$ & 170.6 & 195 & $-24.4$ ($-13$\,\%)\\
$C_{12}$ & 93.8 & 107 & $-13.2$ ($-12$\,\%)\\
$C_{13}$ & 102.1 & 113 & $-10.9$ ($-10$\,\%)\\
$C_{33}$ & 197.7 & 213 & $-15.3$ ($-7$\,\%)\\
$C_{44}$ & 80.8 & 92 & $-11.2$ ($-12$\,\%)\\
$C_{66}$ & 73.2 & 84 & $-10.8$ ($-13$\,\%)\\
\bottomrule
\end{tabular}
\end{table}

\paragraph{Cost.} The molecular-dynamics and Frenkel--Ladd campaign (volume run, Einstein-spring constants, ten Frenkel--Ladd nodes and NVT runs at 600--1200\,K) took 2\,h 34\,min (NiAl) and 2\,h 17\,min ($\gamma$-TiAl) on 64 cores with 32\,GB of memory each. The cost for the five temperatures 600--1000\,K is read from the LAMMPS run logs (loop time of each segment times 64 ranks, for the volume run, the spring constants, the Frenkel--Ladd integration and the five NVT runs): $4.63\times10^{5}$ core-seconds (128.6 core-hours) for NiAl and $4.11\times10^{5}$ (114.2 core-hours) for $\gamma$-TiAl; for all seven temperatures as run it is $5.92\times10^{5}$ and $5.25\times10^{5}$. The tensor-train calculation over the five temperatures took $1.11\times10^{3}$ (NiAl) and $0.83\times10^{3}$ ($\gamma$-TiAl) core-seconds for one run, read from the run logs (three volumes per temperature; Table~\ref{tab:cost}), a factor of 417 and 495 fewer core-seconds; the mean of three runs from different starting pivots costs three times as much ($3.3\times10^{3}$ and $2.5\times10^{3}$ core-seconds), a factor of 139 and 165. The volumes added for the zero-pressure quantities cost a further 2\,946 and 2\,465 core-seconds at 600--1000\,K and 3\,074 and 2\,295 at 300--500\,K; they are not included in these ratios. The elastic-constant study cost about 2\,350 core-seconds for the amplitude test (46 runs), 12\,291 core-seconds for the 216 production runs and 8\,946 core-seconds for the 144 extra runs at 1000\,K.


\clearpage
\section{Thermodynamic-limit integrals and the 108-atom TiAl cell}\label{sec:bulk}
\paragraph{Integrals.} For $N_c\to\infty$ the sums over wave vectors of the cubic term, of the cross term of $F_{4b}$, of $\operatorname{tr}M^2$ and of the tetrahedron $K_4$ become integrals over the Brillouin zone. Per unit cell, $\Sigma_3=\int\!\!\int d\mathbf k_1d\mathbf k_2\sum_{j}|V(\mathbf k_1j_1,\mathbf k_2j_2,-\mathbf k_1-\mathbf k_2\,j_3)|^2$ and $F_b=-\tfrac\beta{12}\Sigma_3$ (the Brillouin zone has unit volume in fractional wave vectors). With $h_{jl}(\mathbf k_c,\mathbf k_i)=\sum_{ab}V_{jab}V^*_{lab}$ and $Q_{p,j\alpha\beta}(\mathbf k_c,\mathbf k_i)=\sum_{ab}V_{jab}\,\mathrm dG^*_{p,a\alpha}(\mathbf k_i)\,\mathrm dG^*_{p,b\beta}(-\mathbf k_c-\mathbf k_i)$, where $\mathrm dG(\mathbf k)$ is the relative displacement of the neighbor $p$ of the reference atom in the wave vector mode $\mathbf k$, $\operatorname{tr}M^2=\int d\mathbf k_c\,|\int d\mathbf k_i\,h|^2$ and the cross term is $\int d\mathbf k_c\sum_pK^{(4)}_{p,\alpha\beta\gamma\delta}|{\textstyle\int} d\mathbf k_i\,Q|^2$ (indices contracted as in the discrete formula); $K_4=\int d\mathbf k_ad\mathbf k_bd\mathbf k_d\,f$ with $f$ the contraction of four vertices (Section~\ref{sec:loc3}). Squares of inner integrals are estimated with two independent inner points and a common outer point. The pair part of $F_{4a}$ per cell is $\sum_{s,p}\sum_{q:|\mathbf D|\le2.1}K^{(4)}_p\cdot G^{\otimes4}_{pq}\cdot K^{(4)}_q$ with the covariance blocks $G_{pq}$ of the relative displacements, from the $8^3$-cell covariance. The points are the first $2^p$ points of a Sobol sequence plus a fixed random shift ($p=8$ with four shifts for the bubble, $p=6$ with two shifts for the cross term, $\operatorname{tr}M^2$ and $K_4$); the shifts are the same at the three volumes. The standard error from the shifts is 0.6\,\% for the bubble at $p=6$, 0.2\,\% at $p=8$, and 1--2\,\% for the cross term and $K_4$. At $\mathbf k=0$ the three translation modes are given zero amplitude. The pressures are $P=-\Delta F/\Delta V$ per cell at $V(1\pm1\%)$ with isotropic scaling. Per-cell formulas: $F_b=-\beta\sum_{\mathbf k_i,j_i}|V|^2/12$, $F_{4a}=-\beta D_4/48$, $F_{4b}=\beta^2\,\mathrm{cross}/8$, $F_{4c}=-\beta^3(K_4+\tfrac32\operatorname{tr}M^2)/24$.

\paragraph{Pressure of the infinite crystal.} For a finite cell the bubble is a sum over $N_c^2/12$ pairs of wave vectors, which grows with the cell. For a cell with $N_c\to\infty$ the sums over wave vectors become integrals over the Brillouin zone, and the vertex $V(\mathbf k_1j_1,\mathbf k_2j_2,\mathbf k_3j_3)$ is evaluated at arbitrary wave vectors from the force constants (the modes at $\mathbf k$ are the eigenvectors of $D(\mathbf k)=\sum_{\mathbf R}\Phi(\mathbf R)e^{i\mathbf k\cdot\mathbf R}$). The bubble per unit cell is then a six-dimensional integral of $\sum_j|V|^2$, and the cross term of $F_{4b}$ and $\operatorname{tr}M^2$ are nine-dimensional integrals of $Q(\mathbf k_c,\mathbf k_i)Q^*(\mathbf k_c,\mathbf k_i')$ (the square of an inner integral over one wave vector), as is the tetrahedron $K_4$. We estimate all of them with digital nets (Sobol points with fixed random shifts, four shifts for the bubble and two for the others, 1\,024 and 128 points per integral), the same points at the three volumes, so that the finite differences in volume are taken between strongly correlated estimates. The real-space sum of $F_{4a}$, $\sum D^{(4)}_{\rm pair}{}^2$ per cell, runs over the pairs within 2.1 cells of a reference pair, since the covariance of relative displacements falls off as $r^{-2}$ and enters four times: raising the radius from 2.1 to 2.5 cells changes the sum by 0.1\,\%, lowering it to 1.5 cells by 4\,\%. The integrals reproduce the sums of the dense code when evaluated on a mesh: for $3^3$ cells the cross term, $\operatorname{tr}M^2$ and the pair $D^{(4)}$ sum agree to 4--6 digits, and the bubble integral gives $-2.181\pm0.003$\,meV per atom at 800\,K, against the extrapolation of the finite cells, $-2.17\pm0.04$\,meV. For the $T^3$ sum the integrals give $-0.198$ and $-0.202$\,meV per atom for two different sets of shifts, against $-0.201\pm0.003$ from the cells (Table~\ref{tab:loc3}), with the three terms $F_{4a}=-0.115$, $F_{4b}=+0.18$ and $F_{4c}=-0.26$\,meV per atom, and the $T^3$ pressure is $+0.0287$ and $+0.0294$\,GPa, against $+0.029$\,GPa for 128 atoms. Table~\ref{tab:bulk} gives the pressure of both alloys at all six temperatures (settings in Section~\ref{sec:bulk}). The calculation is done at one $8^3$-cell covariance grid, which fixes only the range of the force constants, and its cost does not depend on any cell size: 41--90\,s on three cores for one temperature, 2--4 core-minutes. The result agrees with molecular dynamics for 1\,024 atoms within the combined error of the two (the cross noise of 0.004\,GPa and the MD error of 0.002--0.006\,GPa) at all temperatures, with the largest differences $-0.49$\,\% for NiAl at 300\,K ($-0.010$\,GPa) and $+0.17$\,\% for TiAl at 300\,K.

\begin{table}[!htbp]\centering\small
\caption{\textbf{Full pressure of the two alloys in the thermodynamic limit, deterministic calculation against molecular dynamics.} Pressure in GPa at fixed density (volume per atom of the relaxed 0\,K crystal). Ours: self-consistent Gaussian ($P_{\rm HF}$), cubic term ($P_b$) and the three terms of order $T^3$ ($P_{T^3}$), each by differencing at $V(1\pm1\%)$, with the wave-vector sums replaced by integrals over the Brillouin zone and the real-space $T^3$ sum truncated at 2.1 cells; $P$ is their sum. MD: LAMMPS NVT, 20\,ps and 100\,ps, 1\,024 atoms ($\pm1\sigma$ in units of the last digit); $\Delta P$: ours relative to MD in \%. Core-s: processor time for one temperature, ours (three volumes) and one MD run, in s on one core of the same workstation. The self-consistent reference was computed with a $8^3$-cell covariance grid (for NiAl at 800\,K $P_{\rm HF}+P_b$ is 5.082--5.086\,GPa for $8^3$ and 5.085\,GPa for $12^3$ cells, in the same time), the numerical settings are given in the paragraph on the integrals above.}\label{tab:bulk}
\setlength{\tabcolsep}{4pt}
\begin{tabular}{@{}r|rrrrr|r|rr@{}}\toprule
$T$ [K] & MD & $P_{\rm HF}$ & $P_b$ & $P_{T^3}$ & $P$ & $\Delta P$ [\%] & ours & MD\\\midrule
\multicolumn{9}{@{}l}{\emph{B2 NiAl}}\\
100 & 0.6700(21) & 0.6679 & 0.0028 & 0.0001 & 0.6707 & $+0.11$ & 221 & 98\\
200 & 1.3213(20) & 1.3083 & 0.0122 & 0.0007 & 1.3211 & $-0.01$ & 225 & 106\\
300 & 1.9735(22) & 1.9341 & 0.0274 & 0.0022 & 1.9637 & $-0.49$ & 174 & 109\\
400 & 2.6048(41) & 2.5479 & 0.0478 & 0.0049 & 2.6006 & $-0.16$ & 225 & 124\\
600 & 3.8678(40) & 3.7515 & 0.1013 & 0.0143 & 3.8672 & $-0.02$ & 259 & 127\\
800 & 5.1254(55) & 4.9195 & 0.1662 & 0.0291 & 5.1148 & $-0.21$ & 260 & 130\\
\midrule
\multicolumn{9}{@{}l}{\emph{$\gamma$-TiAl}}\\
100 & 0.5057(6) & 0.5032 & 0.0032 & 0.0000 & 0.5065 & $+0.16$ & 224 & 69\\
200 & 1.0090(13) & 0.9975 & 0.0124 & 0.0004 & 1.0103 & $+0.13$ & 209 & 75\\
300 & 1.5116(21) & 1.4858 & 0.0269 & 0.0015 & 1.5142 & $+0.17$ & 125 & 71\\
400 & 2.0189(23) & 1.9705 & 0.0465 & 0.0034 & 2.0204 & $+0.07$ & 128 & 86\\
600 & 3.0284(45) & 2.9240 & 0.0978 & 0.0105 & 3.0323 & $+0.13$ & 130 & 95\\
800 & 4.0423(28) & 3.8628 & 0.1633 & 0.0222 & 4.0482 & $+0.15$ & 136 & 97\\
\bottomrule
\end{tabular}
\end{table}

\paragraph{The 108-atom TiAl cell with the local code.} The local code was extended to $\gamma$-TiAl, which in its two-atom primitive tetragonal cell (in-plane lattice constant $a/\sqrt2$, $c$) has the same two-sublattice index structure as B2, and to arbitrary periodic supercells, whose Gaussian is built from the dynamical matrices on the mesh of the supercell. For the $3\times3\times3$ conventional cells (108 atoms) the local code gives the pressure of the self-consistent Gaussian plus the cubic term (bubble as an explicit sum over the 54 wave vectors of the supercell) within 0.0036\,GPa of the dense code at all six temperatures (3.97419 against 3.97673\,GPa at 800\,K, 0.50158 against 0.50157\,GPa at 100\,K; Table~\ref{tab:tial108}), and the energy per atom to $2\times10^{-5}$\,eV. For a $3\times3\times3$ primitive cell of 54 atoms the Gaussian pressure agrees with the dense code to 0.0026\,GPa (3.8372 against 3.8346\,GPa at 800\,K). One temperature of the 108-atom cell takes 45\,s on three cores.

\begin{table}[!htbp]\centering\small
\caption{\textbf{$\gamma$-TiAl, 108-atom cell: local code against the dense code.} Pressure $P$ of the self-consistent Gaussian plus the cubic term, in GPa. The local code builds the bubble as an explicit sum over the 54 wave vectors of the supercell; about 45\,s per temperature on three cores.}\label{tab:tial108}
\begin{tabular}{@{}rrrr@{}}\toprule
$T$ [K] & local & dense & difference\\\midrule
100 & 0.50158 & 0.50157 & $+0.00001$\\
200 & 1.00044 & 0.99991 & $+0.00053$\\
300 & 1.49820 & 1.49794 & $+0.00026$\\
400 & 1.99875 & 1.99517 & $+0.00358$\\
600 & 2.99029 & 2.98856 & $+0.00173$\\
800 & 3.97419 & 3.97673 & $-0.00254$\\
\bottomrule
\end{tabular}
\end{table}

\clearpage
\begingroup\sloppy

\section{Sublattice factorisation of the harmonic configurational integral of B2 alloys}\label{app:A}
This appendix contains the statement and proof of the sublattice factorisation used in the Letter (``What is proved'').


For the embedded-atom (EAM/alloy) potential
\[
U=\sum_iF_{\alpha_i}(\rho_i)+\frac12\sum_{i\neq j}\phi_{\alpha_i\alpha_j}(r_{ij}),\qquad\rho_i=\sum_{j\neq i}f_{\alpha_j}(r_{ij}),\qquad \hat e_{ij}=\frac{r_i-r_j}{r_{ij}},
\]
the Hessian at the relaxed B2 structure is $H=H^{\rm dir}+H^{\rm emb}$ with
\begin{gather}
H^{\rm dir}_{ij}=-\Big[h_{ij}\,\hat e_{ij}\hat e_{ij}^{\top}+\frac{g_{ij}}{r_{ij}}\big(\mathbb 1_3-\hat e_{ij}\hat e_{ij}^{\top}\big)\Big]\quad(i\neq j),\qquad H^{\rm dir}_{ii}=-\sum_{j\neq i}H^{\rm dir}_{ij},\label{eq:A11dir}\\
g_{ij}=\phi'_{\alpha_i\alpha_j}+F'_{\alpha_i}(\rho_i)f'_{\alpha_j}+F'_{\alpha_j}(\rho_j)f'_{\alpha_i},\qquad
h_{ij}=\phi''_{\alpha_i\alpha_j}+F'_{\alpha_i}(\rho_i)f''_{\alpha_j}+F'_{\alpha_j}(\rho_j)f''_{\alpha_i},\nonumber\\
H^{\rm emb}=\sum_mF''_{\alpha_m}(\rho_m)\,\nabla\rho_m\,\nabla\rho_m^{\top},\qquad
\nabla_{r_j}\rho_m=-f'_{\alpha_j}(r_{mj})\,\hat e_{mj}\ (j\neq m),\quad \nabla_{r_m}\rho_m=\sum_{l\neq m}f'_{\alpha_l}(r_{ml})\,\hat e_{ml},\label{eq:A11emb}
\end{gather}
all functions being evaluated at the lattice distances. In particular $H^{\rm emb}_{jk}\ni F''_{\alpha_m}(\rho_m)f'_{\alpha_j}(r_{mj})f'_{\alpha_k}(r_{mk})\,\hat e_{mj}\hat e_{mk}^{\top}$ for distinct $j,k,m$: the embedding curvature couples all pairs of neighbours of a common centre. $H$ is symmetric, satisfies the acoustic sum rule (three translational zero modes), and is positive semi-definite at the minimum.

\begin{theorem}[Sublattice factorisation of the harmonic configurational integral of B2 NiAl]\label{thmA:NiAl}
Remove the translations by fixing one Al atom, order the remaining $d=3N-3$ degrees of freedom by sublattice, and write
\[
H=\begin{bmatrix}A&C\\C^{\top}&B\end{bmatrix},\qquad A=H_{\rm Ni,Ni},\quad B=H_{\rm Al,Al},\quad C=H_{\rm Ni,Al},\qquad
M=A^{-1/2}CB^{-1}C^{\top}A^{-1/2}.
\]
Assume that the lattice is mechanically stable, $H\succ0$ on these coordinates. Then:
\begin{enumerate}[label=(\alph*)]
\item $A\succ0$, $B\succ0$, the eigenvalues $\mu_j$ of $M$ lie in $[0,1)$, and
\begin{equation}\label{eq:A11det}
\det H=\det A\,\det B\,\det(\mathbb 1-M);
\end{equation}
\item the harmonic configurational integral factorises exactly as
\begin{gather}
Z_{\rm h}=e^{-\beta U_0}\,Z_A\,Z_B\,\det(\mathbb 1-M)^{-1/2},\label{eq:A11Z}\\
Z_A=\Big(\frac{2\pi}{\beta}\Big)^{\frac{3N_{\rm Ni}}2}(\det A)^{-\frac12},\qquad
Z_B=\Big(\frac{2\pi}{\beta}\Big)^{\frac{3N_{\rm Al}-3}2}(\det B)^{-\frac12},\nonumber
\end{gather}
i.e.\ $F_{\rm h}=U_0+F_A+F_B+\frac{\kB T}{2}\ln\det(\mathbb 1-M)$ with $F_{A,B}=-\kB T\ln Z_{A,B}$;
\item the coupling term satisfies
\begin{equation}\label{eq:A11bound}
\operatorname{tr}M\;\le\;-\ln\det(\mathbb 1-M)\;\le\;\operatorname{tr}M+\frac{\|M\|_F^2}{2\,(1-\|M\|_2)},
\end{equation}
so that $Z_{\rm h}=e^{-\beta U_0}Z_AZ_B\,\big[1+\tfrac12\operatorname{tr}M+\mathcal O\big((\operatorname{tr}M)^2\big)\big]$ when $\operatorname{tr}M\ll1$;
\item each of $Z_A$ and $Z_B$ is the integral of a rank-one tensor, in the eigenbasis of $A$ and of $B$ respectively, while the full harmonic Boltzmann tensor is of rank one in the eigenbasis of $H$;
\item the harmonic internal energy is additive over the sublattices and takes the equipartition value, $\avg{U}_{\rm h}-U_0=E_A+E_B=\frac d2\kB T$ with $E_{A,B}=-\partial_\beta\ln Z_{A,B}$; this holds for every $M$, because $\det(\mathbb 1-M)$ does not depend on $\beta$ (for the 128-atom NiAl cell $M$ is not small: $\mu_{\max}=0.987$, Remark~\ref{remA:NiAl});
\item the harmonic pressure is not additive:
\begin{equation}\label{eq:A11P}
P_{\rm h}=-U_0'(V)+P_A+P_B-\frac{\kB T}{2}\,\partial_V\ln\det(\mathbb 1-M),\qquad P_{A,B}=-\partial_VF_{A,B}.
\end{equation}
\end{enumerate}
\end{theorem}

\begin{proof}
\emph{Step 1: Hessian.} Differentiating $U$ twice gives $H=\sum_m\big[F''_{\alpha_m}\nabla\rho_m\nabla\rho_m^{\top}+F'_{\alpha_m}\nabla^2\rho_m\big]+\frac12\sum_{i\neq j}\nabla^2\phi(r_{ij})$. For a radial function $\psi(r_{ij})$ the $(i,j)$ block of $\nabla^2\psi$ is $-\big[\psi''\hat e\hat e^{\top}+(\psi'/r)(\mathbb 1-\hat e\hat e^{\top})\big]$ and the diagonal blocks follow from translation invariance. Collecting the pair-type terms $\phi$ and $F'_{\alpha_i}f_{\alpha_j}+F'_{\alpha_j}f_{\alpha_i}$ gives \eqref{eq:A11dir}, and the remaining term is \eqref{eq:A11emb}. Each term is invariant under a uniform translation, which gives the zero modes.

\emph{Step 2: block structure and Schur complement.} $A$ and $B$ are principal submatrices of $H\succ0$ and hence positive definite. The block factorisation
\[
H=\begin{bmatrix}\mathbb 1&0\\C^{\top}A^{-1}&\mathbb 1\end{bmatrix}\begin{bmatrix}A&0\\0&B-C^{\top}A^{-1}C\end{bmatrix}\begin{bmatrix}\mathbb 1&A^{-1}C\\0&\mathbb 1\end{bmatrix}
\]
gives $\det H=\det A\,\det(B-C^{\top}A^{-1}C)$, and the Schur complement $B-C^{\top}A^{-1}C$, the effective stiffness matrix of the Al sublattice after the Ni displacements have been integrated out, is positive definite. With $X=A^{-1/2}CB^{-1/2}$, $B-C^{\top}A^{-1}C=B^{1/2}(\mathbb 1-X^{\top}X)B^{1/2}$, so $\det(B-C^{\top}A^{-1}C)=\det B\,\det(\mathbb 1-X^{\top}X)$. By Sylvester's determinant identity, which allows the two factors of the product to be exchanged, $\det(\mathbb 1-X^{\top}X)=\det(\mathbb 1-XX^{\top})=\det(\mathbb 1-M)$. Positive definiteness of $\mathbb 1-X^{\top}X$ gives $\sigma_j(X)<1$, and the eigenvalues of $M=XX^{\top}$ are $\mu_j=\sigma_j(X)^2\in[0,1)$. This proves (a).

\emph{Step 3: factorisation.} $Z_{\rm h}=e^{-\beta U_0}(2\pi/\beta)^{d/2}(\det H)^{-1/2}$; inserting \eqref{eq:A11det} and distributing the $d=3N_{\rm Ni}+(3N_{\rm Al}-3)$ factors $(2\pi/\beta)^{1/2}$ gives \eqref{eq:A11Z}.

\emph{Step 4: bounds.} For $\mu\in[0,1)$, $-\ln(1-\mu)=\sum_{k\ge1}\mu^k/k$, so $\mu\le-\ln(1-\mu)\le\mu+\frac12\sum_{k\ge2}\mu^k=\mu+\frac{\mu^2}{2(1-\mu)}\le\mu+\frac{\mu^2}{2(1-\mu_{\max})}$. Summing over $j$ gives \eqref{eq:A11bound}, since $\sum_j\mu_j=\operatorname{tr}M$ and $\sum_j\mu_j^2=\|M\|_F^2$. Exponentiating $-\frac12\ln\det(\mathbb 1-M)$ gives the expansion in (c).

\emph{Step 5: rank one per sublattice.} $Z_A=\int e^{-\frac\beta2u_{\rm Ni}^{\top}Au_{\rm Ni}}\,\mathrm du_{\rm Ni}$, and in the eigenbasis of $A$ the integrand is a product of one-dimensional Gaussians; likewise for $B$ and for $H$.

\emph{Step 6: energy and pressure.} $\avg{U}_{\rm h}=-\partial_\beta\ln Z_{\rm h}$. The factor $\det(\mathbb 1-M)$ does not depend on $\beta$, so $\avg{U}_{\rm h}=U_0-\partial_\beta\ln Z_A-\partial_\beta\ln Z_B=U_0+\frac{3N_{\rm Ni}}2\kB T+\frac{3N_{\rm Al}-3}2\kB T$, which is (e). The factor does depend on $V$ through $A$, $B$ and $C$; differentiating $F_{\rm h}$ with respect to $V$ gives \eqref{eq:A11P}. \end{proof}

\begin{remark}[Numerical evaluation for B2 NiAl]\label{remA:NiAl}
For the 128-atom B2 NiAl cell with the potential of Ref.~\cite{PurjaPunMishin2009} at the lattice parameter of Section~\ref{sec:results}, the blocks of Theorem~\ref{thmA:NiAl} give:
\begin{itemize}
\item $\|C\|_2/\sqrt{\|A\|_2\|B\|_2}=0.76$;
\item $\|M\|_2=\mu_{\max}=0.987$, $\operatorname{tr}M=54.5$ and $-\ln\det(\mathbb 1-M)=85.7$, from 189 non-zero eigenvalues;
\item a coupling free energy $\frac{\kB T}2\ln\det(\mathbb 1-M)=-2.88$\,meV per atom at 100\,K, proportional to $T$;
\item a coupling pressure $-\frac{\kB T}{2}\partial_V\ln\det(\mathbb 1-M)=+0.095$\,GPa at 100\,K, also proportional to $T$, i.e.\ about $14\%$ of the molecular-dynamics pressure at the same $N$, $V$ and $T$.
\end{itemize}
The inter-sublattice coupling is therefore not small. In B2 all eight nearest neighbours of every atom belong to the other sublattice, so $C$ contains the stiffest bonds, and the long-wavelength acoustic modes, in which both sublattices move together, drive $\mu_j$ towards one. By Theorem~\ref{thmA:NiAl}(e,f), a description by two independent sublattice integrals reproduces the harmonic energy exactly but misses the pressure term \eqref{eq:A11P}. The eigenbasis of the full $H$, or of the self-consistent $\Phi^{\star}$ in the Letter, contains the factor $\det(\mathbb 1-M)$ automatically.
\end{remark}


\section{Free-energy expansion: results used and the terms of order $T^3$}\label{app:B}

This appendix states the expansion that is carried out, quotes the classical results on which it rests, and derives the terms that are specific to this work: the three terms of order $T^3$. The Gaussian identities are due to Isserlis, Stein \cite{Stein1981} and Price \cite{Price1958}; the Gibbs--Bogoliubov bound is classical \cite{Bogoliubov1958,Feynman1972}; the self-consistent harmonic approximation (SCHA) goes back to Hooton \cite{Hooton1955} and Werthamer \cite{Werthamer1970}, with its stochastic form due to Errea and co-workers \cite{Errea2014,Bianco2017,Monacelli2021}; the cubic correction is the bubble term of anharmonic lattice dynamics \cite{Maradudin1961,Wallace1972}. These results are quoted without proof.

\subsection{The expansion used in this work, in one place}\label{secB:expansion}

This subsection states exactly which expansion is carried out. Everything else in this Supplement is the proof, or the evaluation, of one of the steps below.

\begin{enumerate}
\item \emph{Coordinates.} The displacements $u$ are measured from the minimum of the potential, the relaxed lattice $x_0$. Every site is a centre of inversion, so $\nabla U(x_0)=0$ and there is no linear term. The axes are not the Cartesian ones: they are the eigenvectors of a trial matrix $\Phi$, scaled by the thermal widths (whitened coordinates $\xi$, $u=S\xi$), and for the embedding part further the density-adapted coordinates $\eta$ of Proposition~\ref{prop:adapted}.
\item \emph{Exact identity, no truncation.} For any trial $\Phi\succ0$ write $U=U_0+\tfrac12u^{\top}\Phi u+W$. Then $Z=Z_\Phi\,\avg{e^{-\beta W}}_\Phi$ exactly. $W$ is the whole difference between the potential and its Gaussian; the potential is \emph{not} expanded in a Taylor series, and $e^{-\beta U}$ is not expanded either (a cubic term in an exponent would make the Gaussian average diverge).
\item \emph{Choice of the Gaussian.} $\Phi=\Phis(V,T)$, the self-consistent (SCHA) Hessian of Theorem~\ref{thmB:scha}. It depends on temperature and volume. With this choice the first and second Hermite components of $W$ vanish (Proposition~\ref{propB:W12}).
\item \emph{Cumulant expansion of $\ln Z$.} 
\[
F=\underbrace{F_\Phi+\avg{W}}_{F_{GB}}\;-\frac\beta2\kappa_2+\frac{\beta^2}{6}\kappa_3-\frac{\beta^3}{24}\kappa_4+\dots,
\]
with $\kappa_n$ the cumulants of $W$ under the SCHA Gaussian (Theorem~\ref{thmB:bubble}). In terms of $Z$ this is a \emph{product} of factors,
$Z=Z_\Phi\,e^{-\beta\avg W}\,e^{\beta^2\kappa_2/2}\,e^{-\beta^3\kappa_3/6}\cdots$, and not a sum of integrals; the product keeps $\ln Z$ proportional to $N$ at every order, which a truncated Taylor sum of $\avg{e^{-\beta W}}$ does not.
\item \emph{Order in $T$ and truncation.} $F_{GB}$ contains $\avg W$ to all orders. The first term that remains is the cubic bubble $F_b=-\tfrac\beta2\|W_3\|^2\propto T^2$ (Theorem~\ref{thmB:bubble}). The next three terms are of order $T^3$ (Proposition~\ref{propB:T3}). The series is cut after them, $F=F_{GB}+F_b+F_{4a}+F_{4b}+F_{4c}+\mathcal O(T^4)$. The series is asymptotic, and the size of the neglected part is estimated, not bounded (Table~\ref{tab:ratio}).
\item \emph{How each quantity is evaluated (embedded-atom potentials).} Every ingredient is a Gaussian average of a derivative of $U$ of order two, three or four. The pair part is evaluated by a three-dimensional Gauss--Hermite rule (Proposition~\ref{prop:pairrule}); the embedding remainder is a tensor train of rank $6$ (rank $10$ for its volume derivative), built by Gaussian-weighted TT-cross, and its Hermite moments are contracted exactly from the cores (Lemma~\ref{lem:contract}). No random sampling is used. Monte Carlo averages appear only as validation.
\item \emph{Not done.} (a) A direct tensor-train quadrature of $e^{-\beta W}$ itself, at a chosen rank, in the manner of the single-element work \cite{Truong2025}. We tested it on 54-atom NiAl at 300\,K (Table~\ref{tab:zero} and Sec.~\ref{secB:zero}); in the eigenbasis it converges with rank, but with the pivots of the reference volume reused a $+2$\,\% pressure bias remains, with pivots rebuilt at every volume the pressure scatters by 4 to 15\,\% between seeds, and the pivot matrices become ill-conditioned; it is not used for any result in the Letter. (b) A Taylor expansion of $U$ or of $e^{-\beta U}$ about the lattice. (c) Any treatment of several minima or non-Gaussian wells: the method assumes one well, whose thermal ellipsoid is the SCHA Gaussian.
\end{enumerate}

\subsection{Classical results used}

Throughout this appendix $\mu=\mathcal N(0,C)$ on $\R^{d}$ with $C=SS^{\top}\succ0$ (the three translations removed), $u=S\xi$ with $\xi\sim\mathcal N(0,I_d)$, and $\mathrm{He}_\alpha(\xi)=\prod_k\mathrm{He}_{\alpha_k}(\xi_k)$ are the probabilists' Hermite polynomials, the orthogonal polynomials of the standard normal weight.

\begin{lemma}[Gaussian identities \cite{Stein1981,Price1958}]\label{lemB:stein}
For $g$ with polynomially bounded weak derivatives:
\begin{enumerate}[label=(\alph*)]
\item (Isserlis--Wick) odd moments of $\mu$ vanish and $\avg{u_{i_1}\cdots u_{i_{2n}}}=\sum_\pi\prod_{\{a,b\}\in\pi}C_{i_ai_b}$, the sum running over the pairings $\pi$ of the $2n$ factors;
\item (Stein) $\avg{u\,g(u)}=C\,\avg{\nabla g(u)}$ and, for every multi-index $\alpha$ and $f(\xi)=g(S\xi)$, $\avg{\mathrm{He}_\alpha(\xi)\,f(\xi)}=\avg{\partial^\alpha f(\xi)}$; consequently $\avg{\nabla^2g}=S^{+\top}\avg{(\xi\xi^{\top}-I)f}S^{+}$ and $\avg{\nabla^3g}=\avg{\mathrm{He}_3(\xi)f}\times_1S^{+\top}\times_2S^{+\top}\times_3S^{+\top}$;
\item (Price) for a symmetric change $\mathrm dC$ of the covariance, $\mathrm d\avg{g}_C=\tfrac12\operatorname{tr}\big(\mathrm dC\,\avg{\nabla^2g}_C\big)$.
\end{enumerate}
\end{lemma}

\begin{lemma}[Hermite (Wiener) chaos \cite{Wiener1938,CameronMartin1947}]\label{lemB:chaos}
$\avg{\mathrm{He}_\alpha\mathrm{He}_\gamma}=\alpha!\,\delta_{\alpha\gamma}$, and every $f\in L^2(\mu)$ has the orthogonal expansion $f=\sum_{n\ge0}f_n$, $f_n=\sum_{|\alpha|=n}\avg{\mathrm{He}_\alpha f}\,\mathrm{He}_\alpha/\alpha!$. If $f$ has polynomially bounded derivatives up to order $n$, then $\|f_n\|^2=\frac1{n!}\sum_{i_1,\dots,i_n}\avg{\partial_{i_1}\cdots\partial_{i_n}f}^2$.
\end{lemma}

\begin{theorem}[Gibbs--Bogoliubov inequality \cite{Bogoliubov1958,Feynman1972}]\label{thmB:GB}
For a trial matrix $\Phi\succ0$, $C=\kB T\,\Phi^{-1}$ and $U_\Phi(u)=U_0+\tfrac12u^{\top}\Phi u$,
\begin{equation}\label{eq:B-FGB}
F\le F_{GB}(\Phi)=F_\Phi+\avg{U-U_\Phi}_\Phi=\avg{U}_C-\frac{\kB T}{2}\ln\det(2\pi e\,C),
\end{equation}
with equality if and only if $U-U_\Phi$ is constant $\mu$-almost everywhere.
\end{theorem}

\begin{theorem}[SCHA equation \cite{Hooton1955,Werthamer1970,Errea2014}]\label{thmB:scha}
In the variable $C$, $\mathrm dF_{GB}=\tfrac12\operatorname{tr}\big(\mathrm dC\,[\avg{\nabla^2U}_C-\kB T\,C^{-1}]\big)$, so that $F_{GB}$ is stationary if and only if $\Phi=\avg{\nabla^2U}_{C}$ with $C=\kB T\Phi^{-1}$.
\end{theorem}

\noindent Whether the fixed-point map $\Phi\mapsto\Phi^{*}(\Phi)=\avg{\nabla^2U}_{\kB T\Phi^{-1}}$ is a contraction near the stationary point is not proved; numerically each iteration reduces the error by a factor of about $0.087$ (Section~\ref{sec:stationarity}). The iteration is symmetrised by projecting onto the lattice-invariant matrices that obey the acoustic sum rule; this leaves the invariant fixed points unchanged and removes round-off symmetry breaking.

\begin{proposition}[Pressure and energy at the optimum]\label{propB:HF}
Let $F^{\star}(V,T)=F_{GB}(\Phis(V,T);V,T)$. Then
\[
P=-\frac{\mathrm dF^{\star}}{\mathrm dV}=-\Big\langle\frac{\partial U(x_0(V)+u)}{\partial V}\Big|_u\Big\rangle_{\Phis},\qquad
\frac{\partial(\beta F^{\star})}{\partial\beta}=\avg{U}_{\Phis}.
\]
\end{proposition}

\begin{proof}
Both identities follow from the envelope theorem: $\partial F_{GB}/\partial C=0$ at $\Phis$, so only the explicit dependences contribute. At fixed $C$, $V$ enters only through $x_0(V)$ in $\avg{U}_C$, and $\beta F_{GB}=\beta\avg U_C-\tfrac12\ln\det(2\pi eC)$ has the explicit $\beta$-derivative $\avg U_C$. \end{proof}

\subsection{The bubble and the terms of order $T^3$}

At $\Phis$ let $W(\xi)=U(x_0+S\xi)-U_0-\frac1{2\beta}|\xi|^2$, so that $F=F_{GB}-\kB T\ln\avg{e^{-\beta(W-\avg W)}}$ exactly (reweighting identity with trial matrix $\Phis$).

\begin{proposition}[First and second chaos vanish]\label{propB:W12}
$W_1=0$ by force balance: every site is a centre of inversion, so $\avg{\nabla U}=0$. $W_2=0$ by Theorem~\ref{thmB:scha}. Indeed $\avg{\partial_k\partial_lW}=\avg{\partial_k\partial_l(U\circ S)}-\kB T\delta_{kl}=(S^{\top}\Phis S)_{kl}-\kB T\delta_{kl}=0$.
\end{proposition}

\begin{theorem}[The bubble term]\label{thmB:bubble}
With $\kappa_n$ the cumulants of $W$,
\[
F=F_{GB}-\frac\beta2\kappa_2+\frac{\beta^2}6\kappa_3-\frac{\beta^3}{24}\kappa_4+\dots,\qquad
\kappa_2=\operatorname{Var}W=\sum_{n\ge3}\|W_n\|^2,
\]
and the leading term is
\begin{equation}\label{eq:B-Fb}
F_b=-\frac\beta2\|W_3\|^2=-\frac{\kB T}{12}\sum_{klm}X_{klm}^2
=-\frac{(\kB T)^2}{12}\sum_{klm}\frac{\Psi^{*}(q_k,q_l,q_m)^2}{\phi_k\phi_l\phi_m},
\end{equation}
where $X=\beta\Psi^{*}[S,S,S]$ and $\Psi^{*}(q_k,q_l,q_m)=\sum_{abc}\Psi^{*}_{abc}(q_k)_a(q_l)_b(q_m)_c$ in the eigenvectors of $\Phis$. The remaining terms are $\mathcal O(T^3)$.
\end{theorem}

\begin{proof}
The cumulant series of $\ln\avg{e^{-\beta Y}}$ with $Y=W-\avg W$ gives the first line. By Proposition~\ref{propB:W12} and Lemma~\ref{lemB:chaos}, $\operatorname{Var}W=\sum_{n\ge3}\|W_n\|^2$. By Lemma~\ref{lemB:chaos} with $n=3$, and since the quadratic part of $W$ has no third derivative,
\[
\|W_3\|^2=\frac16\sum_{klm}\avg{\partial_{klm}(U\circ S)}^2=\frac1{6\beta^2}\sum X^2 .
\]
Multiply by $-\beta/2$. Substituting $S_{ak}=\sigma_k(q_k)_a$ with $\sigma_k^2=\kB T/\phi_k$ gives the last form.

\emph{Orders.} In $\xi$, $\partial^n(U\circ S)\sim\Psi^{(n)}\sigma^n\propto T^{n/2}$. Hence $\|W_n\|^2\propto T^n$, $\beta\|W_3\|^2\propto T^2$, $\beta\|W_4\|^2\propto T^3$, and the leading part of $\beta^2\kappa_3$, of the form $\beta^2\avg{W_3W_3W_4}$, is $\propto T^{-2}T^{5}=T^3$. The fourth cumulant of $W_3$ enters as $\beta^3\kappa_4(W_3)\propto T^{-3}T^{6}=T^3$ as well, so that three terms are of order $T^3$ (Proposition~\ref{propB:T3}). \end{proof}

\begin{proposition}[The three terms of order $T^3$]\label{propB:T3}
Let $D^{(3)}_{klm}=\avg{\partial_{klm}(U\circ S)}$ (so that $X=\beta D^{(3)}$) and $D^{(4)}_{klmn}=\avg{\partial_{klmn}(U\circ S)}$ be the Gaussian-averaged third and fourth derivatives of the potential in the whitened coordinates $\xi$, in eV. Then
\begin{align}
F&=F_{GB}+F_b+F_{4a}+F_{4b}+F_{4c}+\mathcal O(T^4),\nonumber\\
F_{4a}&=-\frac\beta2\|W_4\|^2=-\frac{\beta}{48}\sum_{klmn}\big(D^{(4)}_{klmn}\big)^2,\qquad
F_{4b}=\frac{\beta^2}{2}\avg{W_3^2W_4}=\frac{\beta^2}{8}\sum D^{(3)}_{aij}D^{(3)}_{apq}D^{(4)}_{ijpq},\label{eq:T3ab}\\
F_{4c}&=-\frac{\beta^3}{24}\kappa_4(W_3)=-\frac{\beta^3}{24}\Big[K_4+\tfrac32\operatorname{tr}M^2\Big],\nonumber\\
K_4&=\sum D^{(3)}_{abc}D^{(3)}_{ade}D^{(3)}_{bdf}D^{(3)}_{cef},\qquad M_{cd}=\sum_{ab}D^{(3)}_{abc}D^{(3)}_{abd}.\label{eq:T3c}
\end{align}
Here $W_n$ is the component of order $n$ of the residual $W$ (the $n$-th Hermite component, Lemma~\ref{lemB:chaos}), $\kappa_4(W_3)=\avg{W_3^4}-3\avg{W_3^2}^2$ is the fourth cumulant (the part of the fourth moment that is not a consequence of the variance) of $W_3$, and $K_4$ is the ``tetrahedron'' contraction in which each of four cubic tensors is joined to each of the other three by exactly one index.
\end{proposition}

\begin{proof}
By Theorem~\ref{thmB:bubble} and orthogonality, $\kappa_2=\|W_3\|^2+\|W_4\|^2+\dots$, $\kappa_3=3\avg{W_3^2W_4}+\dots$ (the term $\avg{W_3^3}$ vanishes by parity), and $\kappa_4=\kappa_4(W_3)+\dots$. Averages of products of Hermite components are evaluated with Wick's theorem (the rule that the Gaussian average of a product of Hermite polynomials is the sum over all pairings of their indices in which no two indices of the same factor are paired). For $\avg{W_3^2W_4}$ the only admissible pairing pattern has one line between the two $W_3$ and two lines from each $W_3$ to $W_4$: there are $3\cdot3\cdot4!=216$ such pairings, and $216/(3!\,3!\,4!)=\tfrac14$. For $\kappa_4(W_3)$ the pairings in which the four factors split into two disconnected pairs are removed by the subtraction $3\avg{W_3^2}^2$; the connected ones are the tetrahedron ($6^4=1296$ pairings) and the pattern in which two pairs of factors are joined by double lines and the two pairs are joined by single lines ($1944$ pairings); division by $(3!)^4=1296$ gives the coefficients $1$ and $3/2$. All three closed forms were also checked against exact Gauss--Hermite quadrature for random symmetric tensors (agreement to $10^{-15}$ relative). The orders in $T$ are those of Theorem~\ref{thmB:bubble}.
\end{proof}

\begin{remark}[Phonon form and derivatives]
With lattice translations, $\Psi^{*}(q_k,q_l,q_m)$ vanishes unless the wave vectors of the three modes (their Bloch vectors) add up to a reciprocal-lattice vector, i.e.\ to zero up to the periodicity of the lattice. \eqref{eq:B-Fb} is then the classical bubble diagram evaluated with the self-consistent force constants in place of the harmonic ones. It is \emph{not} stationary with respect to $\Phi$, so its derivatives include the implicit dependence of $\Phis$ on $V$ and $T$:
\[
P_b=-\frac{\mathrm dF_b}{\mathrm dV},\qquad
E_b=\frac{\mathrm d(\beta F_b)}{\mathrm d\beta}=-F_b\,\frac{\mathrm d\ln\sum X^2}{\mathrm d\ln T}.
\]
In this work both derivatives are evaluated by finite differences of the computed $(V,T)$ data, with $\Phis$ re-solved at every point.
\end{remark}

\subsection{Direct tensor-train quadrature of $e^{-\beta W}$ at several ranks}\label{secB:zero}

The exact identity $F=F_\Phi-\kB T\ln\avg{e^{-\beta W}}_\Phi$ can also be evaluated without a cumulant series, by a tensor-train cross approximation of $e^{-\beta W}$ itself at a chosen rank $r$. We tested this on B2 NiAl, 54 atoms, 300\,K, with pivots chosen adaptively (maximum volume, teneva) and three random starting pivot sets per rank. The pressure is $-\mathrm dF/\mathrm dV$ from the volumes $V(1\pm0.004)$ with the pivots of $V_0$ reused. Three coordinate systems were used: (A) Cartesian displacements from the minimum; (B) the eigenbasis of the point Hessian $H$ at the minimum; (C) the eigenbasis of the SCHA Hessian $\Phis$, with $W$ measured from the SCHA Gaussian. The references are the Monte Carlo value of $\ln\avg{e^{-\beta W}}$ on the same Gaussian ($\ge4\times10^4$ antithetic samples), which gives a pressure of $1.957\pm0.012$\,GPa (B) and $1.957\pm0.011$\,GPa (C), and molecular dynamics at the same $N$ and $V$ (1.9625\,GPa).

\begin{table}[tbp]\centering\small
\caption{Direct tensor-train quadrature of $e^{-\beta W}$ for 54-atom B2 NiAl at 300\,K (means over three seeds; two for arm A where one run failed). Arms: (A) Cartesian displacements; (B) eigenbasis of the point Hessian; (C) eigenbasis of the SCHA Hessian. $\Delta\ln\Phi$ is the difference from the Monte Carlo value on the same Gaussian; $P-P_{\rm MD}$ is relative to $1.9625$\,GPa; ``spread'' is the seed-to-seed standard deviation of $P$ as a percentage of $P$; ``cond.'' is the largest condition number of a pivot matrix; ``core-s'' is the cost per run. NaN means that no value could be formed.}\label{tab:zero}
\begin{tabular}{llrrrrr}
\hline\hline
arm & rank & $\Delta\ln\Phi$ & $P-P_{\rm MD}$ (\%) & spread (\%) & cond. & core-s\\
\hline
A & 1 & $-88.8$ & $+4361$ & 68 & 1 & 56\\
A & 2--6 & $\sim-80$ & NaN & --- & $13$ to $1.5\times10^{6}$ & 213--1757\\
B & 1 & $+0.764$ & $+11.5$ & 0.00 & 1 & 64\\
B & 2 & $+0.125$ & $+2.9$ & 0.97 & $1.6\times10^{3}$ & 182\\
B & 3 & $+0.074$ & $+2.8$ & 0.31 & $3.7\times10^{3}$ & 443\\
B & 4 & $+0.045$ & $+2.2$ & 0.76 & $2.6\times10^{4}$ & 935\\
B & 6 & $+0.007$ & $+2.0$ & 0.55 & $5.1\times10^{5}$ & 2124\\
C & 1 & $+0.780$ & $+10.9$ & 0.00 & 1 & 53\\
C & 2 & $+0.041$ & $+2.6$ & 2.04 & $1.1\times10^{3}$ & 193\\
C & 3 & $-0.028$ & $+2.9$ & 1.24 & $4.5\times10^{3}$ & 542\\
C & 4 & $-0.070$ & $+2.2$ & 0.33 & $2.3\times10^{4}$ & 785\\
C & 6 & $-0.068$ & $+2.0$ & 0.40 & $3.1\times10^{5}$ & 2061\\
\hline\hline
\end{tabular}
\end{table}

In Cartesian axes (A) the train diverges at every rank: $\ln\Phi$ is wrong by about 80 and the pressure is NaN or $44$ times the MD value; in three runs the pivots of $V_0$ gave an exactly singular pivot matrix at the neighbouring volumes, which we count as divergence. In the eigenbasis of the point Hessian (B) the train converges with rank: $\Delta\ln\Phi$ falls from $+0.76$ to $+0.007$ and the pressure error from $+11.5$\,\% to $+2.0$\,\%. At rank 6 the pivot matrices have a condition number of $5\times10^{5}$, and at rank 4 of $2.6\times10^{4}$. In the SCHA eigenbasis (C), $|\Delta\ln\Phi|$ is $0.04$ and $0.03$ at ranks 2 and 3, where the seed spread of the pressure is 2.0\,\% and 1.2\,\%; at ranks 4 and 6 it is $0.07$, with condition numbers of $2\times10^{4}$ and $3\times10^{5}$. The pressure keeps a $+2$ to $+3$\,\% bias at ranks $\ge2$ in both eigenbases, although the Monte Carlo reference on the same Gaussians agrees with MD to $0.3$\,\%. The bias is therefore a property of the tensor-train volume derivative of $\ln\Phi$ (about $0.05$ in $\ln\Phi(V_+)-\ln\Phi(V_-)$), and not of the Gaussian or of the identity; Table~\ref{tab:zeroV} examines whether it depends on reusing the pivots of $V_0$. For comparison, the cumulant route (SCHA plus bubble, without the $T^3$ terms) gives $1.9425$\,GPa ($-1.0$\,\%) for this system. The 42 completed runs cost $2.99\times10^{4}$ core-seconds (53 to 2124 core-s each).

\begin{table}[tbp]\centering\small
\begin{tabular}{llrrrrrr}
\hline\hline
arm & step & pivots & $P$ (GPa) & $P-P_{\rm MD}$ (\%) & $P-P_{\rm MC}$ (\%) & spread (\%) & cond.\\
\hline
B & $\pm0.4$\,\% & $V_0$ reused & $2.009$ & $+2.4$ & $+2.7$ & 0.85 & $1.1\times10^{5}$\\
B & $\pm0.4$\,\% & fresh & $2.018$ & $+2.8$ & $+3.1$ & 15.4 & $6.1\times10^{4}$\\
B & $\pm1.0$\,\% & $V_0$ reused & $2.019$ & $+2.9$ & $+2.7$ & 0.83 & $2.0\times10^{5}$\\
B & $\pm1.0$\,\% & fresh & $1.967$ & $+0.2$ & $+0.1$ & 7.3 & $5.6\times10^{4}$\\
C & $\pm0.4$\,\% & $V_0$ reused & $2.006$ & $+2.2$ & $+2.5$ & 0.55 & $2.2\times10^{5}$\\
C & $\pm0.4$\,\% & fresh & $1.828$ & $-6.9$ & $-6.6$ & 14.0 & $8.5\times10^{4}$\\
C & $\pm1.0$\,\% & $V_0$ reused & $2.032$ & $+3.6$ & $+3.3$ & 3.9 & $5.3\times10^{5}$\\
C & $\pm1.0$\,\% & fresh & $1.919$ & $-2.2$ & $-2.5$ & 3.9 & $1.2\times10^{5}$\\
\hline\hline
\end{tabular}
\caption{Direct tensor-train quadrature of $e^{-\beta W}$ at rank 4, 54-atom B2 NiAl, 300\,K: pivots of $V_0$ reused at $V_0(1\pm\delta)$ against pivots rebuilt by a new cross at each of the three volumes. Means over five seeds (TTSEED 11--15), paired seed by seed; arms and references as in Table~\ref{tab:zero}. The Monte Carlo pressures on the same Gaussians are $1.957$\,GPa (both arms, $\pm0.4$\,\%) and $1.966$ and $1.967$\,GPa (B and C, $\pm1.0$\,\%). ``cond.'' is the largest pivot condition number over the three volumes. The cost per run is about 700\,core-s with reused pivots and 2200--2600\,core-s with fresh pivots.}\label{tab:zeroV}
\end{table}

We then ran the same calculation at rank 4 with a new cross, and hence new pivots, at each of the three volumes, for both arms, five seeds and steps $\pm0.4$ and $\pm1.0$\,\%. The fresh crosses reproduce $\ln\Phi(V_0)$ exactly. At $V_0(1\pm\delta)$ they share only 0.7 to 1.2\,\% of their left pivot index tuples with the pivots of $V_0$. For seed 11, arm B and $\pm0.4$\,\%, replacing the pivots of $V_0$ by fresh ones changes $\ln\Phi$ by $+0.071$ at $V_0(1-0.004)$ and by $-0.142$ at $V_0(1+0.004)$. The relative RMS difference between the train and the exact integrand, on 200 points drawn from the quadrature weights, is 0.82 to 1.00 at every volume and for both arms. With reused pivots the seed-to-seed spread of the pressure is 0.6 to 3.9\,\% and the mean lies $+2.5$ to $+3.3$\,\% above the Monte Carlo value in all four cases (Table~\ref{tab:zeroV}); with fresh pivots the spread is 3.9 to 15.4\,\% and the mean differs from the Monte Carlo value by $+3.1$, $+0.1$, $-6.6$ and $-2.5$\,\%, so its sign and size change from case to case. The seed standard deviation of $\ln\Phi$ at $V_0(1\pm\delta)$ is 0.05 to 0.13 with fresh pivots and 0.02 to 0.09 with reused pivots. With five seeds we cannot say whether reusing the pivots of $V_0$ causes the $+2$\,\% bias seen above. The runs cost $3.98\times10^{4}$ core-seconds in all.


\section{Tensor-train algorithms}\label{app:C}

This appendix gives the tensor-train steps of Section~\ref{sec:tt} that are specific to this work: the exact Gaussian integral and Hermite moments of a train, the Gaussian-weighted two-site cross with structured evaluation, the tensors that are formed, and the rank facts used. The tensor-train (TT) format and TT-SVD are those of Oseledets \cite{Oseledets2011}, TT-cross is that of Oseledets and Tyrtyshnikov \cite{OseledetsTyrtyshnikov2010}, skeleton approximations and the maximum-volume (maxvol) rule are those of Goreinov, Tyrtyshnikov and Zamarashkin \cite{Goreinov1997,Goreinov2001}, and the two-site (DMRG-type) cross is that of Savostyanov and Oseledets \cite{Savostyanov2011}; none of these is restated here. The maxvol routine is that of the \texttt{teneva} package \cite{teneva} with $\epsilon=0.02$.

\subsection{Conventions and the Gaussian integral of a tensor train}

Every tensor in this work is a function sampled on a product grid of Gauss--Hermite nodes, $\mathsf A(i_1,\dots,i_d)=f\big(x^{(1)}_{i_1},\dots,x^{(d)}_{i_d}\big)$, written as a tensor train $\mathsf A(i_1,\dots,i_d)=\mathcal G_1(i_1)\cdots\mathcal G_d(i_d)$ with cores $\mathcal G_k(i)\in\R^{r_{k-1}\times r_k}$, $r_0=r_d=1$, and ranks $r_k$ equal to the ranks of the unfoldings of $\mathsf A$ (the matrices with row index $(i_1,\dots,i_k)$ and column index $(i_{k+1},\dots,i_d)$). The rank across a cut counts the independent couplings between the coordinates on its two sides.

\begin{proposition}[Gaussian integral of a tensor train]\label{propC:int}
For the tensor-product Gauss--Hermite rule with weights $w^{(k)}$ (summing to one on each axis),
\begin{equation}\label{eq:C-int}
\sum_{i_1,\dots,i_d}\Big(\prod_kw^{(k)}_{i_k}\Big)\mathsf A(i_1,\dots,i_d)=E_1E_2\cdots E_d,\qquad E_k=\sum_{i}w^{(k)}_i\,\mathcal G_k(i)\in\R^{r_{k-1}\times r_k},
\end{equation}
at cost $\mathcal O(\sum_kn_kr_{k-1}r_k)$. The weight tensor is itself a rank-one train: a Gaussian is a product of one-dimensional Gaussians in the eigenbasis of its stiffness matrix.
\end{proposition}

\begin{proof}
The sum factorises through the matrix product, $\sum_{i_1\dots i_d}\prod_kw_{i_k}\mathcal G_k(i_k)=\prod_k\big(\sum_iw_i\mathcal G_k(i)\big)$. \end{proof}

\subsection{Gaussian-weighted two-site TT-cross with structured evaluation}

The cores are found by the two-site cross \cite{Savostyanov2011} listed in \hyperref[alg:C1]{Algorithm~C.1}. The one-site interpolative pass that fills the cores after the sweeps is the stable form of the cross formula: each fibre core is reshaped to a matrix, orthogonalised by a QR factorisation $\mathsf C_k=QR$, $r_k$ rows of $Q$ are chosen by maxvol, and $\mathcal G_k=Q\,Q(\mathrm{piv},:)^{-1}$, whose entries are bounded by $1+\epsilon$; this equals $\mathsf C_k\hat{\mathsf A}_k^{-1}$ of the cross formula without forming the possibly ill-conditioned inverse.

\begin{keybox}[title={\textbf{Algorithm C.1: two-site weighted TT-cross}}]
\phantomsection\label{alg:C1}\small
\emph{Purpose:} builds, from a small number of function values, the tensor train of a function multiplied by the square root of the Gaussian quadrature weights (Proposition~\ref{prop:weighted}), with ranks adapted to the function.\\[2pt]
\textbf{Input:} function $f$; nodes and weights $\{x^{(k)},w^{(k)}\}$; $r_{\max}$; tolerance $\tau$; number of sweeps $n_{\rm sw}$.\\
\textbf{Weights:} $s^{(k)}_i=\sqrt{w^{(k)}_i/\max_jw^{(k)}_j}$; the tensor crossed is $g(i)=f(x_i)\prod_ks^{(k)}_{i_k}$.\\
\textbf{Init:} all pivot sets are rank one at the central node ($x=0$).\\
\textbf{Repeat $n_{\rm sw}$ times:}
\begin{enumerate}[leftmargin=14pt]
\item \emph{Left to right,} $k=1,\dots,d-1$:
\begin{enumerate}
\item evaluate the supercore $\mathsf B(\alpha,i,j,\gamma)=g(\mathcal I_{k-1}[\alpha],i,j,\mathcal J_{k+1}[\gamma])$, of size $r_{k-1}\times n_k\times n_{k+1}\times r_{k+1}$;
\item reshape to an $(r_{k-1}n_k)\times(n_{k+1}r_{k+1})$ matrix and compute its SVD $U\Sigma V^{\top}$;
\item set $r_k=\min\big(r_{\max},\#\{\sigma_j>\tau\sigma_1\}\big)$;
\item choose rows by maxvol on $U(:,1{:}r_k)$ and form $\mathcal I_k=\{(\mathcal I_{k-1}[\alpha],i)\}$.
\end{enumerate}
\item \emph{Right to left,} $k=d-1,\dots,1$: the same, with maxvol on $V(:,1{:}r_k)$, forming $\mathcal J_k=\{(j,\mathcal J_{k+1}[\gamma])\}$.
\end{enumerate}
\textbf{Assemble} the cores by the one-site interpolative pass of the text, with the final $\mathcal J_k$.\\
\textbf{Un-weight:} $\mathcal G_k(i)\leftarrow\mathcal G_k(i)/s^{(k)}_i$.\\
\textbf{Cost:} about $2\sum_kr_{k-1}n_kn_{k+1}r_{k+1}\le2d\,n^2r^2$ evaluations per sweep.
\end{keybox}

\noindent The SVD step determines the rank from the data. If the supercore has numerical rank $r$, the SVD reveals it, and maxvol applied to the leading singular vectors yields a quasi-optimal skeleton of that rank. Nearly rank-deficient pivot matrices are thereby excluded by construction.

\paragraph{Role of the weights.} The justification is Proposition~\ref{prop:weighted}. The cross interpolates the tensor supplied to it, with an error controlled in the maximum norm by the maximal-volume bound of skeleton approximations: for a rank-$r$ skeleton with maximal-volume pivots the entrywise error is at most $(r+1)\sigma_{r+1}$ \cite{Goreinov2001}. Weighting by $\prod_k\sqrt{w^{(k)}}$ converts this into control in $L^2(\mu_{\rm grid})$, the norm in which the error of Gaussian moments is measured \eqref{eq:momerr}. The weighting also places the maximum-volume pivots in the bulk of the Gaussian.

\paragraph{Structured evaluation.} The site function has the form $f(\eta)=h(M\eta)$, with $M\in\R^{q\times d}$ the matrix of relative neighbour displacements ($q=3n_{\rm nb}$). Each left pivot $\alpha\in\mathcal I_k$ carries its partial feature sum and weight:
\[
\mathcal L_k[\alpha]=\sum_{j\le k}M_{:,j}\,x^{(j)}_{\alpha_j}\in\R^q,\qquad
\ell_k[\alpha]=\prod_{j\le k}s^{(j)}_{\alpha_j}.
\]
These satisfy the nested updates
\[
\mathcal L_k[(\alpha,i)]=\mathcal L_{k-1}[\alpha]+M_{:,k}x^{(k)}_i,\qquad
\ell_k[(\alpha,i)]=\ell_{k-1}[\alpha]\,s^{(k)}_i,
\]
and similarly $\mathcal R_k[\gamma]$ and $\varrho_k[\gamma]=\prod_{j>k}s^{(j)}_{\gamma_j}$ for the right sets. A supercore entry is then
\[
g(\alpha,i,j,\gamma)=h\big(\mathcal L_{k-1}[\alpha]+M_{:,k}x^{(k)}_i+M_{:,k+1}x^{(k+1)}_j+\mathcal R_{k+1}[\gamma]\big)\,\ell_{k-1}[\alpha]\,s^{(k)}_is^{(k+1)}_j\,\varrho_{k+1}[\gamma],
\]
at cost $\mathcal O(q)$ per entry.

\subsection{Exact Hermite moments of a TT, and their lift to Cartesian space}

\noindent The Hermite moments of orders 0--3 of a tensor train are computed by \hyperref[alg:C2]{Algorithm~C.2}.

\begin{keybox}[title={\textbf{Algorithm C.2: moments of order 0--3 from TT cores}}]
\phantomsection\label{alg:C2}\small
\emph{Purpose:} computes the Gaussian averages of a tensor train weighted by Hermite polynomials of orders 0--3 (Lemma~\ref{lem:contract}) by exact contraction of its cores, without forming the full tensor.\\[2pt]
\textbf{Transfer matrices}, for each core $k$ ($r_{k-1}\times r_k$):
\[
E_k=\sum_iw_i\mathcal G_k(i),\quad A_k=\sum_iw_ix_i\mathcal G_k(i),\quad B_k=\sum_iw_i(x_i^2-1)\mathcal G_k(i),\quad D_k=\sum_iw_i(x_i^3-3x_i)\mathcal G_k(i).
\]
\textbf{Environments:} $\mathcal L_1=1$, $\mathcal L_{k+1}=\mathcal L_kE_k$ (row vectors); $\mathcal R_d=1$, $\mathcal R_{k-1}=E_k\mathcal R_k$ (column vectors). The mean is $\mathcal L_{d+1}$.\\
\textbf{First order:} $H_1[k]=\mathcal L_kA_k\mathcal R_k$.\\
\textbf{Second order:} $H_2[k,k]=\mathcal L_kB_k\mathcal R_k$. For $l>k$, $H_2[k,l]=\mathcal L_kA_kY_k[:,l]$, with the column blocks
\[
Y_{d}=\emptyset,\qquad Y_k=\big[A_{k+1}\mathcal R_{k+1},\;E_{k+1}Y_{k+1}\big]\quad(r_k\times(d-k)).
\]
\textbf{Third order:} carry the row block $W_l=[\mathcal L_kA_kE_{k+1}\cdots E_{l-1}]_{k<l}$ ($(l-1)\times r_{l-1}$), with $W_{l+1}=[W_lE_l;\,\mathcal L_lA_l]$. Then
\begin{align*}
H_3[k,l,m]_{k<l<m}&=(W_lA_l)\,Y_l[:,m], & H_3[k,l,l]_{k<l}&=W_lB_l\mathcal R_l,\\
H_3[l,l,m]_{l<m}&=\mathcal L_lB_lY_l[:,m], & H_3[l,l,l]&=\mathcal L_lD_l\mathcal R_l .
\end{align*}
Fill the symmetric tensor from the ordered entries.\\
\textbf{Cost:} $\mathcal O(dnr^2)$ for the transfer matrices, $\mathcal O(d^2r^2)$ for $H_2$, and $\mathcal O(d^3r)$ for $H_3$ (one matrix product per $l$).
\end{keybox}

\noindent By Lemma~\ref{lemB:stein}(b), $H_1$, $H_2$ and $H_3$ are, on the grid, the averaged first, second and third derivatives of the site function with respect to the density-adapted coordinates $\eta$. They are lifted to Cartesian displacements in three steps:
\begin{enumerate}[label=(\roman*)]
\item \emph{Rotate back:} $\xi=V\eta$, so $\avg{\nabla^n_\xi}=V^{\otimes n}\avg{\nabla^n_\eta}$.
\item \emph{Un-whiten:} $u=S\xi$, so $\avg{\nabla_u^n}=(S^{+\top})^{\otimes n}\avg{\nabla^n_\xi}$ on the non-translational subspace, with $S^{+\top}=Q\,\mathrm{diag}(\sigma)^{-1}$. In the implementation this matrix is stored as \texttt{A}$=Q/\sigma$ and applied as \texttt{AV}$=S^{+\top}V$.
\item \emph{Orbit sum:} $\sum_{t\in\mathcal T}P_t^{\otimes n}(\cdot)$ over the lattice translations (Lemma~\ref{lem:orbit}).
\end{enumerate}
The Hessian contribution is $\Phi^{*}_{\rm rem}=\sum_{m\in{\rm reps}}\frac{|\mathcal O_m|}{|\mathcal T|}\sum_tP_t(\mathrm{AV})H_2(\mathrm{AV})^{\top}P_t^{\top}$, and $\Psi^{*}_{\rm rem}$ is analogous with three factors.

\subsection{The tensors used in this work}

Table~\ref{tab:Ctensors} lists every tensor that is formed, entry by entry, together with its contraction.

\begin{table}[!htbp]\centering\small
\renewcommand{\arraystretch}{1.25}
\caption{Explicit definition of every tensor of the method. Left: the physical object; middle: the tensor, i.e.\ the function sampled on the product grid of Gauss--Hermite nodes, written entry by entry; right: the contraction that turns it into a physical quantity and the algorithm used. $\mathcal N(m)$ is the set of neighbours of site $m$, $M_j$ the matrix that maps density-adapted coordinates to the relative displacement of atoms $m$ and $j$, and $x^{(k)}_{i}$ the $i$-th node of coordinate $k$.}\label{tab:Ctensors}
\begin{tabular}{@{}>{\raggedright\arraybackslash}p{0.2\textwidth}>{\raggedright\arraybackslash}p{0.36\textwidth}>{\raggedright\arraybackslash}p{0.38\textwidth}@{}}
\toprule
object & tensor & contraction\\
\midrule
trial Gaussian & $\mathsf W(i_1,\dots,i_d)=\prod_kw_{i_k}$ in the eigenbasis of $\Phi$ (rank 1) & $Z_\Phi$ in closed form:\newline $e^{-\beta U_0}\prod_k\sqrt{2\pi\kB T/\phi_k}$\\
pair $p=(i,j,R)$ & $\mathsf P_p(g_1,g_2,g_3)=\psi_p\big(L_p(x_{g_1},x_{g_2},x_{g_3})^{\top}\big)$, size $n_g^3$ & $\langle\mathsf P_p,w\otimes w\otimes w\rangle$, and the Stein moments \eqref{eq:pair2}--\eqref{eq:pairV}\\
site $m$ & $\mathsf F_m(i_1,\dots,i_d)=\tilde F_m\Big(\sum_{j\in\mathcal N(m)}f_j\big(|R_{mj}+M_j\eta_{\mathbf i}|\big)-\bar\rho_m\Big)$, with $\eta_{\mathbf i}=(x^{(1)}_{i_1},\dots,x^{(d)}_{i_d})$ & TT by \hyperref[alg:C1]{Algorithm~C.1}, ranks $\le r_{\max}$; moments by \hyperref[alg:C2]{Algorithm~C.2}\\
site $m$, pressure & $\mathsf G_m(\mathbf i)=\big(F'(\rho_m)-F'(\bar\rho_m)\big)\,\partial_V\rho_m$ at the same points & its mean, by \hyperref[alg:C2]{Algorithm~C.2}\\
bubble & $\Psi^{*}\in\R^{3N\times3N\times3N}$ (dense, assembled) & $F_b=-\tfrac{\beta}{12}\langle\Psi^{*},C^{\otimes3}\Psi^{*}\rangle=-\tfrac{\kB T}{12}\sum X^2$ (Lemma~\ref{lem:wick})\\
\bottomrule
\end{tabular}
\end{table}

Here $M_j=(S_{j,:}-S_{m,:})V\in\R^{3\times d}$ maps density-adapted coordinates to the relative displacement $w_{mj}=u_j-u_m$. The grids have 9 nodes on the 10 leading cores and 5 on the others, and $d=3N-3$ is 159 (NiAl, 54 atoms), 321 (TiAl, 108) or 381 (NiAl, 128).

\subsection{Rank facts used in the construction}

\begin{proposition}[Rank of all-to-all bilinear couplings]\label{propC:bilinear}
Let $f(\xi)=a(\xi_L)+b(\xi_R)+\xi_L^{\top}B\,\xi_R$ across a cut $L|R$. Then
\[
\operatorname{rank}B\;\le\;\operatorname{rank}f^{\langle L\rangle}\;\le\;\operatorname{rank}B+2 .
\]
For a generic dense $B$, as arises for couplings between delocalised modes, the rank is therefore $\min(|L|,|R|)$ up to an additive constant of at most 2.
\end{proposition}

\begin{proof}
Upper bound: $f^{\langle L\rangle}=a\otimes\mathbf 1+\mathbf 1\otimes b+\sum_r(U\Sigma)_{:,r}^{\top}\xi_L\otimes V_{:,r}^{\top}\xi_R$ from the SVD $B=U\Sigma V^{\top}$.
Lower bound: apply $\partial_{\xi_l}\partial_{\xi_m}$ for $l\in L$, $m\in R$. This annihilates $a$ and $b$ and maps the unfolding to $B$; mixed differentiation (or the corresponding finite differences on a grid) cannot increase the rank. \end{proof}

\begin{proposition}[Exact TT of a function of a separable sum]\label{propC:binom}
The cores \eqref{eq:binomcore} of Lemma~\ref{lem:sepsum}(a) satisfy
\[
e_0^{\top}\mathcal G_1(\eta_1)\cdots\mathcal G_k(\eta_k)=\big(1,s_k,s_k^2,\dots,s_k^p\big),\qquad s_k=\sum_{j\le k}h_j(\eta_j).
\]
For $\tilde F$ quadratic ($p=2$), the TT has rank 3 at every bond.
\end{proposition}

\begin{proof}
Induction on $k$ with the binomial theorem, as in the proof of Lemma~\ref{lem:sepsum}. \end{proof}

\noindent Figure~\ref{fig:ttprofile} illustrates these properties for the computed tensors. At every bond of the site tensor (NiAl, 128 atoms, 300\,K) the three leading singular values carry almost all of the $L^2(\mu)$ norm, consistent with $\tilde F\approx\tfrac12F''\delta\rho^2$ and Proposition~\ref{propC:binom}. The remaining singular values sit at the $10^{-2}$ level and represent the non-separable remainder $R$ of Proposition~\ref{prop:adapted}. It is this part that a low-rank representation truncates. Beyond the first $\sim230$ density-adapted coordinates the tensor is independent of the remaining coordinates, and the rank falls to one.

\begin{figure}[!htbp]\centering
\includegraphics[width=\textwidth]{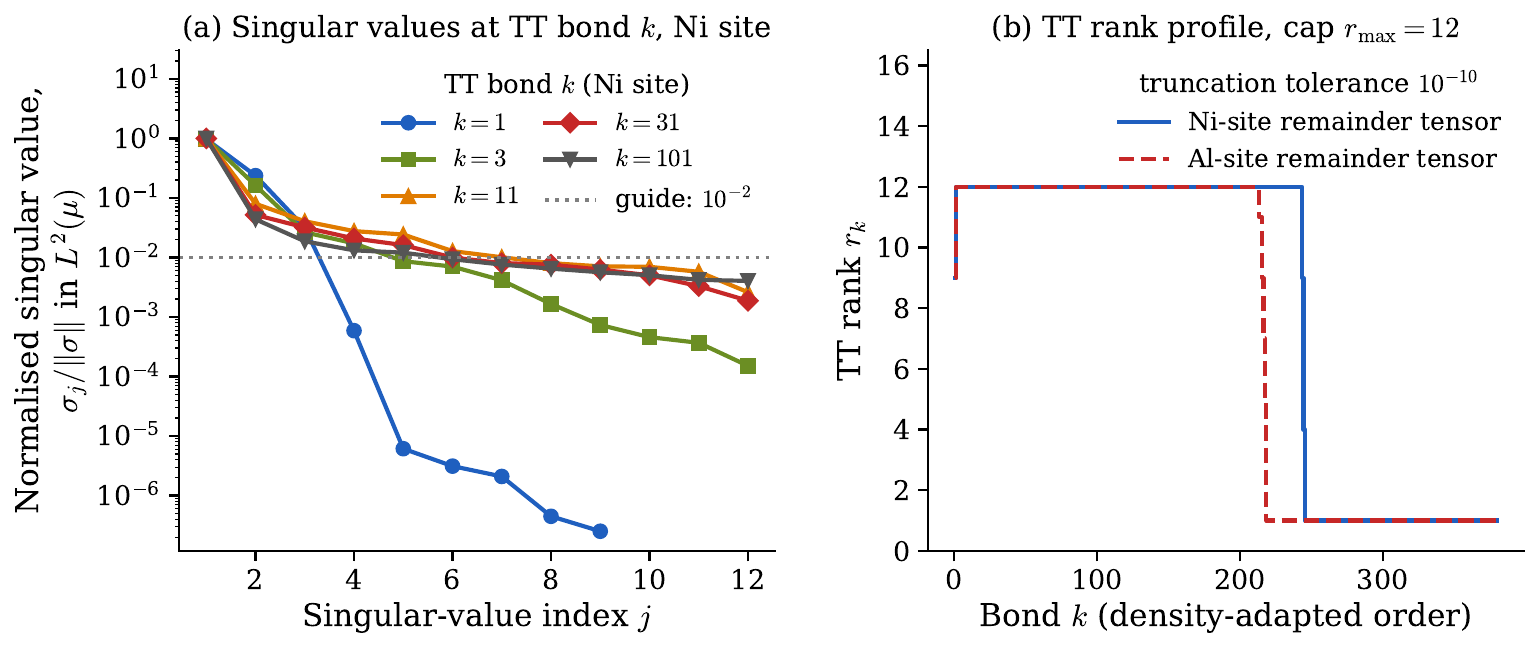}
\caption{The embedding-remainder tensor of the Ni and Al sites in NiAl (128 atoms, 300\,K, $d=381$), built with $r_{\max}=12$ and truncation tolerance $10^{-10}$. These are the settings of this analysis only; the production calculations of Section~\ref{sec:results} use $r_{\max}=6$ ($10$ for the pressure integrand) and tolerance $10^{-6}$. (a) Normalised singular values in $L^2(\mu)$ of the Ni-site tensor at the bonds $k=1,3,11,31,101$, after orthogonalisation of the TT; the dotted line is a guide at $10^{-2}$. (b) TT ranks $r_k$ along the train, in density-adapted order, for the Ni site (solid blue) and the Al site (dashed red).}
\label{fig:ttprofile}
\end{figure}

\subsection{Complexity of one TT-SCHA evaluation}

The cost of the individual steps is summarised in Table~\ref{tab:cost}.

\begin{table}[!htbp]\centering\small
\caption{Computational cost of the thermodynamic-limit calculation for B2 NiAl at 800\,K ($n_g=24$, $r_{\max}=6$), one processor core, per volume; the three volumes run in parallel. Columns: step of the algorithm; number of operations or potential evaluations as a function of the dimension $d$, the number $n_g$ of quadrature nodes per axis, the TT rank $r$, the number $n$ of nodes per TT core, the number $n_{\rm sw}$ of cross sweeps, the number $n_{\rm orb}$ of site orbits and the number $n_{\rm nb}$ of neighbours per site; measured cost. The deterministic linear algebra is negligible except for the dense third-order steps of the bubble.}\label{tab:cost}
\begin{tabular}{@{}lll@{}}
\toprule
step & operations & NiAl, 800\,K, per volume\\
\midrule
eigen-decomposition of $\Phi$ & $\mathcal O(d^3)$ & $<0.1$\,s\\
pair rule & $n_{\rm orb}n_{\rm nb}n_g^3$ evaluations & 3\,s\\
site TT-cross, pivots stored & one-site pass of $d$ cores & 12\,s for the self-consistent iteration\\
vertex train (third moments) & $\mathcal O(d\,r^2)$ per vertex & 2.4--4\,s\\
bubble, $6$-D integral of $\sum|V|^2$ & $2^{13}\times4$ vertices & 21\,s\\
$T^3$: pair sum $D^{(4)}$ & pairs within 2.1 cells & 8\,s\\
$T^3$: cross term and $\operatorname{tr}M^2$ & $9$-D integral & 6\,s\\
$T^3$: tetrahedron $K_4$ & $9$-D integral & 10\,s\\
\bottomrule
\end{tabular}
\end{table}


\endgroup

\section*{Code and data availability}\addcontentsline{toc}{section}{Code and data availability}
The Python implementation and the data underlying all tables and figures are available from the corresponding author on request; the files are listed in Table~\ref{tab:code}.
\begin{table}[!htbp]\centering\small
\caption{Files of the implementation and of the data. Left: file name; right: content, with a reference to the statement of this Supplemental Material that the code implements.}\label{tab:code}
\begin{tabular}{@{}p{0.27\textwidth}p{0.68\textwidth}@{}}
\toprule
file & content\\\midrule
\texttt{eam\_numpy.py} & batched re-implementation of LAMMPS \texttt{eam/alloy} (energy, forces, fixed-$u$ volume derivative), validated to $10^{-12}$\,eV\\
\texttt{tt\_gauss.py} & Gaussian from $\Phi$; pair rule (Prop.~\ref{prop:pairrule}); density-adapted site functions (Prop.~\ref{prop:adapted}); Hermite contraction (Lemma~\ref{lem:contract})\\
\texttt{tt\_cross.py} & Gaussian-weighted two-site TT-cross with structured evaluation (Prop.~\ref{prop:weighted})\\
\texttt{tt\_scha.py} & orbit reduction and assembly (Lemma~\ref{lem:orbit}), deterministic SCHA iteration, TT bubble (Lemma~\ref{lem:wick})\\
\texttt{tt\_quartic.py} & terms of order $T^3$: fourth-order Hermite moments of a tensor train, pair fourth derivatives, scalable contractions (Proposition~\ref{propB:T3})\\
\texttt{tt\_local.py}, \texttt{tt\_local3.py} & local formulation: force constants of an atom and its neighbours, covariance by Fourier transform, local self-consistent iteration (cost independent of $N$), local cubic and $T^3$ terms; \texttt{run} accepts stored cross pivots\\
\texttt{tt\_vertex.py}, \texttt{tt\_kint.py}, \texttt{tt\_fast.py} & tensor-train vertex $V(\mathbf k_1j_1,\mathbf k_2j_2,\mathbf k_3j_3)$ by one left-to-right pass, tetrahedron by Sobol net, symmetry-reduced sunset sum for the bubble\\
\texttt{fixes/fast\_pressure.sh}, \texttt{fast\_run.py}, \texttt{fast\_combine.py} & full pressure at three volumes in parallel with stored cross pivots (\texttt{fixes/loc/piv\_n4.pkl}); results in \texttt{fixes/fast/}\\
\texttt{fixes/fl\_node.py}, \texttt{fl\_combine.py} & Frenkel--Ladd integration from an Einstein crystal (independent check of the free energy on the 16-atom cell)\\
\texttt{fixes/t3\_*.py} & drivers, checks of the closed forms against exact quadrature, comparison with the direct free-energy integration (Tables~\ref{tab:t3free}--\ref{tab:t3small})\\
\texttt{run\_tt.py} & driver: \texttt{python run\_tt.py NiAl 4}, \texttt{python run\_tt.py TiAl 3}\\
\texttt{convergence\_study.py}, \texttt{mc\_site.py} & validation against Monte Carlo, convergence in $n_g$ and $r_{\max}$\\
\texttt{results/tt\_*.json}, \texttt{conv\_*.json} & all numbers in Tables~\ref{tab:val}--\ref{tab:diag}\\
\texttt{results/ttscha\_Phi\_*.npy} & converged self-consistent Hessians at every $(T,V)$\\
\bottomrule
\end{tabular}
\end{table}
